\documentclass[12pt,aps,prd,showpacs,amsmath,amssymb,floatfix,nofootinbib,superscriptaddress,reprint]{revtex4-1}
\usepackage{graphicx} 
\usepackage{multirow,bigdelim}
\usepackage[dvipsnames]{xcolor}
\usepackage[colorlinks,linkcolor=blue,citecolor=Green]{hyperref}

\usepackage{xpatch}

\usepackage[sicmds,freestanding]{hepunits}

\DeclareSIUnit{\fm}{\femto\metre}

\newcommand{\olsi}[1]{\,\overline{\!{#1}}} 

\newcommand{\LHCb}{\textsc{LHC}b\,}

\usepackage{latexsym}
\usepackage{amsfonts}
\usepackage{amsmath,amssymb}
\usepackage{slashed}
\usepackage{color}

\usepackage{supertabular} 
\usepackage{epsfig}
\usepackage{lipsum}
\usepackage{orcidlink}
\usepackage{accents}
\usepackage[inline]{enumitem}

\usepackage{etoolbox} 
\AtBeginEnvironment{pmatrix}{\everymath{\displaystyle}}

\newcommand{\mcov}{\mathring{m}_{\chi_{c1}}}

\newcommand{\mcot}{\mathring{m}_{\chi_{c2}}}

\newcommand{\mco}{\mathring{m}_{\chi}}

\newcommand{\mcos}{\mathring{m}_{\chi_{c0}}}

\newcommand{\drmd}[1]{m_{D,#1}}
\newcommand{\drmds}[1]{m_{D^\ast,#1}}

\newcommand{\ac}{\olsi{c}}

\newcommand{\QM}{\text{QM}}
\newcommand{\cm}{\text{cm}}

\newcommand{\uestc}{\affiliation{School of Physics, University of Electronic Science and Technology of China, Chengdu 611731, China}}

\newcommand{\ific}{\affiliation{Instituto de F\'isica Corpuscular (centro mixto CSIC-UV), \\
Institutos de Investigaci\'on de Paterna, Apartado 22085, 46071, Valencia, Spain}}

\newcommand{\itp}{\affiliation{CAS Key Laboratory of Theoretical Physics, Institute of Theoretical Physics, \\Chinese Academy of Sciences, Beijing 100190, China}}

\newcommand{\ucas}{\affiliation{School of Physical Sciences, University of Chinese Academy of Sciences, Beijing 100049, China}}

\newcommand{\peng}{\affiliation{Peng Huanwu Collaborative Center for Research and Education, Beihang University, Beijing 100191, China}}

\newcommand{\scnt}{\affiliation{Southern Center for Nuclear-Science Theory (SCNT), Institute of Modern Physics,\\ Chinese Academy of Sciences, Huizhou 516000, China}}

\usepackage{gentium}
\usepackage[T1]{fontenc}
\usepackage[cochineal,bigdelims,vvarbb]{newtxmath}

\usepackage[normalem]{ulem}

\begin{document}

\title{\boldmath $P$-wave charmonium contribution to hidden-charm states from reanalysis of lattice QCD data}

\author{Pan-Pan Shi\orcidlink{0000-0003-2057-9884}} \email{Panpan.Shi@ific.uv.es}
\ific\itp\ucas

\author{Miguel Albaladejo\orcidlink{0000-0001-7340-9235}}\email{Miguel.Albaladejo@ific.uv.es}
\ific 

\author{Meng-Lin~Du\orcidlink{0000-0002-7504-3107}}\email{du.ml@uestc.edu.cn} \uestc

\author{Feng-Kun~Guo\orcidlink{0000-0002-2919-2064}}
\email{fkguo@itp.ac.cn}
\itp\ucas\peng\scnt

\author{Juan Nieves\orcidlink{0000-0002-2518-4606}} \email{jmnieves@ific.uv.es}
\ific 

\newcommand{\miguel}[1]{{\color{Red} [\textbf{MA:} #1]}}
\begin{abstract}

We reanalyze, considering the contribution of $P$-wave charmonia, lattice data for the $D \olsi{D}$--$D_s \olsi{D}_s$ coupled-channel of S. Prelovsek et al. [JHEP 06, 035 (2021)] and $D \olsi{D}{}^\ast$ systems of S. Prelovsek et al. [Phys. Rev. Lett. 111, 192001 (2013)]  with $m_{\pi}\simeq 280$ and $266$ MeV, and $L=24a/32a$ ($a\simeq 0.09$ fm) and $L=16a$ ($a\simeq0.1239(13)$ fm), respectively. The hidden-charm states with  $J^{PC}=0^{++}$, $1^{++}$, and $2^{++}$ quantum numbers are then searched for. For $0^{++}$, the analysis reveals three poles in the $D\olsi{D}$--$D_s \olsi{D}_s$ coupled-channel amplitude, corresponding to three states. Two of these poles, located near the $D\olsi{D}$ and $D_s \olsi{D}_s$ thresholds, can be interpreted as mostly molecular states. A third pole above the $D_s \olsi{D}_s$ threshold is originated from the $P$-wave $\chi_{c0}(2P)$ charmonium state. 
The number of poles found in the $D\olsi D$--$D_s \olsi D_s$ system is the same as that found in the original lattice analysis though the position of the third pole changes sizeably.
In the $1^{++}$ sector, we find two poles in the complex energy plane. The first one is related to the molecular $X(3872)$ state, with a compositeness exceeding $90\%$, while the second one, stemming from the $\chi_{c1}(2P)$ charmonium, appears above the $D \olsi{D}{}^\ast$ threshold and it likely corresponds to the recently discovered $\chi_{c1}(4010)$ state.
In the $2^{++}$ sector, we also report two poles and find that the dressed $\chi_{c2}(2P)$ is lighter than the  $D^*\olsi{D}{}^\ast$ molecular state, with the dynamics of the latter closely related to that of the heavy-quark spin-symmetry partner of the $X(3872)$. Our exploratory study of the $1^{++}$ and $2^{++}$ sectors offers valuable insights into their dynamics, but given that the fits that we carry out are underconstrained, more lattice data are required to draw robust conclusions.

\end{abstract}

\maketitle

\section{Introduction}

Quantum Chromodynamics (QCD) describes the strong interactions between quarks and gluons. However, the phenomenon of \textit{quark confinement}, one of the fundamental properties of QCD, precludes the direct observation of quarks, which are confined into colourless hadrons. Nonetheless, insights into quark confinement can be gained by exploring the properties of hadrons produced in high-energy reactions. In particular, in the last decades many hadrons have been discovered which do not fit into the traditional picture of mesons and baryons as objects composed of a quark-antiquark pair and three quarks, respectively. These new structures are therefore candidates for exotic hadrons, in a broad sense. The investigation of exotic states not only sheds light on the mechanism of the low energy interaction governed by QCD but also provides a unique window into the quark confinement phenomenon. So far, many experimental and theoretical studies have been performed, in a huge endeavour to bolster our understanding of the nature of the exotic states.

Hidden-charm exotic states have rightfully attracted significant attention from both experimental and theoretical perspectives. The discovery by the Belle Collaboration in 2003  of the $X(3872)$~\cite{Choi:2003ue}, also known as $\chi_{c1}(3872)$~\cite{ParticleDataGroup:2024cfk}, was quickly confirmed by other collaborations~\cite{Acosta:2003zx,Abazov:2004kp,Aubert:2004ns}. It marked the beginning of an ever-increasing list of so-called $XYZ$ states, which have revolutionized the charmonium-like spectroscopy (for recent reviews, see Refs.~\cite{Hosaka:2016pey,Esposito:2016noz,Lebed:2016hpi,Ali:2017jda,Olsen:2017bmm,Guo:2017jvc,Albuquerque:2018jkn,Liu:2019zoy,Guo:2019twa,Brambilla:2019esw,Chen:2022asf,ParticleDataGroup:2024cfk,Liu:2024uxn}). The abundance of these charmonium-like states and of the processes in which they are involved offer a wealth of observables which can help in unraveling their nature. Despite this abundance, distinguishing the exotic states from conventional charmonia is a complicated task. First, all quantum numbers (total spin $J$, parity $P$, and charge conjugation $C$) of conventional quark model states can always be produced by exotic state models. Second, the masses of exotic and conventional states also frequently lie close to each other or near a hadronic threshold, which complicates the analysis of the experimental spectrum. Consequently, further investigations are required to understand the nature of these states. 

According to the Review of Particle Physics (RPP)~\cite{ParticleDataGroup:2024cfk}, the $X(3872)$ mass and width are $(3871.64 \pm 0.06)\ \text{MeV}$ and $(1.19 \pm 0.21)\ \text{MeV}$, respectively, and the \LHCb Collaboration~\cite{Aaij:2011sn} has determined its quantum numbers, $J^{PC}=1^{++}$. 
One should notice that the RPP values of the $X(3872)$ mass and width come from averaging values using the Breit-Wigner parameterization, which, however, is questionable for resonances so close to strongly coupled thresholds.\footnote{Extractions of the $X(3872)$ pole using the more proper Flatt\'e parameterization have been done by the LHCb~\cite{LHCb:2020xds} and BESIII~\cite{BESIII:2023hml} Collaborations, which led to an imaginary part of the pole position much smaller than the width quoted in RPP.}
Recently, the Belle Collaboration reported a hint of an isoscalar structure in 
the $\gamma \psi(2S)$ invariant mass distribution via a 
two-photon process~\cite{Belle:2021nuv}. This new structure, with a mass of $(4014.3 \pm 4.0 \pm 1.5)~ \text{MeV}$, a decay width of $(4\pm 11 \pm 6)~ \text{MeV}$ and a global significance of $2.8\sigma$, is located near the $D^*\olsi{D}{}^\ast$ threshold. This proximity suggests its potential as a candidate for a $D^*\olsi{D}{}^\ast$ shallow bound state with quantum numbers $J^{PC}=2^{++}$. Previous works~\cite{Tornqvist:1993ng,Swanson:2005tn,Liang:2010ddf,Nieves:2012tt,Hidalgo-Duque:2012rqv,Sun:2012zzd,Hidalgo-Duque:2013pva,Guo:2013sya,Albaladejo:2013aka,Albaladejo:2015dsa,Cincioglu:2016fkm,Baru:2016iwj,Ortega:2017qmg,Wang:2020dgr,Dong:2021juy} already predicted the existence of such a state, whereas the measured width aligns well with theoretical predictions, despite some uncertainties in these predictions~\cite{Albaladejo:2015dsa,Baru:2016iwj,Shi:2023mer}. Hence, the discovery of this narrow structure raises the possibility of a $2^{++}$ charmonium-like state, whose dynamics would be closely related to that of the heavy-quark spin-symmetry (HQSS) partner of the $X(3872)$~\cite{Nieves:2012tt, Hidalgo-Duque:2012rqv, Albaladejo:2013aka,Guo:2013sya,Cincioglu:2016fkm,Baru:2016iwj,Ji:2022uie,Wu:2023rrp,Shi:2023ntq,Zheng:2024eia}.
In addition, experimental efforts have been devoted to the search for the $X(3872)$ scalar partner,\footnote{Strictly speaking, in the heavy quark limit, there exist two  $S$-wave $D^{(*)}\olsi D{}^{(*)}$ spin multiplets, corresponding to the total angular momentum of the light degrees of freedom to be $s_\ell = 0$ and $1$, respectively. While the $X(3872)$ is in the multiplet with $s_\ell=1$, the $D\olsi D$ pair is a superposition of $s_\ell=0$ and $s_\ell=1$ states (see, e.g., Refs.~\cite{Hidalgo-Duque:2012rqv,Guo:2017jvc}). That is, the $D\olsi D$ interaction cannot be completely fixed by solely using the $X(3872)$ as input.} and some hints suggest its possible existence. Near the $D\olsi{D}$ threshold, a bump reported by the Belle Collaboration in the $e^+e^-\to J/\psi D \olsi{D}$ reaction~\cite{Belle:2007woe} was interpreted as a bound state~\cite{Gamermann:2007mu,Wang:2019evy}. The enhancement just above the $D\olsi{D}$ threshold, observed by the Belle and BaBar Collaborations in the reaction $\gamma\gamma\to D\olsi{D}$~\cite{Belle:2005rte,BaBar:2010jfn}, might also hint at the existence of a possible structure near such threshold.
However, the broad bump above the $D\olsi D$ threshold in the Belle and BaBar data for $\gamma\gamma\to D\olsi D$ could be due to a broad resonance with a mass about 3.84~GeV interpreted as $\chi_{c0}(2P)$~\cite{Guo:2012tv}. As for the bump in the $e^+e^-\to J/\psi D \olsi{D}$ reaction, the $\chi_{c0}(2P)$ was also suggested to be the origin~\cite{Chao:2007it}, and Belle reported the $\chi_{c0}(3860)$ from their analysis~\cite{Belle:2017egg}. Moreover, the BESIII Collaboration unsuccessfully searched for the scalar partner of $X(3872)$ via the reaction $\psi(3770)\to \gamma \eta \eta^{\prime}$, with no significant signals observed in the $\eta\eta^{\prime}$ invariant mass distribution at the $90\%$ confidence level~\cite{BESIII:2023bgk}.
For the $D_s^+D_s^-$ system, the \LHCb collaboration reported a new structure, referred to as $X(3960)$ with a mass of $(3956 \pm 5 \pm 10)$ MeV, decay width of $(43 \pm 13 \pm 8)$ MeV and quantum numbers $J^{P C} = 0^{++}$~\cite{LHCb:2022aki}. This state is a candidate of $D_s^+D_s^-$ molecule~\cite{Dong:2021juy,Ji:2022uie,Ji:2022vdj}.

Lattice QCD (LQCD) can provide first-principle calculations concerning these charmonium-like states. This would prove especially useful given the limited availability of experimental results for the study of the spin-0 and spin-2 partners of the $X(3872)$. In Ref.~\cite{Bali:2011rd}, $c\olsi c$ and $D\olsi{D}{}^\ast$ interpolating operators were employed to calculate the charmonium spectrum with a pion mass about 280~MeV, and a finite-volume energy level was found to be significantly below the $D\olsi D{}^{*}$ threshold, with a splitting of $(88\pm 26)\,\MeV$. However, we should mention that no scattering analysis was performed.
In the subsequent calculation carried out in Ref.~\cite{Prelovsek:2013cra} with a pion mass of $(266\pm4)\,\MeV$, several operator combinations are employed to obtain energy levels. For the results excluding the $J/\psi \omega$ operator (Fig. (1b) of Ref.\,\cite{Prelovsek:2013cra}), an elastic
$S$-wave $D\olsi{D}{}^\ast$ phase shift analysis was performed, and a binding energy of $(11\pm 7)\,\MeV$ for the $X(3872)$ was extracted.
In Ref.~\cite{Padmanath:2015era}, the authors calculated the LQCD energy levels using $c\olsi c$, $D\olsi{D}{}^\ast$ and diquark-antidiquark interpolating operators with the same pion mass as in Ref.~\cite{Prelovsek:2013cra}, and remarkably found that the lattice candidate for the $X(3872)$ vanished when the $c\olsi c$ operator was absent. Besides, the $D\olsi{D}{}^\ast$ operator had a more significant effect on the pole position than the diquark-antidiquark operator.
The $0^{++}$ hidden-charm sector was studied in Ref.~\cite{Lang:2015sba} using $c\olsi c$, $D\olsi{D}$ and $J/\psi\omega$ interpolating operators with pion masses about 266~MeV and 156~MeV. This study revealed a virtual state\footnote{This virtual state was not mentioned in Ref.~\cite{Lang:2015sba}. In Ref.~\cite{Prelovsek:2020eiw}, the authors supplemented a discussion about this state.} below the $D\olsi{D}$ threshold by about $20\,\text{MeV}$. Subsequently, the $D\olsi{D}$--$D_s \olsi{D}_s$ coupled-channel system was calculated in Ref.~\cite{Prelovsek:2020eiw} using  $c\olsi c$, $D\olsi{D}$, $D_s \olsi{D}_s$ and $J/\psi\omega$ interpolators with a pion mass about 280~MeV, and three $0^{++}$ states were found: a bound state below the $D\olsi{D}$ threshold, and two resonances below and above the $D_s\olsi{D}_s$ threshold, respectively.
In Refs.~\cite{Wilson:2023hzu, Wilson:2023anv}, using charmonium and more than other 20 multi-hadron operators with a pion mass about 391~MeV, the Hadron Spectrum Collaboration found a single scalar and one tensor resonance around 4~GeV.
There is also a recent LQCD calculation of the isoscalar $1^{++}$ sector with both $c\olsi c$ and $D\olsi D{}^*$ operators with pion masses ranging from $250$ to $417\,\MeV$ \cite{Li:2024pfg}. They reported a shallow bound state corresponding to the $X(3872)$ below the $D\olsi D{}^*$ threshold. 

Following Ref.~\cite{Hanhart:2019isz}, three hypotheses have been proposed to unravel the nature of charmonium-like states
near the relevant open-charmed thresholds. In view I, charmonia are supposed to exist solely below the first relevant $S$-wave threshold of two charmed mesons, while charmonium-like states close to or above the relevant open-charmed threshold are considered as hadronic molecules. 
View II suggests that, in the Fock space, the wave function of the physical charmonium-like state is constructed by different components, including charmonia, molecular structures, compact tetraquarks, and more. The contributions of these components can influence the masses of charmonium-like states and the relative couplings. Still, the number of these states is the same as that of charmonia.\footnote{Here the charmonium-like states denote states that share the same quantum numbers as charmonia. }
Finally, view III explores the coexistence of hadron-molecular and quark model states.

Focusing on the hidden-charm states with positive charge conjugation, the contributions of the $2P$ charmonia cannot be neglected. Moreover, the LQCD calculation carried out in Ref.~\cite{Padmanath:2015era} showed that the $X(3872)$ disappears when the $c\olsi c$ interpolator is not considered. On the other hand, some of the $2P$ charmonium states predicted by different quark models~\cite{Godfrey:1985xj, Ebert:2002pp, Zeng:1994vj, Workman:2022ynf} still have not been confirmed or identified experimentally. These facts motivates us to understand the nature of the charmonium-like states and to distinguish between the three different hypotheses mentioned above.

In this work, we consider simultaneously the two charmed-meson rescattering and the $2P$ charmonia contributions, and investigate the possible poles in the hidden-charm region with  $J^{PC}=0^{++},~1^{++}$ and $2^{++}$ quantum numbers. Specifically, we construct effective interactions between charmed mesons and  $c\olsi{c}(2P)$ states, respecting HQSS within a formalism similar to that introduced in Ref.~\cite{Cincioglu:2016fkm} (for similar approaches, see also Refs.~\cite{Hanhart:2011jz, Guo:2016bjq, Zhou:2017dwj, Xiao:2023lpv, Wang:2023ovj}). 
The low-energy constants (LECs) of the effective interactions will be fitted performing a reanalysis of the $0^{++}$ and $1^{++}$ LQCD energy levels, determined in  Refs.~\cite{Prelovsek:2020eiw} and~\cite{Prelovsek:2013cra}, respectively, using relatively low pion masses. Detailed information about lattice calculations is listed in Table~\ref{tab:lat_detail}. 
We use L\"uscher's method~\cite{Luscher:1985dn,Luscher:1986pf,Luscher:1990ux}, advantageously reformulated in Refs.~\cite{Doring:2011vk,Doring:2012eu} to analyze LQCD data in finite boxes and to connect with physical observables in the infinite volume. The formalism was generalized for the case of a moving frame  in Ref.~\cite{Rummukainen:1995vs}. Then, with the $D\olsi{D}$--$D_s \olsi{D}_s$ coupled-channel and the $D\olsi{D}{}^\ast$ scattering amplitudes constrained by LQCD, poles in the infinite-volume scattering amplitudes will be studied.  In addition, we will predict a pole in the $D^*\olsi{D}{}^\ast$ system with quantum numbers $2^{++}$.

This manuscript is organized as follows. In Sec.~\ref{Sec:Formalism},
we present the scattering matrix between two charmed mesons in the presence of charmonium states, both for the infinite and finite volume cases. The scattering information for the $0^{++}$ $D\olsi{D}$--$D_s\olsi{D}_s$ coupled-channel system is analyzed in Sec.~\ref{Sec:0++}, while the $D\olsi{D}{}^\ast$ system is discussed in Sec.~\ref{Sec:1++}. In Sec.~\ref{Sec:2++}, we predict the energy levels of the $2^{++}$ $D^*\olsi{D}{}^\ast$ system and give also details of a pole that we find in this sector. Finally, the main conclusions of our work are compiled in Sec.~\ref{Sec:Summary}, while the $D\olsi{D}$--$D_s\olsi{D}_s$ coupled-channel correlation matrix is given for completeness in Appendix \ref{sec:appendix}.

\begin{table}[tbp]
    \centering
    \renewcommand\arraystretch{1.6}
    \caption{Lattice details for  $D\olsi D{}^{*}$~\cite{Prelovsek:2013cra} and  $D \olsi D$--$D_s\olsi D_s$~\cite{Prelovsek:2020eiw} systems, including the quantum numbers $I(J^{PC})$,  total squared momenta $|\vec P|^2$, irreducible representations $\Lambda^{(P)C}$, time-space volumes $T/a\times (L/a)^3$, pion masses $m_{\pi}$, lattice spacings $a$, and interpolating operator types in the used lattice calculations. Since the isovector $D\olsi D{}^{*}$ scattering is not involved in our work, in the last column, we only list the lattice details for isoscalar $D\olsi D{}^{*}$ scattering. In this work, we use the energy levels without the contribution of the $J/\psi \omega$ operator (Fig.~1(b) of Ref.~\cite{Prelovsek:2013cra}). \label{tab:lat_detail}}
    \begin{ruledtabular}
    \begin{tabular}{lcc}
    & $D\olsi D$--$D_s \olsi D_s$~\cite{Prelovsek:2020eiw} & $D\olsi D{}^{*}$~\cite{Prelovsek:2013cra}\\
    \hline
    $I(J^{PC})$ & $0(0^{++}/2^{++})$  & $0(1^{++})$\\
    $|\vec P|^2$ [$(2\pi/L)^2$] & 0, 1, 2 & 0\\
    $\Lambda^{(P)C}$ & $A_1^{++}$, $A_1^+$, $B_1^+$ & $T_1^{++}$ \\
    $T/a\times (L/a)^3$ & $128\times 24^3/96\times 32^3$ & $32\times 16^3$ \\
   $m_{\pi}$ [MeV]  & $280(3)$ & $266(4)$ \\
    $a$ [fm] & $0.08636(98)(40)$ & $0.1239(13)$\\
    Interpolators  & $c\olsi c$, $D\olsi D$, $D_s \olsi D_s$, $D^* \olsi D{}^{*}$, $J/\psi \omega$ & $c \olsi c$, $D \olsi D{}^{*}$\\
    \end{tabular}
    \end{ruledtabular}
\end{table}

\section{Formalism}\label{Sec:Formalism}

\subsection[Coupled-channel $T$-matrix]{\boldmath Coupled-channel $T$-matrix }

In this section, we review the general formalism~\cite{Cincioglu:2016fkm} for the S-wave scattering of two charmed mesons, including the contribution from interactions driven by the existence of bare $P$-wave charmonium states. We start from the leading order (LO) two-charmed meson potential in a derivative expansion respecting HQSS~\cite{Nieves:2012tt,Hidalgo-Duque:2012rqv,Guo:2013sya}.

We generally consider an $n+1$ coupled-channel system, where the first $n$ channels stand for the charmed-anticharmed pair of mesons, and the additional one accounts for the $P$-wave charmonium state. The $S$-wave scattering amplitude at a given total energy $E$ in the center-of-mass (c.m.) frame is expressed through the on-shell unitary $T$-matrix, as detailed in Ref.~\cite{Cincioglu:2016fkm}, 
\begin{align}
\langle p |T(E) | p' \rangle =F_{\Lambda}(p) \left( V(E)^{-1} - G(E) \right)^{-1} F_{\Lambda}(p'),
\label{Eq:T-matrix-coupled}
\end{align}
where $V(E)$ and $G(E)$ denote the potential and two-point function matrices, respectively. The matrix of Gaussian form factors $F_{\Lambda}(p)$ reads:
\begin{align}
F_{\Lambda}(p)=
\begin{pmatrix}
\text{diag}\left[\text{exp}(- p_i^2/ \Lambda^2)\right]_{n\times n} & 0 \\
0  & 1
\end{pmatrix},
\label{Eq:form-factor-coupled}
\end{align}
which is introduced to ensure the exact coupled-channel unitarity for the Gaussian regularized potential to be specified below.
Here $ p_i$ is the nonrelativistic three-momentum in the c.m. frame for the $i$-th channel
\begin{align}
p_i = \sqrt{2\mu_i(E - m_{i,1}-m_{i,2})},
\label{Eq:Momentum_on-shell}
\end{align}
where $m_{i,1}$ and $m_{i,2}$ are the meson masses involved in the channel, and $\mu_i$ the corresponding reduced mass. The matrix potential $V(E)$ is given by
\begin{align}
V(E) = \begin{pmatrix}
\left[V_{\QM}(E)\right]_{n\times n} & \left[V_{c\olsi c}(E)\right]_{n\times 1} \\
\left[V_{c\olsi c}(E)\right]_{1\times n} & 0
\end{pmatrix},
\end{align}
where $V_{\QM}(E)$ represents the two charmed-meson effective potential, which will be discussed below in Subsec.~\ref{sec:4H-potential}, and $V_{c\olsi c}(E)$ accounts for the interaction between the charmed mesons and the $P$-wave charmonium state.
The diagonal matrix $G(E)$ includes the Green's function of the $D^{(*)}_{(s)}\olsi{D}{}^{(*)}_{(s)}$ pair and the propagator of the charmonium state, 
\begin{subequations}\label{eq:Gfun}
\begin{equation}
G(E)=\begin{pmatrix}
\text{diag}\left[G^{ii}_{\QM}(E)\right]_{n\times n} & 0\\
0 &  G_{c\olsi{c}}(E)
\end{pmatrix}\,,
\label{Eq:green-coupled}
\end{equation}
with 
\begin{align}
G_{\QM}^{ii}(E) 
& =\int \frac{d^3 \vec q}{(2\pi)^3}\frac{e^{-2 \vec q\,^2/\Lambda^2}}{E -m_{i,1} - m_{i,2} - \vec q^{\,2}/2\mu_i + i\epsilon}\,, \label{eq:Gfun:channel}\\
G_{c\olsi{c}}(E)&=\frac{1}{E-\mco}\,,\label{eq:Gfun:channel-bare}
\end{align}
\end{subequations}
where $\mco$ denotes the $P$-wave charmonium bare mass. We employ a Gaussian form factor to regulate the ultraviolet (UV) divergence of the loop function $G^{ii}_{\QM}(E)$, whose analytical expression is given in 
Eq.~(43) of Ref.~\cite{Cincioglu:2016fkm}, and consider two cutoff values: $\Lambda = 0.5$ and $1.0\,\text{GeV}$.

Resonances, bound states, and virtual states are associated with poles of the $T$-matrix [\textit{cf.} Eq. \eqref{Eq:T-matrix-coupled}]. The positions of these poles in the complex $E$-plane are determined by solving det$[\mathbb{I}-V(E_r)G(E_r)]=0$, where $E_r$ stands for the complex pole position.
Through analytical continuation, the complex $E$-plane is divided into $2^n$ Riemann sheets (RSs), each labeled by $(\pm,...,\pm)$ according to the signs of the imaginary parts of the c.m. momenta in the two-body channels, and the possible poles can be searched for on each RS. The physical RS corresponds to $(+,...,+)$.
For the $i$-th channel, the relation between the different RSs is
\begin{align}
G_{\QM}^{ii,{\rm II}}(E) = G_{\QM}^{ii,{\rm I}}(E) + 2i \frac{\mu_i p_i}{2\pi}e^{-2p_i^2/\Lambda^2},
\end{align}
where the superscripts I and II denote the Green's function on the first and second RSs, respectively, with respect to the $i$-th channel ($i=1,...,n$). Furthermore, the coupling constant between a pole and the $i$-th channel is obtained from the residues of the $T$-matrix in Eq.~\eqref{Eq:T-matrix-coupled}:
\begin{align}
g_{i}g_{j}=\lim_{E\to E_{r}} (E-E_{r})T_{ij}(E),
\label{Eq:coupling}
\end{align}
where $T_{ij}(E)$ is to be computed on the RS where the pole appears. 

The compositeness condition for a shallow bound state coupled to a single channel was given by Weinberg in the pioneering work~\cite{Weinberg:1965zz}. This condition has been discussed and generalized in several ways (see Refs.~\cite{Baru:2003qq,Gamermann:2009uq,Hyodo:2013nka, Aceti:2014ala, Matuschek:2020gqe,Li:2021cue,Albaladejo:2022sux, Kinugawa:2024kwb} and references therein). We start by noting the following sum rule, fulfilled by the amplitude components evaluated at the pole~\cite{Garcia-Recio:2015jsa}: 
\begin{align}
-1 =\sum_{ij}g_ig_j\left[
\delta_{ij} \frac{\partial G^{ii}_{\QM}(E)}{\partial E}  + 
G^{ii}_{\QM}(E)\frac{\partial V_{ij}(E)}{\partial E}G^{jj}_{\QM}(E) \right]_{E=E_r}\,,
\label{Eq:sum-rule}
\end{align}
where the nonrelativistic Green's function $G^{ii}_{\QM}(E)$ should be on the first RS for a bound state and on the unphysical RS for a resonance.
The weight of the molecular component in the $i$-th channel, $P_{i,r}$, may be 
identified with one of the terms in the above sum rule:
\begin{align}
P_{i,r}=\text{Re}~\left(\widetilde{P}_{i,r}\right) = \text{Re}~\left(-g^2_{i}\left[\frac{\partial G^{ii}_{\QM}(E)}{\partial E}\right]_{E=E_r}\right).
\label{Eq:posibility-coupled}
\end{align}
For a shallow bound state, the quantity $\widetilde{P}_{i,r}$ is real and interpreted as the probability of finding this state in the $i$-th channel. For a resonance, $\widetilde{P}_{i,r}$, which represents the integral of the squared wave function in the $i$-th channel, is a complex number, and hence it has no interpretation as a probability. 

\subsection{\boldmath $T$-matrix in a finite volume}

When the two hadrons are confined within a finite box of size $L$, the meson three-momenta are quantized, and thus the two-point function matrix in Eq. \eqref{Eq:green-coupled} should be replaced by:
\begin{align}
\widetilde{G}(E,L)=\begin{pmatrix}
\text{diag}\left[\widetilde{G}^{ii}_{\QM}(E)\right]_{n\times n} & 0\\
0  & G_{c\olsi{c}}(E,L)
\end{pmatrix},
\label{Eq:green-coupled-FV}
\end{align}
where $\widetilde{G}^{ii}_{\QM}(E,L)$ represents the $i$-th channel Green's function in the finite volume. On the other hand, for short-range interactions, the $L$-dependent terms are exponentially suppressed, and thus the potential remains unchanged between the finite and infinite volume cases~\cite{Doring:2011vk,Albaladejo:2012jr,Albaladejo:2013bra}. The $T$-matrix in the finite volume is then expressed as:
\begin{align}
\langle p | \widetilde T(E,L) | p' \rangle=F_{\Lambda}(p)\frac{V(E)}{\mathbb{I}-V(E)\widetilde{G}(E,L)}F_{\Lambda}(p').
\label{Eq:T-matrix-coupled-FV}
\end{align}
Here the poles of the finite-volume $\widetilde T$-matrix correspond to the energy levels $E_n(L)$,
\begin{align}
\text{det}[\mathbb{I}-V(E_n(L))\widetilde{G}(E_n(L),L)]=0,
\label{Eq:determinant}
\end{align}
for which the on-shell infinite-volume amplitude at the finite-volume eigen-energies is recovered as\footnote{Since the $G$ function is regularized (either in the box or in the infinite volume) with a Gaussian regulator, the difference below, which provides the infinite-volume amplitude $\langle p_n | T(E_n)| p'_n \rangle$, depends explicitly on the cutoff $\Lambda$. This remaining nonphysical dependence on $\Lambda$  quickly disappears as the volume increases. Indeed, it is exponentially suppressed as $\exp(-L^2\Lambda^2/8)$, as shown in Appendix B of Ref.~\cite{Albaladejo:2013aka}. The inclusion of a Gaussian regulator is quite an efficient technique, from the computational point of view, to evaluate the L\"uscher zeta function ${\cal Z}_{00}(1,E_n)$ introduced in Ref.~\cite{Luscher:1990ux}.}
\begin{equation}
 \langle p_n | T(E_n)| p'_n \rangle  = F_{\Lambda}(p_n) \frac{1}{\widetilde{G}(E_n,L)-G(E_n)}F_{\Lambda}(p_n')
\label{Eq:TfromElevels}
\end{equation}
where $p^{(\prime)}_n = \sqrt{2\mu_i(E_n - m_{1}-m_{2})}$ is computed using the appropriate meson masses for the initial and final channels, respectively.
Note that the charmonium propagator $G_{c\olsi{c}}(E)$ remains unchanged in the finite-volume case since the one-body propagation does not receive any finite-volume corrections.
Besides, the inclusion of the $c\olsi{c}$ channel in Eq. \eqref{Eq:T-matrix-coupled} aims to consider the contribution of charmonium to the effective potential as discussed in Ref.~\cite{Cincioglu:2016fkm}. Therefore, to obtain the same short-range interaction for the finite and infinite volume cases requires using the same charmonium propagator also in the finite box.

Now we discuss the explicit formula for the Green's function $\widetilde{G}(E,L)$ in both rest and moving frames. Due to the periodic boundary condition, in the rest frame, the three-momentum takes discrete values, specifically:
\begin{subequations}
\begin{equation}
\vec{q}=\frac{2\pi}{L}\,\vec{n},\,  \quad \vec{n}\in \mathbb{Z}^3\,.
\end{equation}
This leads to the Green's function in the $i$-th channel as
\begin{align}
\widetilde{G}_{\QM}^{ii}(E)&=\frac{1}{L^3}\sum_{\vec{n}\in \mathbb{Z}^3} \frac{e^{-2{\vec q\,}^2/\Lambda^2}}{E-w_i(\vec q)},
\label{Eq:two-point-rest}
\end{align}
\end{subequations}
with $w_{i}(\vec q)=m_{i,1}+m_{i,2} + \vec q\,^2/2\mu_i$. 

In the moving frame, the Green's function differs from that in the rest frame due to the violation of the Lorentz symmetry on the lattice. We define the $i$-th channel energy in the moving frame,
\begin{align}
W=\sqrt{\vec{q}^{\,2}_{i,1}+ m_{i,1}^2} + \sqrt{\vec{q}^{\,2}_{i,2}+ m_{i,2}^2}\,,
\label{Eq:momtum-total}
\end{align}
where $\vec{q}_{i,1(2)}$ is the three-momentum of the first (second) charmed meson in the moving frame, and its discrete values are $\vec{q}_{i,1(2)}=2\pi\vec{n}_{i,1(2)}/L$ for $\vec{n}_{i,1(2)} \in \mathbb{Z}^3$. The quantized total momentum of the charmed-meson pair is thus:
\begin{align}
\vec{P}=\vec{q}_{i,1} + \vec{q}_{i,2} = \frac{2\pi}{L}\vec{N}, ~~~~~\vec{N}\in \mathbb{Z}^3.
\end{align}
We boost the charmed mesons from the moving frame to the c.m. frame and obtain the three-momentum~\cite{Rummukainen:1995vs, Gockeler:2012yj,Li:2021mob}:\footnote{Here we use the shorthand notation~\cite{Rummukainen:1995vs}  $\vec{\gamma} \, \vec{a} = \vec{a}_\perp + \gamma \vec{a}_\parallel$, where $\vec{a}_{\perp(\parallel)}$ are the components of the vector $\vec{a}$ perpendicular (parallel) to the velocity $\vec{v}$, $\displaystyle \vec{a}_{\parallel} = \frac{\vec{a} \cdot \vec{v}}{v^2} \vec{v}$ and $\vec{a}_\perp = \vec{a} - \vec{a}_{\parallel}$. Analogously, $\vec{\gamma}^{-1}\,\vec{a} = \vec{a}_{\perp} + \gamma^{-1} \vec{a}_{\parallel}$.}
\begin{align}
\vec{k}_i&= \vec{\gamma}\left({\vec q_{i,1}}-\vec{v}\sqrt{\vec{q}_{i,1}^2+m_{i,1}^2}\right)\nonumber\\
&=\vec{\gamma}^{\,-1}\vec{q}_{i,1}-\frac{1}{2}{\vec \gamma}^{-1}\left(\frac{m_{i,1}^2-m_{i,2}^2}{E_{\cm}^2}+1\right)\vec{P},
\label{Eq:momentum-cm}
\end{align}
where the velocity of the two-meson system in the moving frame is $\vec{v}=\vec{P}/W$,  $\gamma=1/\sqrt{1-\vec{v\,}^2}$ is the Lorentz boost factor, and $E_{\cm}=E$ is the c.m. energy of the two-meson system (the subindex $\cm$ is included here only for clarity). Finally, in the c.m. frame, the Green's function is rewritten as 
\begin{align}
\widetilde{G}_{\QM}^{ii,N}(E)&=\frac{1}{\gamma L^3}\sum_{\vec{n}\in \mathbb{Z}^3} \frac{e^{-2{\vec k\,}_i^2/\Lambda^2}}{E-w_{i}(\vec k)}.
\label{Eq:two-point-mv}
\end{align}
Here, the three-momentum $\vec{k}$ in Eq. \eqref{Eq:momentum-cm} is quantized as
\begin{align}
\vec{k}_i=\frac{2\pi}{L}\vec{\gamma\,}^{-1}\vec{n}^{\,\prime}=\frac{2\pi}{L}\vec{\gamma}^{\,-1}\left[\vec{n}-\frac{\vec{N}}{2}\left(1+\frac{m_{i,1}^2-m_{i,2}^2}{E_{\cm}^2}\right)\right],
\label{Eq:mom_moving}
\end{align}
and its components are:
\begin{align}
\vec{k}_{i,\alpha}=\frac{2\pi}{L}\left[\vec{n}^{\,\prime}_{\alpha}+\left(\frac{1}{\gamma}-1\right)\frac{\vec{v}_{\alpha}(\vec{v}\cdot\vec{n}^{\,\prime})}{\vec{v}^{\,2}}\right],
\end{align}
where the subscript $\alpha$ ($\alpha=x,y,z$) denotes the component of the vector in the $\alpha$-direction, and $\vec n^{\,\prime}$ is 
\begin{align}
\vec n^{\,\prime} =\vec{n}-\frac{\vec{N}}{2}\left(1+\frac{m_{i,1}^2-m_{i,2}^2}{E_{\cm}^2} \right).
\end{align}

\subsection{Interaction between charmed mesons}
\label{sec:4H-potential}

We consider the $S$-wave scattering of a charmed-anticharmed pair of mesons, including the contribution stemming from the presence of a $P$-wave charmonium state relatively close to their threshold. The interactions are constrained~\cite{Nieves:2012tt,Hidalgo-Duque:2012rqv,Cincioglu:2016fkm} by the approximate HQSS of QCD. In the heavy quark limit, the interactions between the charmed mesons are given in terms of contact potentials, consistent with HQSS, which for the finite charm-quark mass case are identified as the LO terms of an effective field theory (EFT)~\cite{AlFiky:2005jd}. The higher-order terms are suppressed by powers of $M^{-1}$, where $M$ denotes the hard scale where new degrees of freedom need to be included explicitly.
In addition, the interaction between charmed mesons and the $P$-wave charmonium multiplet is also constructed respecting HQSS~\cite{Casalbuoni:1992yd,Cincioglu:2016fkm}.
The LECs, which encode the strengths for the interactions between charmonium states and the charmed mesons, as well as the $D^{(*)}_{(s)}\olsi{D}{}^{\, (*)}_{(s)}D^{(*)}_{(s)}\to\olsi{D}{}^{\, (*)}_{(s)}$ contact terms, are currently unknown. To study the scattering of hidden-charm systems, it is essential to determine these LECs through various methods such as experimental data, lattice data, or phenomenological models. 

In this work, we extract the LECs from the lattice results reported in Refs.~\cite{Prelovsek:2013cra,Prelovsek:2020eiw}.
As mentioned in the Introduction, energy levels for the isoscalar $0^{++}$  $D\olsi{D}$--$D_s \olsi{D}_s$ coupled-channel system were obtained in Ref.~\cite{Prelovsek:2020eiw} using  $D\olsi{D}$, $D_s \olsi{D}_s$, $J/\psi \omega$, and $c\olsi c$ interpolators. The LQCD calculation shows that the highest energy level (about $4070$ MeV) in the rest frame is below the $D^\ast \olsi{D}{}^\ast$ threshold (around $4100\,\MeV$). Therefore, the contribution of the $D^\ast \olsi{D}{}^\ast$ channel to the isoscalar $0^{++}$ charmonium-like states should be small and can be safely disregarded. In the isoscalar $1^{++}$ sector, the lattice data from Ref.~\cite{Prelovsek:2013cra}, calculated using the interpolating operators $c\olsi c$ and $D\olsi D{}^*$, are employed to compute the energy levels, as listed in Table \ref{Tab:levels_1++}.
We notice that more interpolating operators were used in the updated lattice calculations in Ref.~\cite{Padmanath:2015era}. However, as shown in Fig. 6~(a) of Ref.~\cite{Padmanath:2015era}, in addition to the contributions of $c\olsi c$ and $D\olsi D{}^{*}$ operators, the energy levels exhibit sizeable overlap with other operators, such as the non-negligible contribution of $J/\psi(0)\omega(0)$ one to the $n=6$ energy level. Therefore, using such data would require to substantially increase the number of coupled channels and thus the number of involved parameters. 
To reduce the number of free parameters, we use the lattice data from Ref.~\cite{Prelovsek:2013cra}. This is reasonable as the $D\olsi D{}^*$ is known to be the most relevant channel since it has the strongest coupling to the $X(3872)$~\cite{ParticleDataGroup:2024cfk}.\footnote{The decay modes that have the largest branching fractions of the $X(3872)$ decays are $X(3872)\to D^0\olsi D{}^{*0}$ ($\text{Br}[X(3872)\to D^0\olsi D{}^{*0}]=(34\pm 12)\%$) and $X(3872)\to D^0 \olsi D{}^{0}\pi^0$ ($\text{Br}[X(3872)\to D^0\olsi D{}^{0}\pi^0]=(45\pm 21)\%$)~\cite{ParticleDataGroup:2024cfk}, despite their tiny phase spaces. One also notices that the $D^{*0}$ mainly decays into $D^0\pi^0$. Therefore, the channel that couples most strongly to the $X(3872)$ should be the $D\olsi{D}{}^\ast$ one.} The contribution of the $D_s \olsi{D}_s^{\ast}$ and $D^{\ast}\olsi{D}{}^\ast$ channels can be also reasonably neglected.\footnote{The second energy level in Table~\ref{Tab:levels_1++} is quite close to the $D^*\olsi D{}^*$ threshold. 
However, for the $D^*\olsi D{}^*$ $S$-wave scattering, the allowed quantum numbers are either $1^{+-}$ or $2^{++}$~[25]. Therefore, for the analysis of the $1^{++}$ system, the contribution of the $D^*\olsi D{}^*$ channel can be safely excluded because of $C$-parity. } 
Consequently, the relevant channels in the different isoscalar $J^{PC}$ sectors are
\begin{align}
J^{PC}=0^{++} &: \left\{D\olsi{D},~D_s\olsi{D}_s, \chi_{c0}(2P)\right\},\nonumber\\
J^{PC}=1^{++} &: \left\{D^* \olsi{D},~\chi_{c1}(2P)\right\},\nonumber\\
J^{PC}=2^{++} &: \left\{D^*\olsi{D}{}^\ast, \chi_{c2}(2P)\right\}.
\end{align}

\begin{table}[t]
	\centering
    \renewcommand\arraystretch{1.6}
 	\caption{Three LQCD energy levels reported in the calculation of Ref.~\cite{Prelovsek:2013cra} and obtained for a box size $L$ of approximately $1.98$ fm. The energies are given with respect to the spin-averaged lattice mass $M_{\text{av}}^L = (m_{\eta_c}+3m_{J/\psi})/4 \simeq 2428 ~\text{MeV}$, and the central values of momenta associated to each energy-level (Eq.~\eqref{Eq:Momentum_on-shell}) are compiled in the last row.}
  \begin{ruledtabular}
  \begin{tabular}{cccc}
				&	$1$ & $2$ & $3$ \\
	\hline
    $(E_n - M_{\text{av}}^L)\, \text{[MeV]}$			&	$785(8)$		& $946(11)$ & $1028(18)$ \\
	$\sqrt{\vec{p}^2}\, \text{[MeV]}$		&	$274 i$ 	    & $480$ & $625$\\	
	\end{tabular}
  \end{ruledtabular}
	\label{Tab:levels_1++}
\end{table}

At LO, the $D\olsi{D}$--$D_s \olsi{D}_s$ coupled-channel matrix potential can be derived from the findings of previous works (see Eqs.~(49), (53) and (54) of Ref.~\cite{Hidalgo-Duque:2012rqv} and \cite{Cincioglu:2016fkm}
\begin{align}
V(0^{++})=\begin{pmatrix}
C_{0a} & \frac{1}{\sqrt{2}}(C_{0a}-C_{1a}) & -\frac{\sqrt{3}}{2}d\\
\frac{1}{\sqrt{2}}(C_{0a}-C_{1a}) & \frac{1}{2}(C_{0a}+C_{1a}) & -\sqrt{\frac{3}{8}}d\\
 -\frac{\sqrt{3}}{2}d &  -\sqrt{\frac{3}{8}}d & 0\\
 \end{pmatrix},
 \label{Eq:potential_0++}
\end{align}
where $C_{0a}$ and $C_{1a}$ denote the potentials for the charmed mesons, and $d$ is the coupling constant between the charmed mesons ($D\olsi{D}$, $D_s\olsi{D}_s$) and the charmonium $\chi_{c0}(2P)$.\footnote{We extend the interaction between the charmed mesons and the $P$-wave charmonium from isospin $SU(2)$ to the flavor $SU(3)$ symmetry using that the $\chi_{c0}$ is an isospin and SU(3) singlet. Thus, its coupling to the charmed meson pairs is given by $\chi_{c0}^\dag  D_a \olsi D_a$ with $a=1,2,3$ the flavor index, and in Ref.~\cite{Hidalgo-Duque:2012rqv}, the singlet is $ D_a \olsi D_a =\sqrt{2}\left[|D\olsi D; I=0,I_3=0\rangle + |D_s\olsi D_s\rangle/\sqrt{2}\right]$. 
}

The LECs $C_{0a}$ and $C_{1a}$ can also be written as~\cite{Hidalgo-Duque:2012rqv}: 
\begin{align}
C_{0a}=\frac{2}{3}C_a^{(1)}+\frac{1}{3}C_{a}^{(8)}\,, \quad C_{1a}=C_a^{(8)}, \label{eq:octet}
\end{align}
where $C_a^{(1)}$ and $C_a^{(8)}$ are the LECs for flavor $SU(3)$ singlet and octet, respectively. 
For the $1^{++}$ and $2^{++}$ sectors, the LO potentials are identical~\cite{Nieves:2012tt,Cincioglu:2016fkm},
\begin{align}
V(1^+|2^+)=\begin{pmatrix}
C_{0X} & d^{\prime} \\
 d^{\prime} &  0 \\
 \end{pmatrix},
 \label{Eq:potential_1++}
\end{align}
where $C_{0X}=(C_{0a}+C_{0b})$ is the coupling constant for the scattering of the isoscalar $D \olsi{D}{}^\ast$ ($D^* \olsi{D}{}^\ast$) mesons and the positive charge conjugation, and $d'$ is the coupling constant between $\chi_{c0}(2P)$ ($\chi_{c2}(2P)$) and $D\olsi{D}{}^\ast$ ($D^*\olsi{D}{}^\ast$). 
The potential of Eq. \eqref{Eq:potential_1++} can be equivalently expressed as a single-channel potential as~\cite{Hanhart:2011jz, Albaladejo:2013aka}:\footnote{The $T$-matrix for $n+1$ channels previously discussed can also be reduced to an $n$-channel $T$-matrix in a similar manner, \textit{i.e.}, by using energy-dependent potentials that include the exchange of the bare charmonium state in the $s$ channel.}
\begin{align}
V(1^+|2^+)=C_{0X}+\frac{d^{\prime 2}}{E-\mco},
\label{Eq:potential_single}
\end{align}
where $\mco$ is the bare mass of the $\chi_{c1}(2P)$ or $\chi_{c2}(2P)$ state. This potential provides an attractive interaction from the $s$-channel charmonium-exchange part when $E< \mco$ and a repulsion for $E>\mco$.

To study the $S$-wave scattering of charmed mesons including charmonium contributions, we see that we have several free parameters at a given cutoff $\Lambda$: $C_{0a}$, $C_{1a}$, $C_{0X}$, $d^{(\prime)}$, and the bare masses of the $P$-wave charmonia in Eq. \eqref{eq:Gfun:channel-bare} ($\mcos$, $\mcov$ and $\mcot$), which need to be determined. The LECs in the $0^{++}$ and $1^{++}$ sectors can be extracted from the LQCD energy levels of the isoscalar $D\olsi{D}- D_s \olsi{D}_s$ coupled-channel~\cite{Prelovsek:2020eiw} and $D\olsi{D}{}^\ast$~\cite{Prelovsek:2013cra} systems.

With the determined LECs we can explore the different observables (infinite volume), such as phase shifts. The $S$-matrix for the $D\olsi{D}$--$D_s\olsi{D}_s$ coupled-channel system in the $S$-wave is parameterized as~\cite{Weinstein:1990gu,Oller:1998hw}:
\begin{align}
S=\begin{pmatrix}
\eta e^{2i\delta_1} & i\sqrt{1-\eta^2}e^{i(\delta_1+\delta_2)}\\
i\sqrt{1-\eta^2}e^{i(\delta_1+\delta_2)} & \eta e^{2i\delta_2}
\end{pmatrix},
\label{Eq:S-matrix}
\end{align}
where $\eta$ is the inelasticity, and $\delta_1$ and $\delta_2$ are the phase shifts for the scattering of $D\olsi{D}\to D\olsi{D}$ and $D_s\olsi{D}_s\to D_s\olsi{D}_s$, respectively. The $S$- and $T$-matrices are related through:
\begin{align}
S_{ij}(E) & = \delta_{ij} - 2i \rho_{ij}(E) T_{ij}(E)\,,\nonumber\\
\rho_{ij}(E) & = \frac{4m_i m_j\sqrt{p_i\,p_j}}{8\pi {(m_i+m_j)}}\,.
\label{Eq:relation_S_T}
\end{align}
where $m_1= m_D$ and $m_2=m_{D_s}$. Note that a factor $4 m_i m_j$ is included in the numerator of the phase-space factor $\rho_{ij}$ to account for the different relativistic and nonrelativistic $T$-matrix normalizations.

In the $1^{++}$ and $2^{++}$ sectors, the $T$-matrix in Eq.~\eqref{Eq:T-matrix-coupled} can be rewritten as the single-channel scattering with the effective potential in Eq. \eqref{Eq:potential_single}. As a result, the $T$-matrix for the $1^{++}$ and $2^{++}$ sectors is parameterized as:\footnote{The scattering matrix in terms of the phase shift $\delta$ in the single-channel case is expressed as  $S(E)=e^{2i\delta(E)}$.}
\begin{align}
  T(E)=-\frac{2\pi}{\mu}\frac{1}{p\, \text{cot}\, \delta - ip},
\label{Eq:s-matrix-single}
\end{align}
where, near the threshold, $p\,\cot\delta$ can be expressed as a function of the scattering length $a$ and effective range $r$ through the effective range expansion (ERE),
\begin{align}
p\,\cot\delta=-\frac{1}{a}+\frac{1}{2}r p^2 + \cdots\,.
\label{Eq:ERE}
\end{align}

\section{\boldmath Results for the $D\olsi{D}$--$D_s\olsi{D}_s$ coupled-channel system}\label{Sec:0++}

\subsection{\boldmath Parameters in the $0^{++}$ sector}

In the $0^{++}$ sector, the LECs in Eq. \eqref{Eq:potential_0++} and the bare mass $\mcos$ are determined by fitting to the $D \olsi{D}$--$D_s \olsi{D}_s$ coupled-channel LQCD energy levels reported in Ref.~\cite{Prelovsek:2020eiw}. There, the authors have simulated four different lattice irreducible representations: $A_1$ with squared total momentum $|\vec P |^2=0, 1, 2$, and $B_1$ with squared total momentum $|\vec P |^2=1$, where the squared momentum is given in  $(2\pi/L)^2$ units. 
This calculation provides 27 energy levels in two finite volumes ($L=24a$ and $32 a$, with $a=0.08636\,\fm$), where 8 levels are calculated in the rest frame, while the rest of them are in moving frames. Among those levels, four are in the $B_1$ representation, solely contributed by the $D$-wave scattering, and hence they are excluded in the present analysis. Additionally, a level with $|\vec{P}|=1$ and $L=24a$, dominated by the $c\olsi{c}[J=2]$ operator, is omitted. Note that the exclusion/inclusion of the $D$-wave contribution does not change the behavior of the $J^{PC}=0^{++}$ poles, as discussed in Ref.~\cite{Prelovsek:2020eiw}. Consequently, we consider the $S$-wave contributions and utilize the remaining 22 energy levels with momentum $\lvert\vec{P}\rvert^2=0$, $1$, and $2$ 
to constrain the LECs, $\mcos$, $d$, $C_{0a}$, and $C_{1a}$. 

The lattice data in our calculation include 8 points obtained in the rest frame and 14 in moving frames. We use two different methods to analyze these energy levels. In the first one, we fit the LECs only to the 8 levels calculated in the rest frame. In the other method,  we incorporate all 22 points from both the rest and moving frames. To address the UV divergence, we adopt the Gaussian form factor introduced in Eq. \eqref{Eq:T-matrix-coupled-FV} with two cutoff values $\Lambda=0.5\,\GeV$ and $\Lambda=1.0\,\GeV$. In summary, we perform four different fits: Fit 1 and Fit 2 employ solely the rest-frame energy levels and examine $\Lambda=0.5\,\GeV$ and $\Lambda=1.0\,\GeV$, respectively; the difference in the resulting bare parameter values reflects their cutoff dependence, which should be there to balance the LO cutoff dependence from the regulator. 
In contrast, Fit 3 and Fit 4 incorporate energy levels from both the rest and moving frames and consider the same two cutoff values. The fits are carried out by minimizing the appropriate $\chi^2$ function with the MINUIT algorithm~\cite{James:1975dr,iminuit,iminuit.jl}. 

\begin{table}[t]
    \centering
\caption{LECs fitted to the lattice energy levels reported in Ref.~\cite{Prelovsek:2020eiw}. In Fit 1 ($\Lambda=0.5\,\GeV$)  and Fit 2 ($\Lambda=1.0\,\GeV$) only  LQCD energy levels obtained in the rest frame have been considered. Energy levels computed in moving frames have been also taken into account in Fit 3 ($\Lambda=0.5\,\GeV$) and Fit 4 ($\Lambda=1.0\,\GeV$). The statistical correlation matrices are given in Appendix~\ref{sec:appendix} for all scenarios. \label{tab:LEC-value}   }
\begin{ruledtabular}
\begin{tabular}{lcccccc}
& $\mcos[\GeV]$ & $d[\fm^{1/2}]$ & $C_{0a}[\fm^2]$ & $C_{1a}[\fm^2]$ & d.o.f. & $\chi^2/\text{d.o.f.}$ \\
\hline
Fit 1 & $3.94(2)$  & $0.88(18)$  & $-0.23(52)$ & $-0.64(28)$ & $4$ & $0.86$ \\
 Fit 2 & $4.03(5)$  & $0.58(28)$  & $-0.44(24)$ & $-0.41(13)$  & $4$ & $1.63$ \\
 Fit 3   & $3.99(1)$  & $0.53(22)$  & $-1.14(29)$ & $-1.03(19)$ & $18$ & $3.26$ \\
 Fit 4   &  $4.08(4)$  & $0.44(15)$  & $-0.47(10)$ & $-0.55(6)$ & $18$ & $3.27$ \\
\end{tabular}
\end{ruledtabular}
\end{table}

The determined LECs from each fit are collected in Table~\ref{tab:LEC-value}, while in Figs.~\ref{Fig:energy-level-rest} and \ref{Fig:energy-level-mv}, we present the obtained dependence of the energy levels on the size $L$ of the finite box. We vary $L$ from $1$~fm to $5.5$~fm and numerically calculate the energy levels with the four sets of LECs listed in Table~\ref{tab:LEC-value}. The left (right) panel in Fig.~\ref{Fig:energy-level-rest} corresponds to the rest-frame energy levels computed with the parameters from Fit 1 (Fit 2), while the energy levels in the left (right) columns of Fig. \ref{Fig:energy-level-mv} are associated to Fit 3 (Fit 4), respectively. Since in Fits 1 and 2, only the rest-frame lattice data are utilized to adjust the LECs, in Fig.~\ref{Fig:energy-level-rest} we exclusively show the rest-frame energy levels, while only the moving-frame energy levels which are considered in Fits 3 and 4 are also shown in Fig.~\ref{Fig:energy-level-mv}. The blue bands depicted in both figures are calculated with the bootstrap method, and correspond to the 1$\sigma$ uncertainty of the fitted LECs.

\begin{figure*}[t]\centering
\includegraphics[scale=0.5]{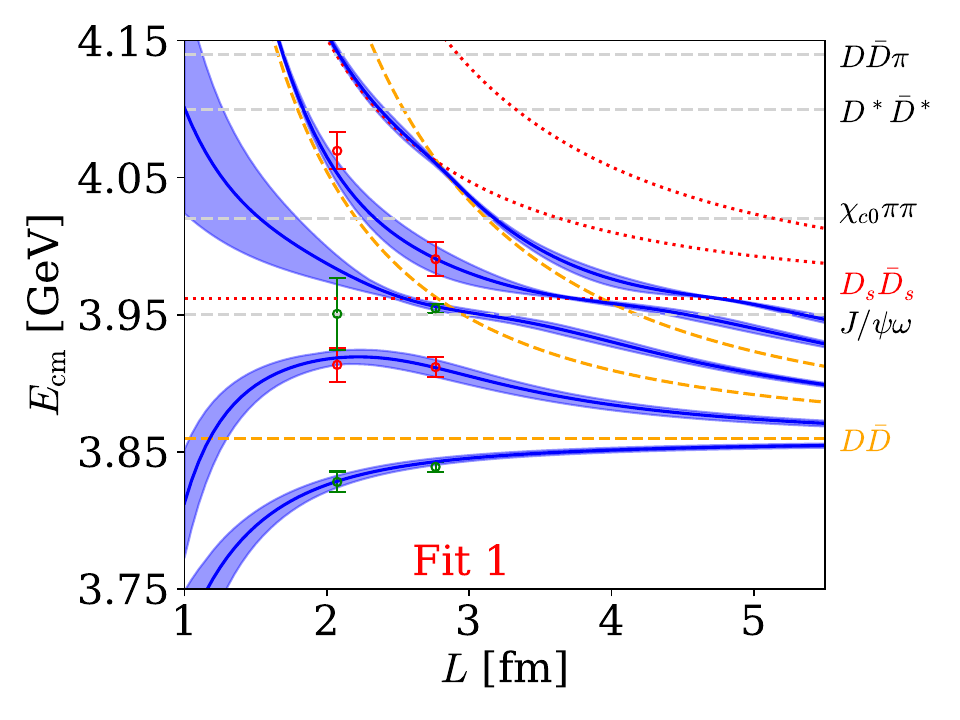}
 \includegraphics[scale=0.5]{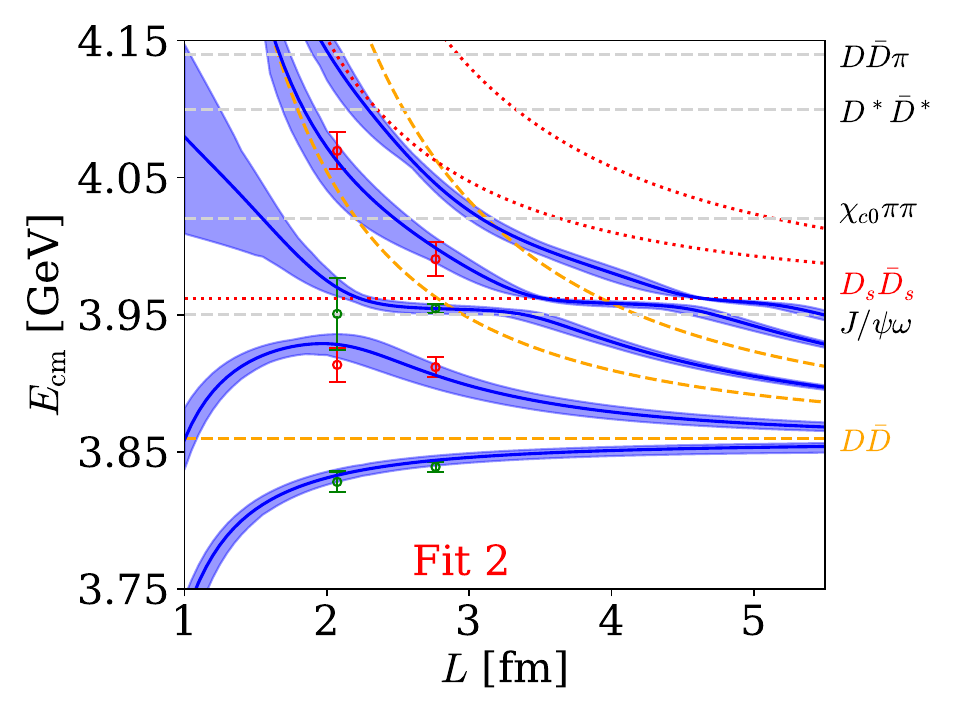}
\caption{Volume dependence of the rest-frame energy levels. The left and right panels correspond to Fit 1 ($\Lambda=0.5\,\GeV$)  and Fit 2 ($\Lambda=1.0\,\GeV$) in Table~\ref{tab:LEC-value}. Data from \cite{Prelovsek:2020eiw} correspond to two box sizes,  $L=24a$ and $32a$ with $a=0.08636\,\fm$. The red and green colors for the LQCD energy levels are used solely for the sake of clarity to help distinguish between neighbor points. The dashed orange and red lines represent the free $D\olsi{D}$ and $D_s\olsi{D}_s$ energy levels, respectively. The solid blue bands, which reflect the $68\%$ confidence level uncertainties in the LECs, stand for the energy levels $E_{n}(L)$ ($n=0,1,2,3,4$).  }
\label{Fig:energy-level-rest}
\end{figure*}

\begin{figure*}[t] 
\centering	
\includegraphics[scale=0.5]{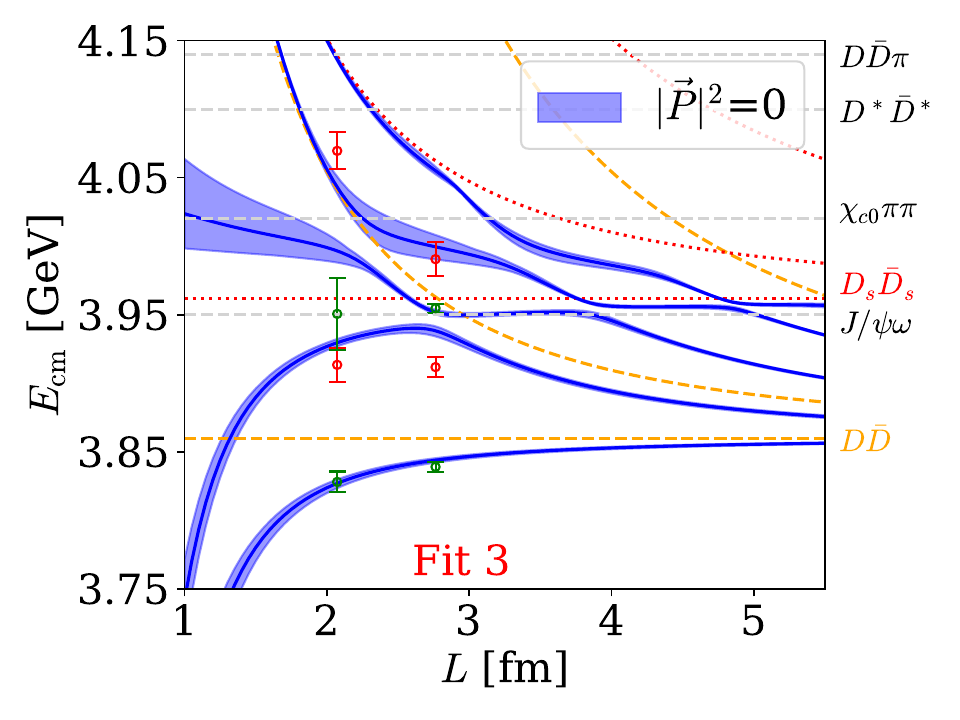}
\includegraphics[scale=0.5]{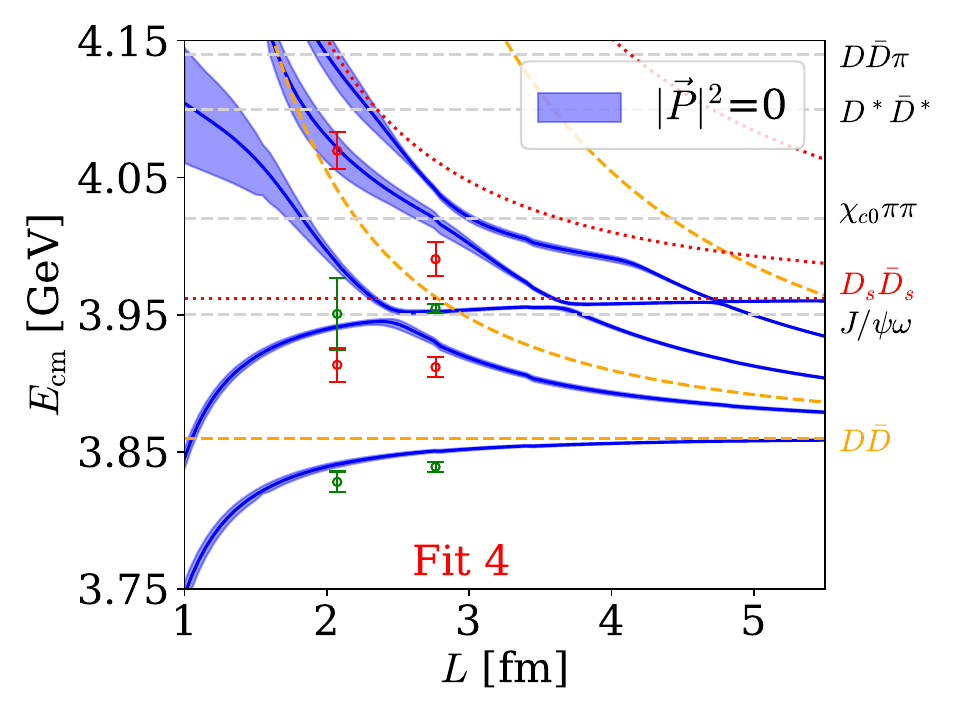}\\
\includegraphics[scale=0.5]{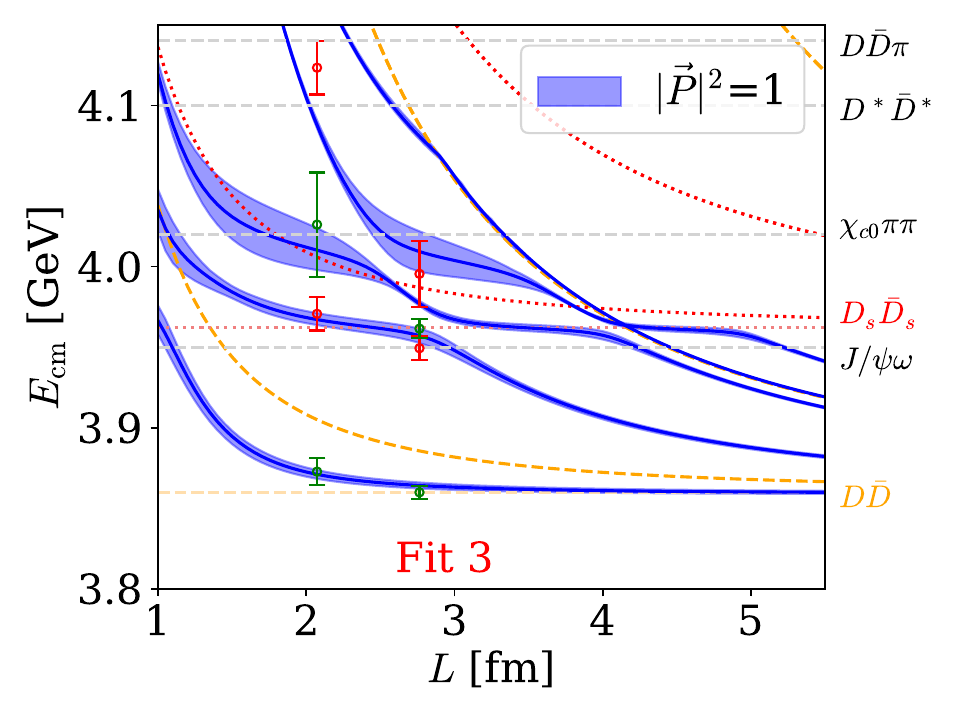}
\includegraphics[scale=0.5]{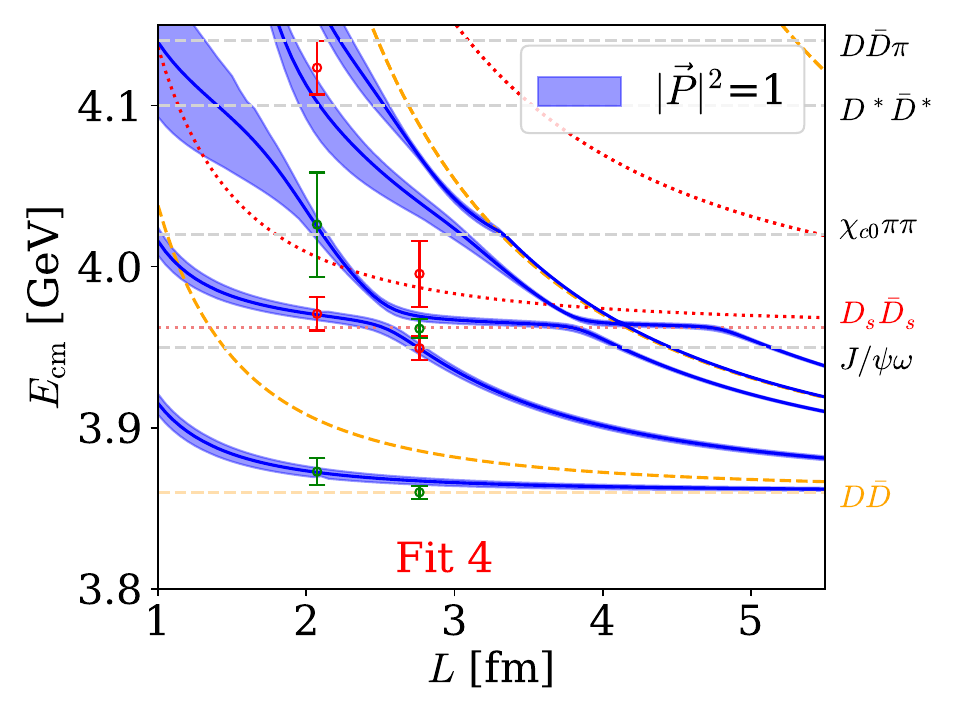}\\
\includegraphics[scale=0.5]{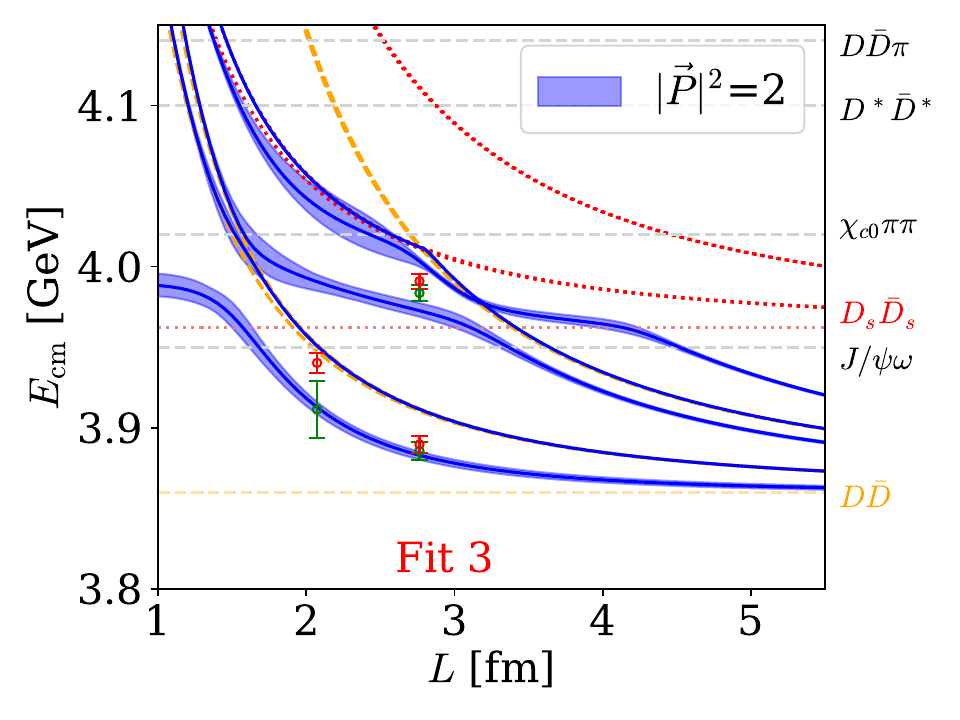}
\includegraphics[scale=0.5]{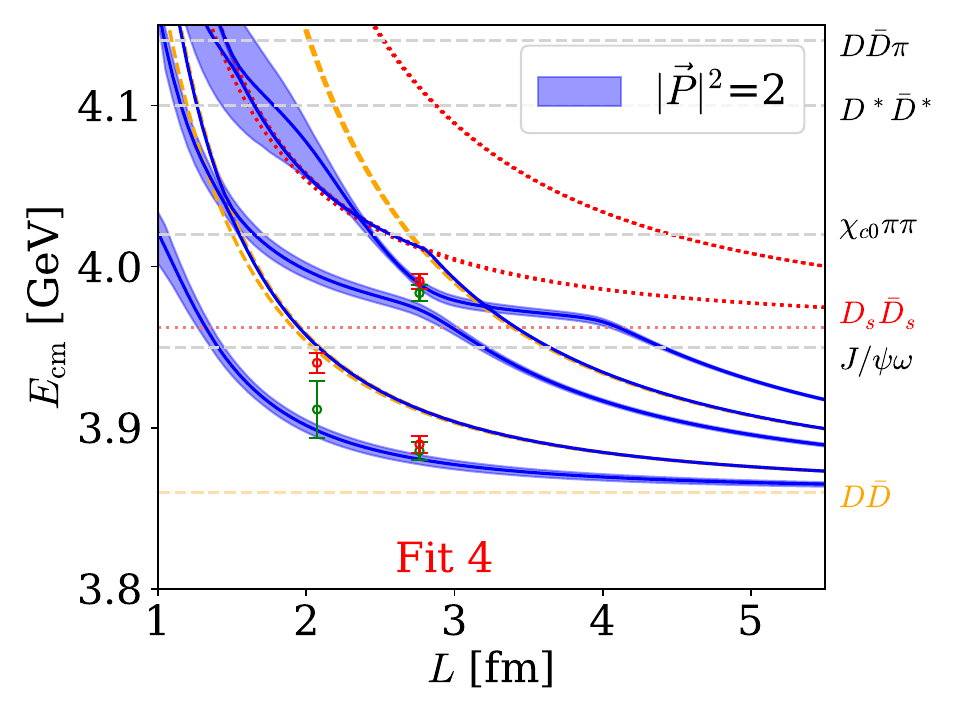}
\caption{Volume dependence of the rest- and moving-frame energy levels obtained from the results of Fit 3 ($\Lambda=0.5\,\GeV$)  and Fit 4 ($\Lambda=1.0\,\GeV$) and shown in the left and right panels, respectively. As in Fig.~\ref{Fig:energy-level-rest}, data from \cite{Prelovsek:2020eiw} correspond to two box sizes,  $L=24a$ and $32a$ and the meaning of the different lines and the blue bands is the same as in that figure. For simplification, an energy level with $|\vec{P}|=1$ and $L=24a$, which is displayed in Figure~2 of Ref.~\cite{Prelovsek:2020eiw}, is omitted here, since it is dominated by the $c\olsi{c}[J=2]$ operator.}
\label{Fig:energy-level-mv}
\end{figure*}

In Table~\ref{tab:LEC-value}, we observe that the charmonium $\chi_{c0}$ bare mass takes similar values in Fit 1 and Fit 3, as well as in Fit 2 and Fit 4. However, the coupling constants between charmed mesons and $\chi_{c0}$ vary significantly among the four scenarios. Nevertheless, comparing LECs for different cutoff values requires some care, as they are cutoff dependent (see for instance discussions in Ref.~\cite{Cincioglu:2016fkm}).

As can be seen in the corresponding plots in Fig.~\ref{Fig:energy-level-mv} for the moving frame with $|\vec P|^2 = 2$, the two lowest energy levels for finite-volume size $L=32a$ are very close to each other. We fail to reproduce this behavior, and thus the $\chi^2/\text{d.o.f.}$ in Fit 3 and Fit 4 are considerably larger than those in Fit 1 and Fit 2.

Taking into account the masses of the heavy-light mesons in the LQCD simulation of Ref.~\cite{Prelovsek:2020eiw}, we see that the values of $\mcos$, which may be effectively interpreted as the bare mass of the charmonium $\chi_{c0}(2P)$, for all four scenarios are above the LQCD $D\olsi{D}$ threshold (3860 MeV). Therefore, the $\chi_{c0}(2P)$ effectively produces an attractive interaction when the energy $E$ is around the $D\olsi{D}$ threshold. Around the LQCD $D_s\,\olsi{D}_s$ threshold (3962 MeV), the $\chi_{c0}(2P)$ generates a repulsive interaction in the case of Fit 1, while the interaction is attractive for the other scenarios. In addition, the mass splitting between $\chi_{c0}(2P)$ and ground-state charmonia might be less affected by the unphysical charm quark mass employed in the LQCD simulation. We follow Ref.~\cite{Prelovsek:2020eiw} and define an averaged ground-state mass 
\begin{align}
M_{\text{av}}=\frac{m_{\eta_c}+3m_{J/\psi}}{4}.
\label{Eq:average_charmonium}
\end{align}
With the LQCD masses of Ref.\,\cite{Prelovsek:2020eiw}, the average mass is approximately $M_{\text{av}}^L=3103\,\MeV$, while the experimental average mass $M_{\text{av}}^{\text{exp}}\simeq 3069\,\MeV$. The splittings between $\mcos$ in Table \ref{tab:LEC-value} and $M_{\text{av}}^L$ are $833$, $924$, $886$, and $974\,\MeV$ for Fit 1, 2, 3 and 4, respectively. Various quark models~\cite{Godfrey:1985xj,Zeng:1994vj,Ebert:2002pp} predict the mass gaps between the $2P$ charmonium and $M_{\text{av}}^{\text{exp}}$ to be about $800$--$900\,\MeV$. It can be seen that the mass gaps we obtain here are similar or somewhat slightly higher than this range.

The contributions of the bare charmonium to the two-charmed meson interaction in the $SU(3)$ flavor-singlet and flavor-octet configurations are different, vanishing obviously in the latter one.  Therefore, the $SU(3)$ octet counterterm $C_{1a}= C_a^{(8)}$ [see Eq.~\eqref{eq:octet}] is expected to be comparable with the results determined without considering the charmonium contributions, as the pion-mass dependence of $C_{1a}$ is neglected and the $L$-dependent terms are exponentially suppressed. 

The line shape of the $D_s^+D_s^-$ invariant mass distribution, in the decay channel $B^{+}\to D_s^+D_s^-K^+$~\cite{LHCb:2022aki}, was fitted in  Ref.~\cite{Ji:2022uie}  and as result $\mathcal{C}_{1 a}=0.36_{-0.26}^{+0.25}\left[-0.31_{-0.05}^{+0.03}\right]\,\fm^2$ was obtained,  with a Gaussian form-factor cutoff $\Lambda=0.5[1.0]\,\GeV$. 
Subsequently in Ref.~\cite{Ji:2022vdj}, a comprehensive analysis of the \LHCb data for $B^+\to K^+ D^+D^-$~\cite{LHCb:2020pxc}, $B^+\to K^+ D_s^+D_s^-$~\cite{LHCb:2022aki}, and the Belle~\cite{Belle:2005rte} and BaBar~\cite{BaBar:2010jfn} data for the $\gamma\gamma\to D\olsi{D}$ in the $D\olsi{D}$--$D_s \olsi{D}_s$--$D^\ast \olsi{D}{}^\ast$--$D^*_s \olsi{D}{}^\ast_s$ coupled-channel framework was conducted, which led to $\mathcal{C}_{1 a}=(-0.33 \pm 0.02)\,\fm^2$ for $\Lambda=1.0\,\text{GeV}$.

In this work, as compiled in Table \ref{tab:LEC-value}, $C_{1a}$ for Fits 1, 2, and 4 turns out to be consistent with the values obtained in Refs.~\cite{Ji:2022uie,Ji:2022vdj}, while its absolute value in Fit 3 is much larger ($\sim 4\sigma$ deviation). Also, both Fit 3 and 4 have very large $\chi^2/{\rm d.o.f.}$ values. 
This discrepancy may stem from the small cutoff ($\Lambda=0.5\,\text{GeV}$) used in analyzing the moving-frame lattice data. The c.m. three-momentum in Eq. \eqref{Eq:mom_moving} is larger than that in the rest frame, requiring probably a larger cutoff to saturate the wave function of the two-charmed-meson system~\cite{Albaladejo:2017blx}.
On the other hand, $|C_{0a}|$ from the Fits 1, 2, and 4, as listed in Table \ref{tab:LEC-value}, are significantly smaller than the values reported in Refs.~\cite{Ji:2022uie,Ji:2022vdj}. This can be attributed to the contribution of the exchange of the bare $2P$ charmonium state, which provides an additional source of attraction in the present framework. The discrepancies in Fit 3 could be also due to the small cutoff used in the analysis of the moving-frame lattice data.

\subsection{Energy levels and phase shifts}

The behavior of energy levels as a function of finite volume size $L$ provides valuable insights that can guide the search of poles. Both in the left and right panels of Fig. \ref{Fig:energy-level-rest}, the lowest energy level $E_0(L)$ remains below the $D\olsi{D}$ threshold. For $L> 5\,\fm$, this energy level approaches to a constant (smaller than the $D\olsi{D}$ threshold), suggesting the presence of a $D\olsi{D}$ bound state. The binding energy is determined by the mass splitting between $E_0(\infty)$ and the $D\olsi{D}$ threshold~\cite{Beane:2003da}.\footnote{Equivalently, one can look for a bound-state pole in the infinite volume amplitude.} The presence of plateaus in the energy levels $E_{n}(L)$ ($n=3,4,5$) as the box size $L$ varies can point to the existence of a narrow resonance near the $D_s\olsi{D}_s$ threshold. 

In the first left panel of Fig. \ref{Fig:energy-level-mv}, the behaviour of the rest-frame lowest energy level $E_0(L)$ for $L > 5\,\fm$ might also hint the presence of a bound state. However, in the right panel (Fit 4) the lowest energy level almost reaches the $D\olsi{D}$ threshold, making it challenging to claim the existence of a very shallow bound state. Coming back to the first ($\lvert\vec P \rvert^2=0$) left panel, one can appreciate two plateaus below and above the $D_s \olsi{D}_s$ threshold, respectively. This points out two narrow resonances in this system~\cite{Wiese:1988qy}. The plateau \textit{above} the $D_s \olsi{D}_s$ threshold can be also observed in the predictions for moving frames.
In the right panels of Fig. \ref{Fig:energy-level-mv} (Fit 4), a plateau is also observed close to the $D_s \olsi{D}_s$ threshold. Note that in Fig.\,\ref{Fig:energy-level-mv}, although the energy levels at specific sizes of the finite volume $L$ are very close, they do not cross, which aligns with the ``avoided level crossing'' phenomenon.

We use the LECs collected in Table \ref{tab:LEC-value} to compute 
phase shifts and inelasticities [see Eqs.~\eqref{Eq:S-matrix} and \eqref{Eq:relation_S_T}]  in the infinite volume limit. They are shown in Figs. \ref{fig:phase-rs} (Fits 1 and 2) and \ref{fig:phase-mv} (Fits 3 and 4). The large inelasticities, above $D_s\olsi{D}_s$ threshold, in Fig.~\ref{fig:phase-rs} indicate a significant mixing between the $D\olsi{D}$ and $D_s \olsi{D}_s$ channels. The behaviour of the $\delta_{D\olsi{D}}$ phase-shift suggests the presence of a bound state below the $D\olsi{D}$ threshold, which is consistent with the observation of the lowest energy levels in Fig. \ref{Fig:energy-level-rest}. In Fig. \ref{fig:phase-mv}, the phase shift $\delta_{D\olsi{D}}$ is also consistent with the existence of a bound state below $D\olsi{D}$ threshold for Fit 3, while for Fit 4 there is not a bound state below $D\olsi{D}$ threshold, but a resonance around the $D_s\olsi{D}_s$ threshold. We will discuss the pole content of the amplitudes in Subsec.~\ref{subsec:poles_DDbar_DsDsbar}.

\begin{figure*}
\centering	
 \includegraphics[scale=0.5]{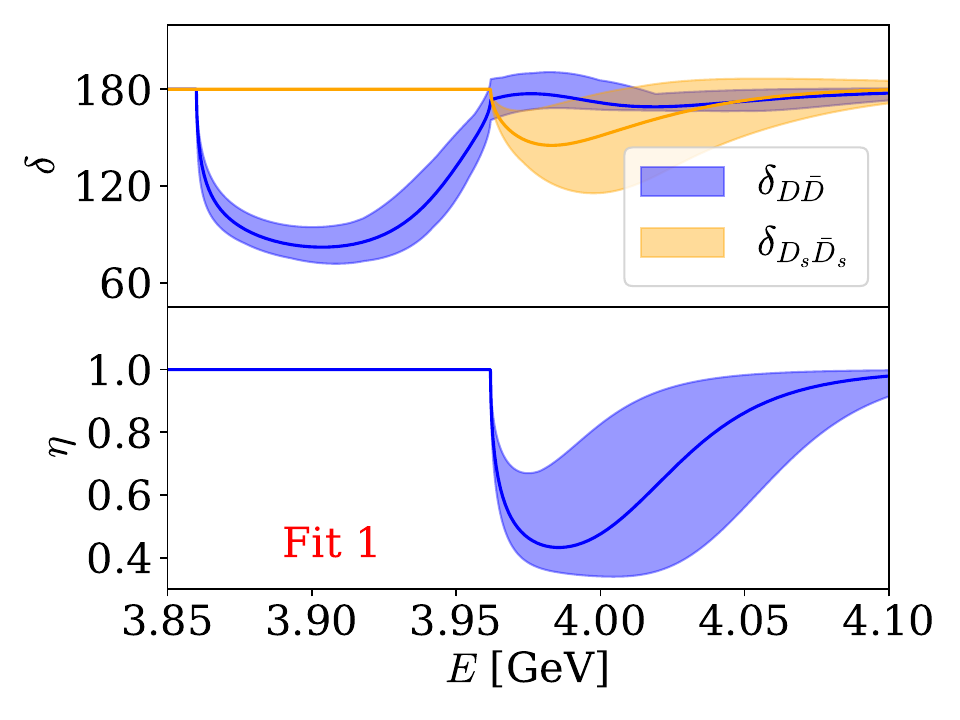}
 \includegraphics[scale=0.5]{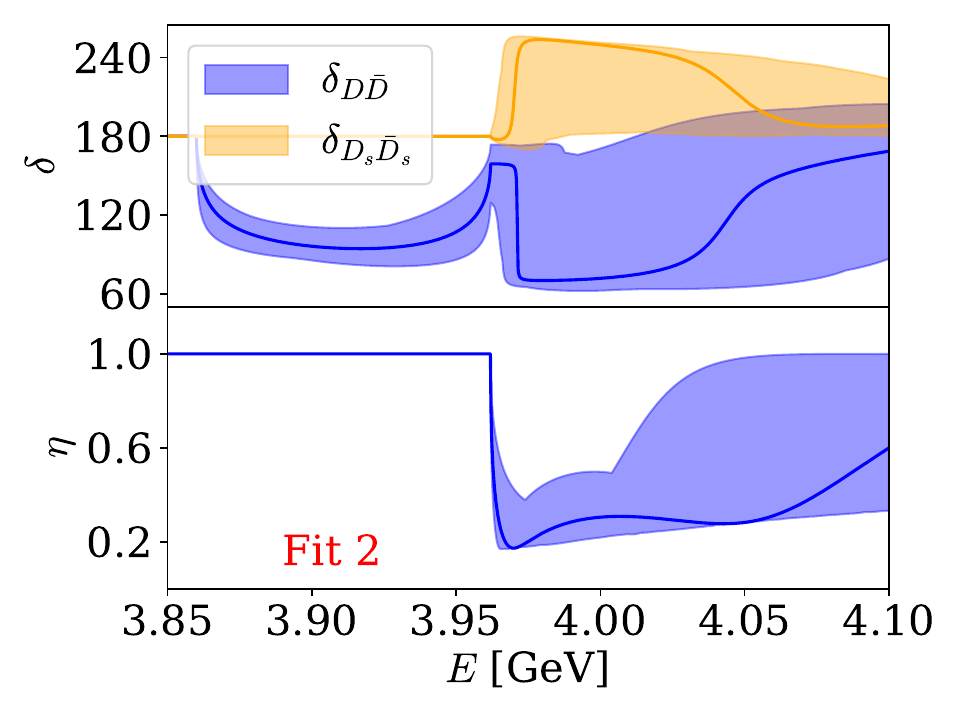}
\caption{Inelasticity ($\eta$) and isoscalar $D\olsi{D}$ and $D_s\olsi{D}_s$  phase-shifts ($\delta_{DD}$ and $\delta_{D_sD_s}$) defined in Eqs.~\eqref{Eq:S-matrix} and \eqref{Eq:relation_S_T} predicted from the LECs determined in  Fit 1 (left) and Fit 2 (right) of Table \ref{tab:LEC-value}. The solid colored bands show $68\%$ confidence level uncertainties inherited from the statistical errors and correlations of the fitted LECs. }
\label{fig:phase-rs}
\end{figure*}

\begin{figure*}
\centering
 \includegraphics[scale=0.5]{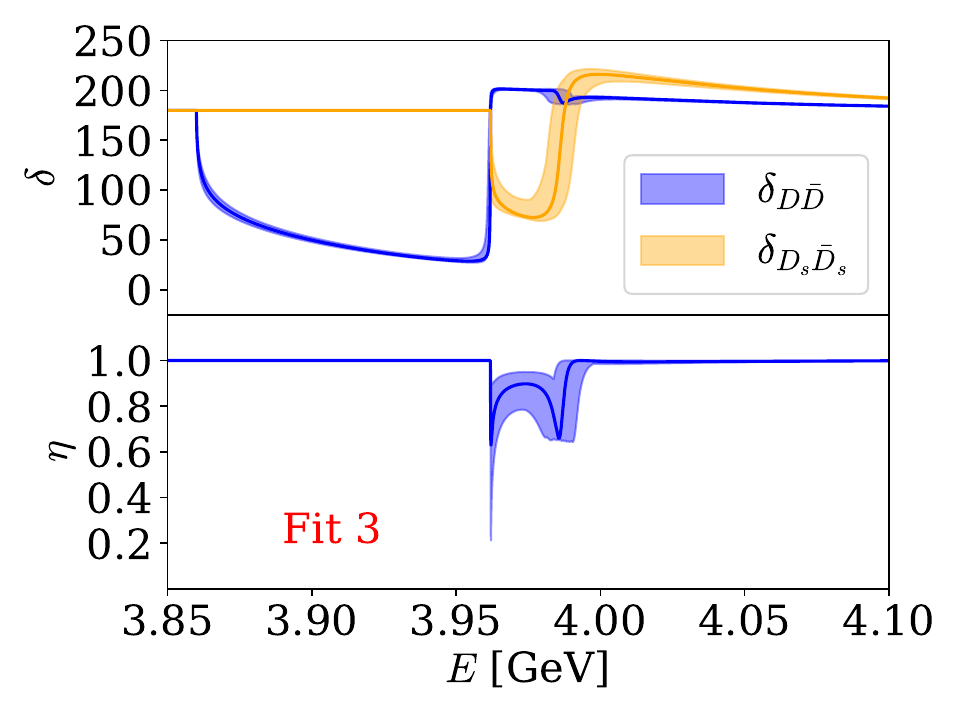}
 \includegraphics[scale=0.5]{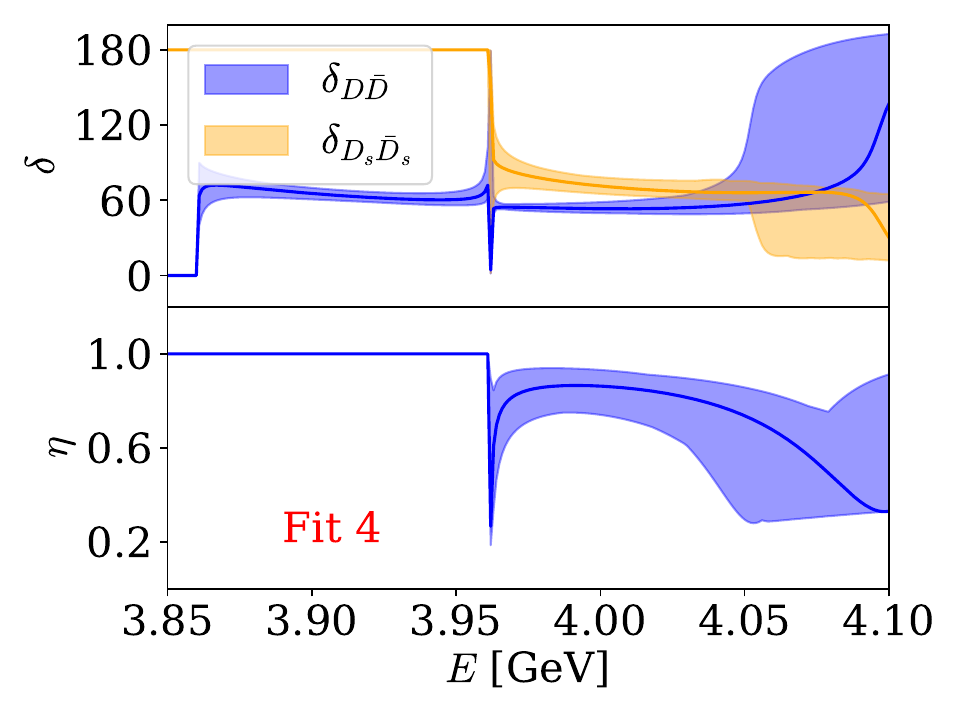}
\caption{Same as in Fig.~\ref{fig:phase-rs} from the LECs determined in  Fit 3 (left) and Fit 4 (right) of Table \ref{tab:LEC-value}. }
\label{fig:phase-mv}
\end{figure*}

We now compare the phase shifts and inelasticity calculated in this work and those obtained by the $K$-matrix parameterization based on LQCD results of Ref.\,\cite{Prelovsek:2020eiw} (\textit{cf.} Fig. 8 in that reference). 
Near the $D_s \olsi{D}_s$ threshold, $\delta_{D_s\olsi{D}_s}$ has a negative slope both in our work (Figs.~\ref{fig:phase-rs} and \ref{fig:phase-mv})\footnote{In the case of Fit 2, this is true only in extremely narrow energy region.} and in Ref.\,\cite{Prelovsek:2020eiw}. This is due to the bound state present slightly below the threshold. The behaviour of this phase shift is however different far above threshold. In particular, in the case of Fit 2, slightly above the $D_s\olsi D_s$ threshold, the real part of $S_{11}$  [Eq.~\eqref{Eq:relation_S_T}] vanishes, which produces the dramatic decrease of $\delta_{D\olsi D}$ which can be appreciated in the right panel of Fig.~\ref{fig:phase-rs}.
In Fit 3, due to the appearance of a narrow third pole $44\,\MeV$ above the $D_s \olsi{D}_s$ threshold (\textit{cf. infra}, Sec.\,\ref{subsec:poles_DDbar_DsDsbar}), there is a strong variation in this phase in our work (Fig.~\ref{fig:phase-mv}, left panel) that is not present in Ref.~\cite{Prelovsek:2020eiw}, where the corresponding pole is broader.
The inelasticity both in Ref.\,\cite{Prelovsek:2020eiw} and our work presents a sharp drop right above the $D_s \olsi{D}_s$ threshold due to the strong $D\olsi{D}$--$D_s\olsi{D}_s$ channel coupling and the nearby resonance (see Sec.~\ref{subsec:poles_DDbar_DsDsbar}). However, the four fits present different shapes and minima for the inelasticity, and these also differ in relation to the analysis results in Ref.\,\cite{Prelovsek:2020eiw}. This can be probably attributed to the different patterns of couplings of both channels to the dynamically generated resonances.  On the other hand, 
the $\delta_{D\olsi{D}}$ phase shift is given in Ref.\,\cite{Prelovsek:2020eiw} above $E \gtrsim 3.9\,\GeV$. The trend is similar to our predictions, although there are sharp differences between both works. As a general observation, the total phase variation from $3.9$ to $4.1\,\GeV$ is larger in Ref.\,\cite{Prelovsek:2020eiw} than in our work.

\subsection{\boldmath Pole analysis for the $D{\olsi{D}}-D_s \olsi{D}_s$ coupled-channel system}\label{subsec:poles_DDbar_DsDsbar}

\begin{table*}[tbp]
    \centering
    \renewcommand\arraystretch{1.6}
    \caption{Pole content of the $D\olsi{D}$--$D_s\olsi{D}_s$ coupled-channel amplitude obtained from the LECs determined in Fit 1 of Table\,\ref{tab:LEC-value}. These dynamically generated states have quantum numbers $J^{PC}=0^{++}$, and show as poles on the RSs specified in the second row. Their positions $E_r=(m_r-i\Gamma_r/2)$, couplings to the relevant channels (Eq.~\eqref{Eq:coupling}), and molecular components (Eq.~\eqref{Eq:posibility-coupled}), in the case of bound states, are also provided. Uncertainties are calculated from the $1\sigma$ errors of the fitted LECs, taking into account their statistical correlations. Meson masses from lattice calculation in Ref.~\cite{Prelovsek:2020eiw} are used. \label{tab:pole-fit1}}
    \begin{ruledtabular}
    \begin{tabular}{l|c|c|c}
        Pole [MeV] & $ 3857^{+2}_{-2}$ & $3952^{+7}_{-13}-i( 24_{-7}^{+9})$ & $4018^{+36}_{-33}-i( 55_{-16}^{+11})$ \\
        \hline
        RS & $(+,+)$ & $(-,+)$ & $(-,-)$  \\ \hline\hline
        Channel&\multicolumn{3}{c}{Coupling $g_{i,r}$ [GeV$^{-1/2}$]}
        \\\hline
        $D\olsi{D}$ & $ 0.78^{+0.16}_{-0.22}$  &$  0.43^{+0.05}_{-0.04}+i( 0.42^{+0.11}_{-0.10})$ &$  0.31^{+0.02}_{-0.06}+i( 0.24^{+0.03}_{-0.04})$  \\ \hline
        $D_s\olsi{D}_s$ & $ 0.39^{+0.14}_{-0.12}$  &$  1.29^{+0.24}_{-0.20}+i( 0.55^{+0.23}_{-0.20})$ &$  0.28^{+0.08}_{-0.13}+i( 0.42^{+0.02}_{-0.04})$ \\ \hline\hline
        Channel&\multicolumn{3}{c}{Molecular component $P_{i,r}$}
        \\\hline
        $D\olsi{D}$ & $ 0.83^{+0.07}_{-0.11}$  & - & - \\ \hline
        $D_s\olsi{D}_s$ & $ 0.01^{+0.00}_{-0.00}$ & - & - \\
    \end{tabular}
    \end{ruledtabular}
\end{table*}

\begin{table*}[!ht]
    \centering
    \renewcommand\arraystretch{1.6}
    \caption{Same as in Table~\ref{tab:pole-fit1}, but using the LECs  determined in Fit 2 of Table\,\ref{tab:LEC-value}.}\label{tab:pole-fit2}
    \begin{ruledtabular}
    \begin{tabular}{l|c|c|c}
        Pole [MeV] & $ 3856^{+3}_{-6}$ & $3970^{+20}_{-6} + i(1^{+33}_{-0}) $ & $4084^{+97}_{-74} - i(76^{+90}_{-55})$ \\
        \hline
        RS & $(+,+)$ & $(+,-)$ & $(-,-)$  \\ \hline\hline
        Channel&\multicolumn{3}{c}{Coupling $g_{i,r}$ [GeV$^{-1/2}$] }
        \\\hline
        $D\olsi{D}$ & $ 0.83^{+0.22}_{-0.31}$  &$  0.24^{+0.80}_{-0.11} + i(0.46^{+0.25}_{-0.43}) $ &$  0.37^{+0.17}_{-0.20} + i(0.39^{+0.07}_{-0.11}) $ \\ \hline
        $D_s\olsi{D}_s$  & $ 0.37^{+0.15}_{-0.16} $  &$  -0.55^{+0.11}_{-0.81} + i(0.79^{+0.69}_{-0.20}) $ &$  0.25^{+0.13}_{-0.16} + i(0.43^{+0.06}_{-0.10}) $ \\ \hline\hline
        Channel&\multicolumn{3}{c}{Molecular component $P_{i,r}$}
        \\\hline
        $D\olsi{D}$ & $ 0.89^{+0.07}_{-0.13} $ & - & -   \\ \hline
        $D_s\olsi{D}_s$ & $ 0.01^{+0.01}_{-0.01}$ & - & - \\
    \end{tabular}
    \end{ruledtabular}
\end{table*}

\begin{table*}[!ht]
    \centering
    \renewcommand\arraystretch{1.6}
    \caption{Same as in Table~\ref{tab:pole-fit1}, but using the LECs determined in Fit 3 of Table\,\ref{tab:LEC-value}. }\label{tab:pole-fit3}
    \begin{ruledtabular}
    \begin{tabular}{l|c|c|c}
        Pole [MeV] & $ 3859.6^{+0.4}_{-0.9}$ & $3961\pm1 - i(1\pm1) $ & $4006^{+20}_{-5} -i(14^{+9}_{-2}) $ \\
        \hline
        RS & $(+,+)$ & $(-,+)$ & $(-,-)$  \\ \hline\hline
        Channel&\multicolumn{3}{c}{Coupling $g_{i,r}$ [GeV$^{-1/2}$] }
        \\\hline
        $D\olsi{D}$ & $ 0.49^{+0.15}_{-0.19}$  &$  0.09^{+0.04}_{-0.04} + i(0.08^{+0.05}_{-0.04} )$ &$  0.22^{+0.05}_{-0.02} + i(0.09^{+0.02}_{-0.01} )$  \\ \hline
        $D_s\olsi{D}_s$ & $ 0.12^{+0.09}_{-0.07}$   &$  0.65^{+0.17}_{-0.22}+ i(0.14^{+0.10}_{-0.07}) $ &$  0.23^{+0.07}_{-0.02} + i(0.30^{+0.03}_{-0.04}) $ \\ \hline\hline
        Channel&\multicolumn{3}{c}{Molecular component $P_{i,r}$}
        \\\hline
        $D\olsi{D}$ & $ 0.99^{+0.01}_{-0.07}$  & - & -  \\ \hline
        $D_s\olsi{D}_s$ & $ 0.5^{+0.9}_{-0.4} \times 10^{-3}$  & - & -  \\
    \end{tabular}
    \end{ruledtabular}
\end{table*}

\begin{table*}[!ht]
    \centering
    \renewcommand\arraystretch{1.6}
    \caption{Same as in Table~\ref{tab:pole-fit1}, but using the LECs, but using the LECs determined in Fit 4 of Table\,\ref{tab:LEC-value}. }\label{tab:pole-fit4}
    \begin{ruledtabular}
    \begin{tabular}{l|c|c|c}
        Pole [MeV] & $ 3859.9^{+0.1}_{-0.9}$ & $3962.0^{+0.5}_{-0.2}- i(0^{+0.4}_{-0.0}) $ & $4102^{+45}_{-47}- i(43^{+26}_{-22}) $ \\
        \hline
        RS & $(-,+)$ & $(-,+)$ & $(-,-)$  \\ \hline\hline
        Channel&\multicolumn{3}{c}{Coupling $g_{i,r}$ [GeV$^{-1/2}$]}
        \\\hline
        $D\olsi{D}$ & $ 0.34^{+0.17}_{-0.86} i$ &$  0.00^{+0.03}_{-0.00}+ i(0.06^{+0.04}_{-0.14}) $ &$  0.37^{+0.09}_{-0.11} + i(0.26^{+0.05}_{-0.05}) $  \\ \hline
        $D_s\olsi{D}_s$ & $0.04^{+0.04}_{-0.12} i$ &$  -0.01^{+0.30}_{-0.46} + i(0.16^{+0.18}_{-0.40}) $ &$  0.28^{+0.06}_{-0.07} + i(0.31^{+0.07}_{-0.06}) $ \\ \hline\hline
        Channel&\multicolumn{3}{c}{Molecular component $P_{i,r}$}
        \\\hline
        $D\olsi{D}$ & $ 0.99^{+0.57}_{-0.62}$ & - & -  \\ \hline
        $D_s\olsi{D}_s$ & $ 3.5^{+8.1}_{-3.5}\times 10^{-3}$ & - &  -  \\
    \end{tabular}
    \end{ruledtabular}
\end{table*}

The energy levels in Figs. \ref{Fig:energy-level-rest} and \ref{Fig:energy-level-mv} (as well as the phase shifts) give hints about the presence of poles in the amplitudes, but it is challenging to directly get the detailed information about poles from the energy levels. For example, confirming the existence of a bound state below the $D\olsi{D}$ threshold solely based on the energy levels in the top panels of Fig. \ref{Fig:energy-level-mv} is not straightforward. Hence, a detailed analysis in the infinite volume is essential to unravel crucial details, including the pole position, couplings, and contributions from different channels. Such a study will provide a more comprehensive understanding of the underlying dynamics which governs the system.

Considering the fact that the potential remains identical in both the finite and infinite volumes up to exponentially suppressed difference~\cite{Doring:2011vk}, we can employ the LECs in Table \ref{tab:LEC-value} to predict the possible charmonium-like states in the $D\olsi{D}-D_s \olsi{D}_s$ coupled-channel system. However, the direct utilization of the physical charmed-meson masses to predict the hidden charm states poses challenges.
On the one hand, the difference in the charmonium spin averaged mass of Eq.~\eqref{Eq:average_charmonium} between the LQCD calculation~\cite{Prelovsek:2020eiw} and the experimental measurements is about $34\,\MeV$. This implies the charm quark mass used in the LQCD simulation is not the physical one, and this also hints probably to the existence of some discretization errors. For simplicity, in what follows, we will refer to an unphysical charm quark mass, but it would comprise both effects. Consequently,
the unphysical charm quark mass affects the value of the bare mass $\mcos$ listed in Table \ref{tab:LEC-value}, rendering the parameter $\mcos$ unsuitable for direct use in physical scenarios. On the other hand,
the gaps of the isospin-average thresholds between LQCD~\cite{Prelovsek:2020eiw} and experimental values~\cite{ParticleDataGroup:2024cfk} of $D\olsi{D}$ ($D_s\olsi{D}_s$) thresholds are\footnote{Excluding the effect of the unphysical charm mass evaluated by the spin averaged mass of charmonium, the splitting of the $D\olsi{D}$ thresholds between LQCD and experiment still remains about $92$~MeV.}
\begin{subequations}
\begin{align}
M_{D\olsi{D}}^{L}-M_{D\olsi{D}}^{\text{exp}} 
& \simeq 126\,\MeV\,,\\
M_{D_s\olsi{D}_s}^{L}-M_{D_s\olsi{D}_s}^{\text{exp}} 
& \simeq 25\,\MeV\,,
\end{align}
\end{subequations}
Such a large difference ($\sim$100 MeV) between the mismatch of the $D\olsi{D}$ and $D_s\olsi{D}_s$ thresholds highlights the challenge of neglecting the light-quark mass (relevant to the pion mass) dependence and directly estimating the pole positions with physical values for  $m_{D}$ and $m_{D_s}$. Alternatively, we can calculate the pole positions and then estimate the mass difference between pole masses and the relevant thresholds using the values for  $m_{D}$ and $m_{D_s}$ obtained in the LQCD simulation of Ref.~\cite{Prelovsek:2020eiw}. The latter splittings are expected to be less sensitive to the effect of the unphysical light quark mass and thus to provide a more accurate estimate of the pole positions. The final masses of the dynamically generated states that will be shown below are obtained by adding these splittings to the physical thresholds constructed with the experimental  $m_{D}$ and $m_{D_s}$ masses.  

The $T$-matrix for the $D\olsi{D}$--$D_s \olsi{D}_s$ coupled-channel system has four RSs, \textit{i.e.} $(+,+)$, $(-,+)$, $(-,-)$, and $(+,-)$,\footnote{For each channel, the corresponding loop function $G^{ii}_\text{QM}(E)$ in Eq.~\eqref{eq:Gfun:channel} can be computed in its respective physical ($+$) or unphysical ($-$) RS. The four RSs of the coupled-channel system are denoted according to the combination of RSs taken for each channel.} which are reached by means of analytical continuation.
Poles can appear on different RSs, and their properties (position, residues, etc.) can be computed directly from the amplitudes on the appropriate RS. We look for relevant poles on different RSs using the LECs determined in each of the four fits collected in Table~\ref{tab:LEC-value}.  Results are listed in Tables \ref{tab:pole-fit1}, \ref{tab:pole-fit2}, \ref{tab:pole-fit3}, and \ref{tab:pole-fit4}. In general, for each of the fits we find three poles in the $D\olsi{D}$--$D_s\olsi{D}_s$ coupled-channel system: two poles close to the $D\olsi{D}$ and $D_s\olsi{D}_s$ thresholds, respectively, and an additional one above the $D_s\olsi{D}_s$ threshold. The various poles for the different fits are summarily displayed in Fig.~\ref{Fig:pole_position}.

\begin{figure}[t!]
    \includegraphics[width=0.5\textwidth,keepaspectratio]{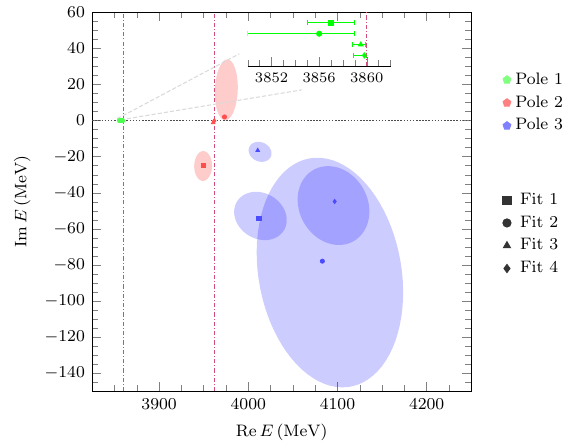} 
    \caption{Pole positions ($J^{PC}=0^{++}$) found in the $D\olsi{D}$-$D_s\olsi{D}_s$ coupled-channel amplitude and compiled in Tables \ref{tab:pole-fit1}-\ref{tab:pole-fit4}. The colors, green, red, and blue, denote the poles close to the $D\olsi{D}$, $D_s\olsi{D}_s$ thresholds and that located in the $(-,-)$ RS, respectively. The shapes, square, circle, triangle, and diamond, denote the poles calculated with the LECs from Fits 1, 2, 3, and  4, respectively. The vertical lines  mark the $D\olsi{D}$ and $D_s\olsi{D}_s$ thresholds. The inset shows the location of the (bound state) pole 1. Note that in the case of pole 2, the results from Fits 3 and 4  overlap and thus are not distinguishable in the plot. 
    \label{Fig:pole_position} }
\end{figure}

A pole below the $D\olsi{D}$ threshold is found for the sets of LECs corresponding to Fits 1, 2, and 3, as listed in the second column of Tables \ref{tab:pole-fit1} - \ref{tab:pole-fit4}. In all three cases, its coupling to $D\olsi{D}$ is much stronger than to $D_s \olsi{D}_s$. Therefore, it can be identified as a  $J^{PC}=0^{++}$ bound state whose wave function is dominated by the $D\olsi{D}$ molecular component, with the contribution of $\chi_{c0}(2P)$ being tiny. In the case of Fit 4, a pole below the $D\olsi{D}$ threshold is also observed, listed in  Table \ref{tab:pole-fit4} as a virtual state on the second [$(-,+)$] RS. Considering $1\sigma$ uncertainty of the LECs, $94\%$ of the bootstrap samples result in a virtual state, while $6\%$ result in a bound state. Because in either case it is so close to the threshold, it becomes quite relevant and should significantly influence the physical scattering.  To reduce the light quark-mass dependence, we evaluate the mass splitting between the position of this pole and the $D\olsi{D}$ threshold $(2m_D)$, \textit{i.e.}, $\Delta E=-3\pm2$, $-4^{+6}_{-3}$, $-0.5^{+0.7}_{-0.4}$, $-0.2^{+0.9}_{-0.2}\,\MeV$, 
for the four fits carried out in this work. Taking into account uncertainties, the four determinations of the binding energy are in excellent agreement with the value reported in the LQCD analysis of Ref.~\cite{Prelovsek:2020eiw}. The binding energy of this $0^{++}$ state is also compatible with the predictions of Refs.\,\cite{Ji:2022uie,Ji:2022vdj}.
 
The second pole, located near the $D_s\olsi D_s$ threshold, is found on the $(+,-)$ RS for Fit 2 and the $(-,+)$ RS for the other three fitting scenarios,
and it has significantly larger couplings to the $D_s \olsi{D}_s$ channel, as can be seen in Tables \ref{tab:pole-fit1} - \ref{tab:pole-fit4}. The mass splittings to the $D_s \olsi{D}_s$ threshold for this second pole are 
$(-10^{+13}_{-7}-24_{-7}^{+9}i)$, 
$(8^{+20}_{-6} - 1^{+33}_{-0} i)$, 
$(-1^{+1}_{-1} - 1^{+1}_{-1} i)$, 
$(0.1^{+0.2}_{-0.5}- 0^{+0.3}_{-0.0} i) \,\MeV$  in each of the four fits, respectively. Within $1\sigma$, our findings for this second pole agree also well to the result quoted in the LQCD work of Ref. ~\cite{Prelovsek:2020eiw}, as well as the predictions in Refs.\,\cite{Ji:2022uie,Ji:2022vdj}. This pole is associated with the $X(3960)$, a hidden-charm state recently observed by the \LHCb Collaboration in the $D_s \olsi{D}_s$ invariant mass distribution~\cite{LHCb:2022aki} (see also the discussion in Ref.\,\cite{Abreu:2023rye}). The small coupling of this pole to the $D \olsi{D}$ channel can also explain the fact that the candidate of the $D_s\olsi{D}_s$ molecule $X(3960)$ has not been yet observed in the $D\olsi{D}$ invariant mass distribution. Besides, the uncertainty of the second pole in Tables \ref{tab:pole-fit1} and \ref{tab:pole-fit2} are too large to confirm the precise pole position. 

\begin{figure}[tb]
    \centering
    \includegraphics[width=\linewidth]{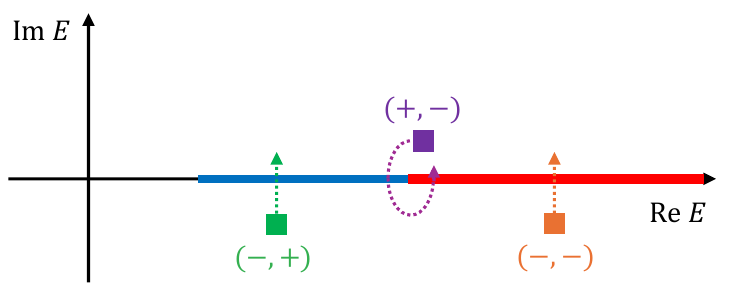}
    \caption{Paths from poles on the $(-,+)$ (green dotted line), $(-,-)$ (orange dotted line), and $(+,-)$ (purple dotted line) RSs to the physical region. Poles are denoted as filled squares. The blue and red thick solid lines denote the cuts from the $D\olsi{D}$ and $D_s\olsi{D}_s$ thresholds, respectively. }
    \label{fig:paths}
\end{figure}
One comment is in order on the pole on the $(+,-)$ RS for Fit 2. 
Because of the Schwarz reflection principle, each resonance pole comes as a complex conjugated pair on the same RS.
The RSs on which the resonance pole can reach the physical region (defined as the upper edge along the cuts on the physical RS), from the pole location by crossing the cuts along a straight path perpendicular to the real axis, are either $(-,+)$ or $(-,-)$. In such a case, one chooses the pole on the lower half plane; see the green and orange squares and the corresponding paths in Fig.~\ref{fig:paths}.
However, for the conjugated pole pair on the $(+,-)$ RS, the one closer to the physical region is the one on the upper half plane, since it can reach the physical region by circling the $D_s\olsi{D}_s$ threshold shown as the purple path in Fig.~\ref{fig:paths}, crossing the cuts twice.
Such a pole is shielded by the $D_s\olsi{D}_s$ threshold and shows up in the line shape of scattering amplitudes as a threshold cusp~\cite{Zhang:2024qkg}, instead of a smooth resonance peak.
The cusp behavior at the $D_s\olsi{D}_s$ threshold in the phase shift can be seen in the right plot of Fig.~\ref{fig:phase-rs}. 
\begin{figure*}[tb]
    \centering
    \includegraphics[width=0.33\linewidth]{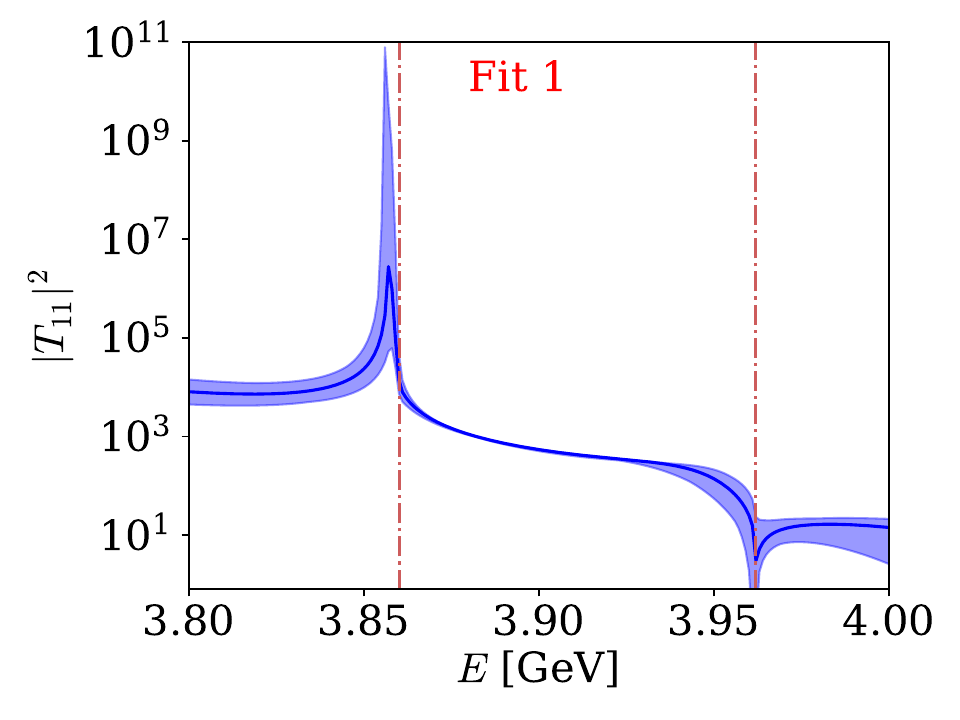}
    \includegraphics[width=0.33\linewidth]{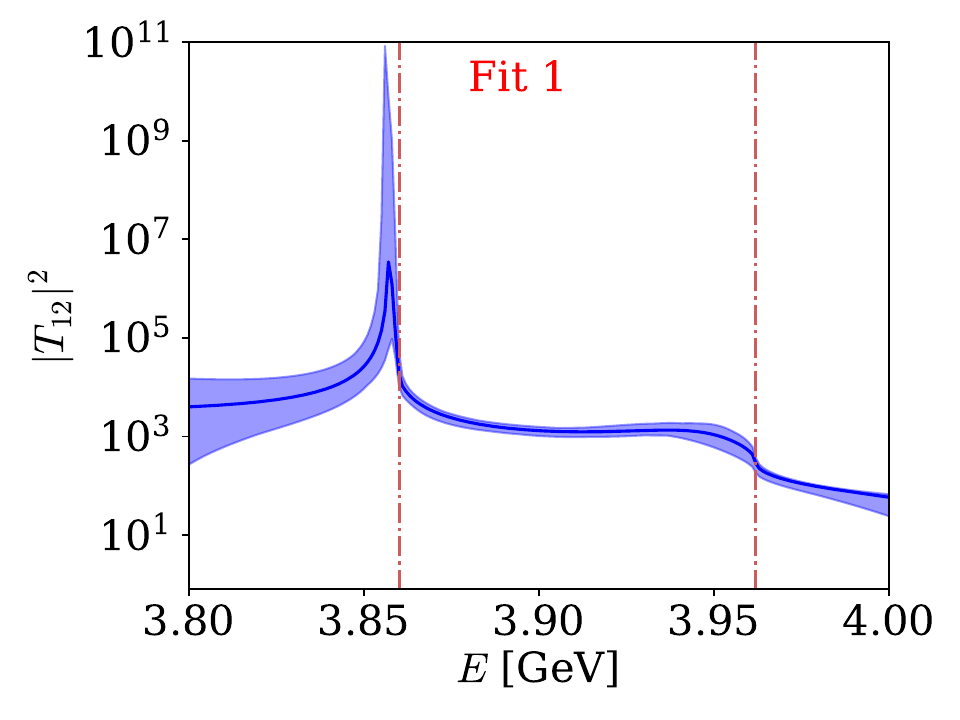}
    \includegraphics[width=0.33\linewidth]{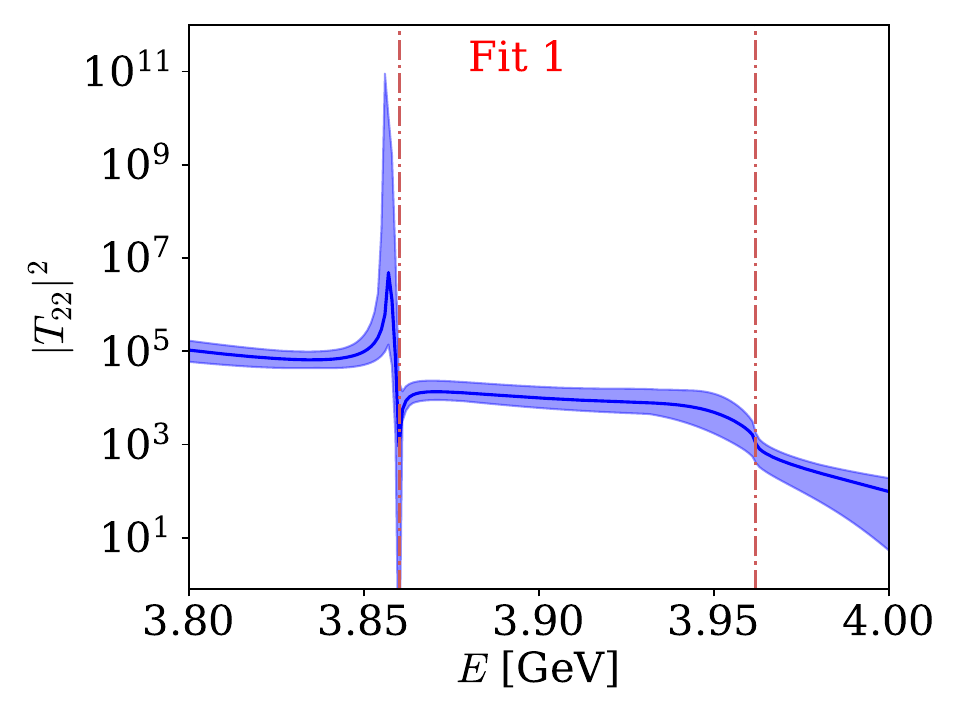}\\
    \includegraphics[width=0.33\linewidth]{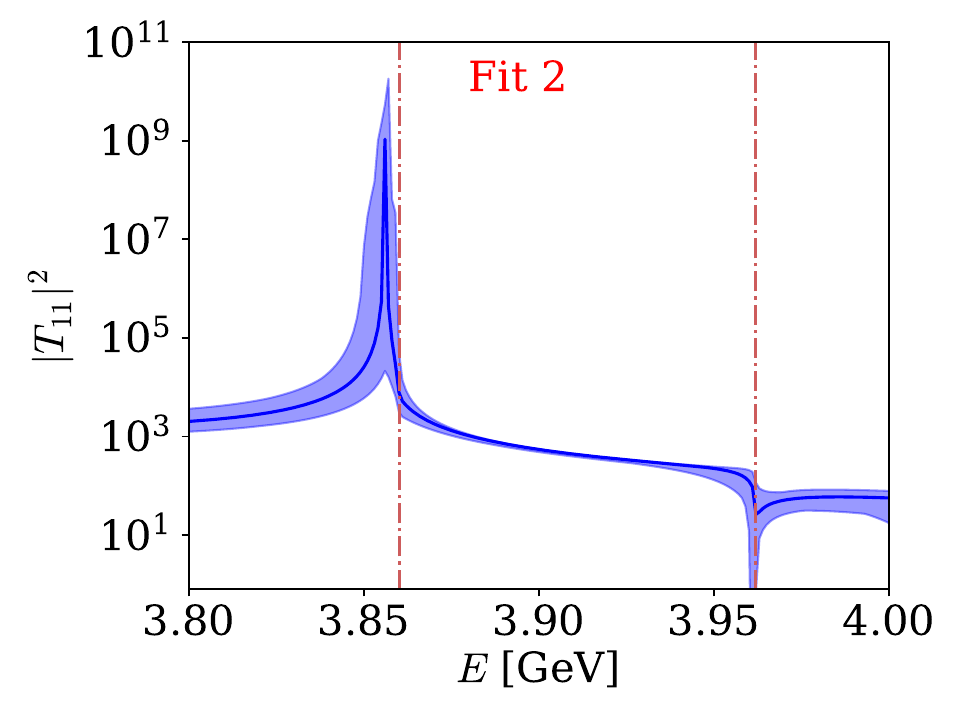}
    \includegraphics[width=0.33\linewidth]{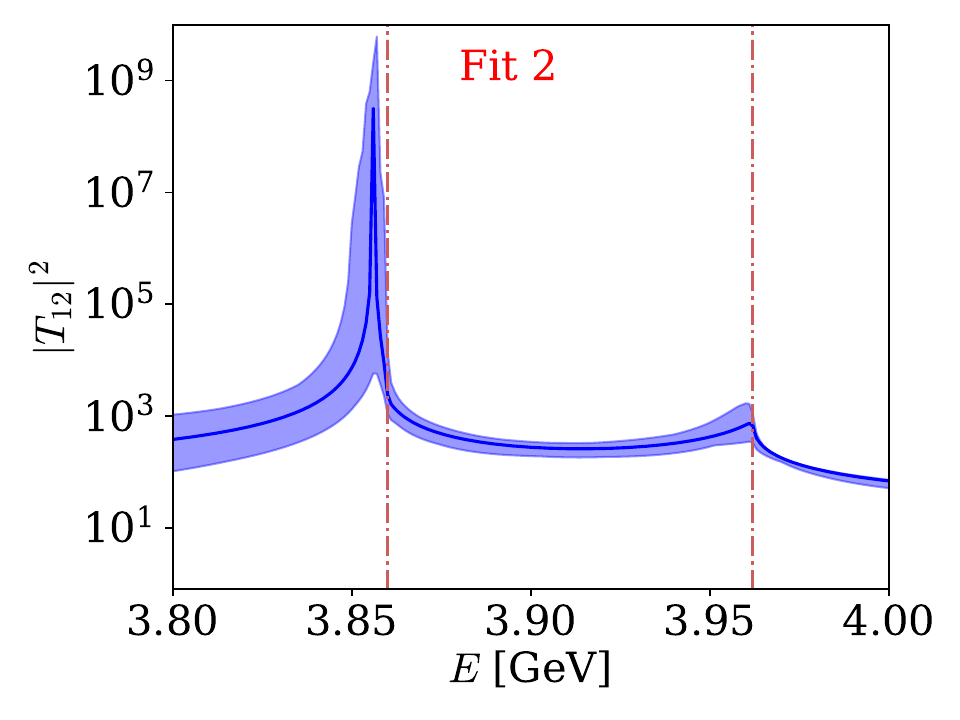}
    \includegraphics[width=0.33\linewidth]{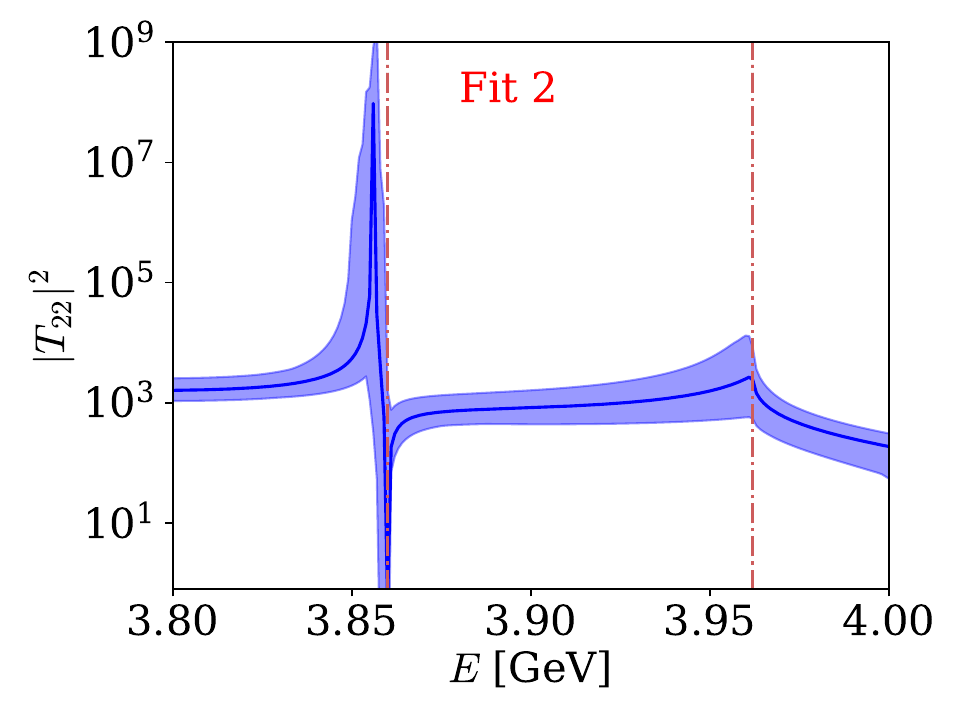}\\
    \includegraphics[width=0.33\linewidth]{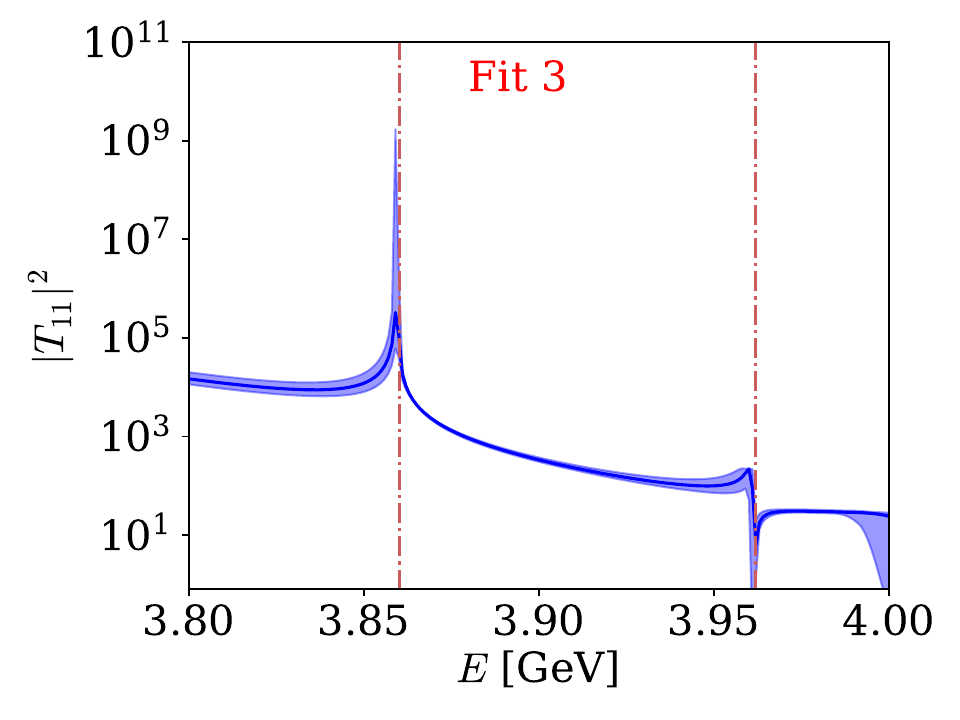}
    \includegraphics[width=0.33\linewidth]{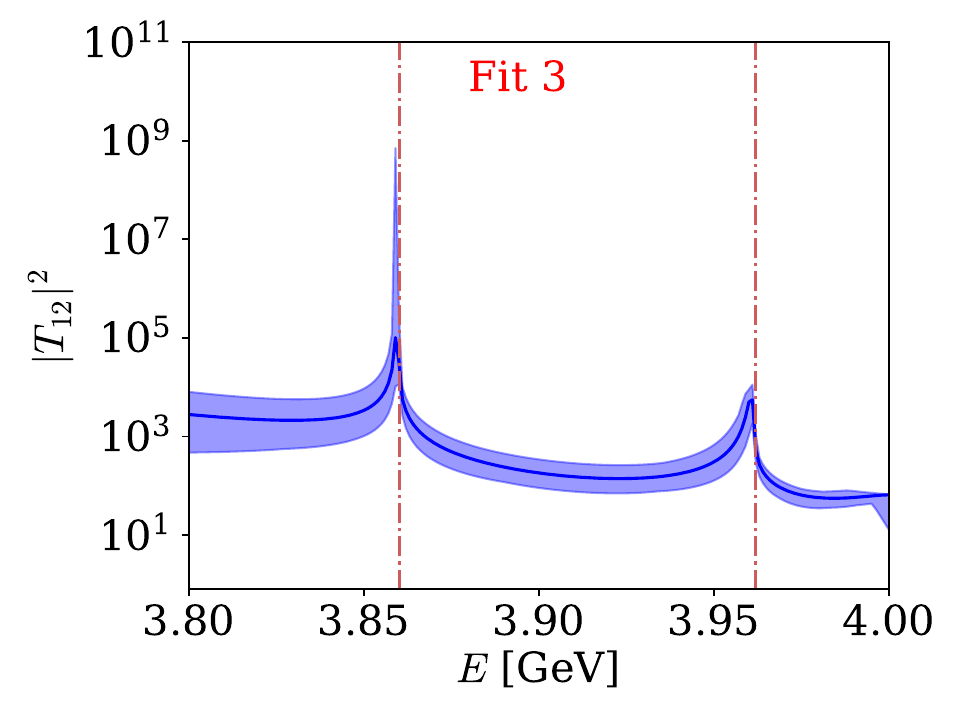}
    \includegraphics[width=0.33\linewidth]{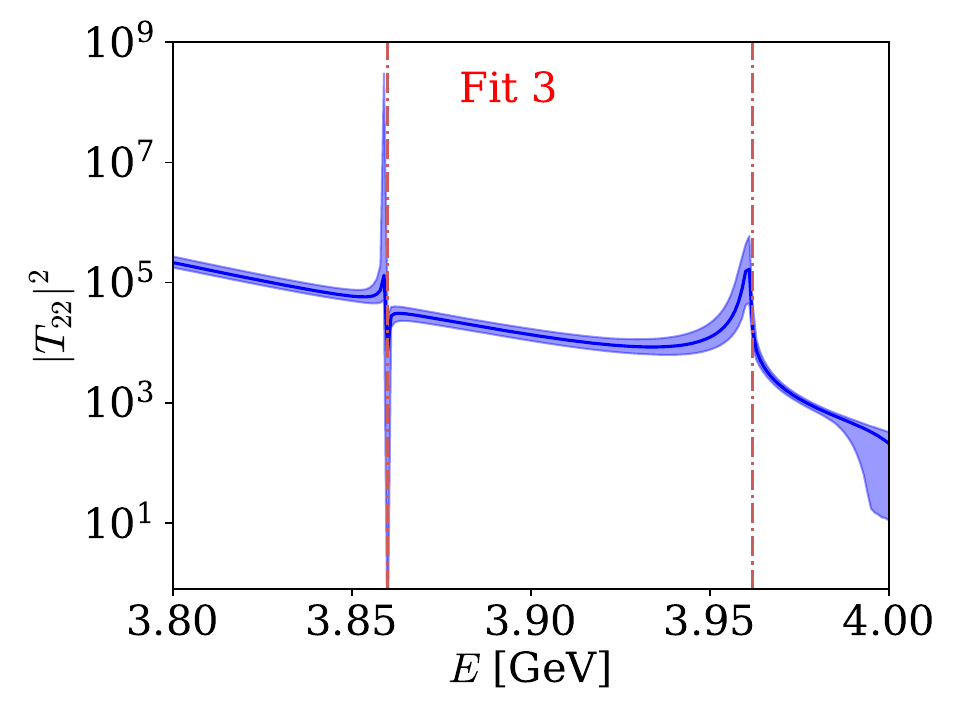}\\
    \includegraphics[width=0.33\linewidth]{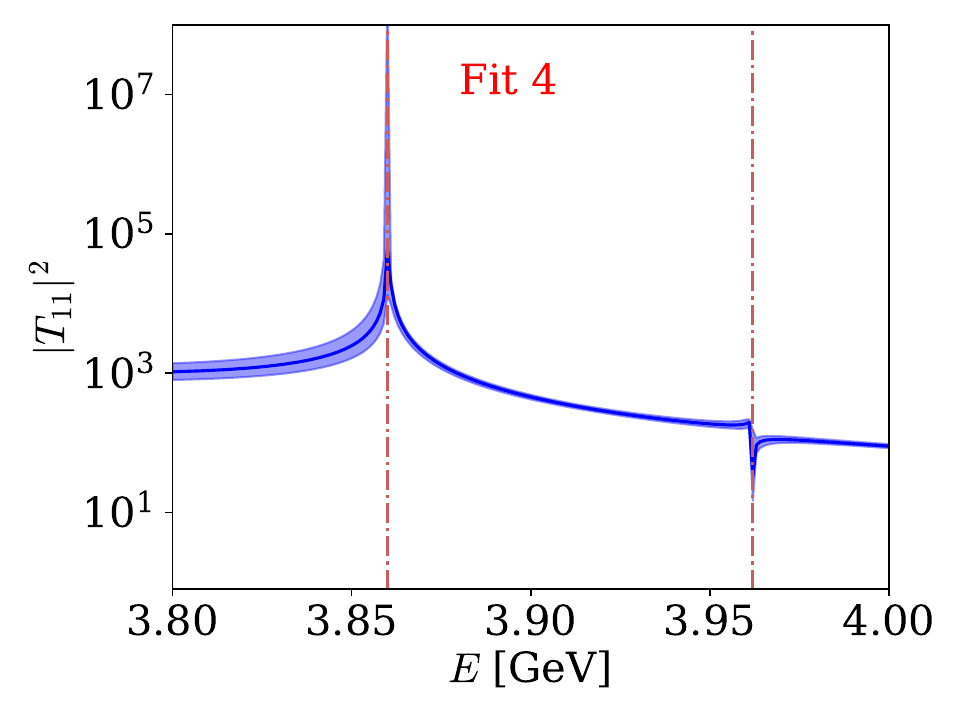}
    \includegraphics[width=0.33\linewidth]{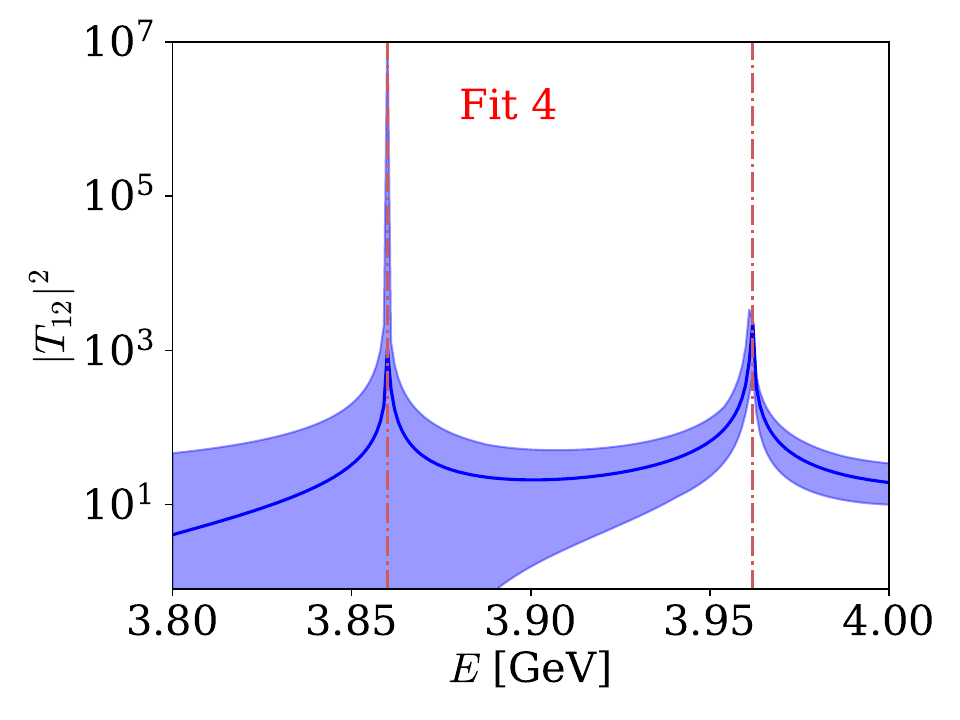}
    \includegraphics[width=0.33\linewidth]{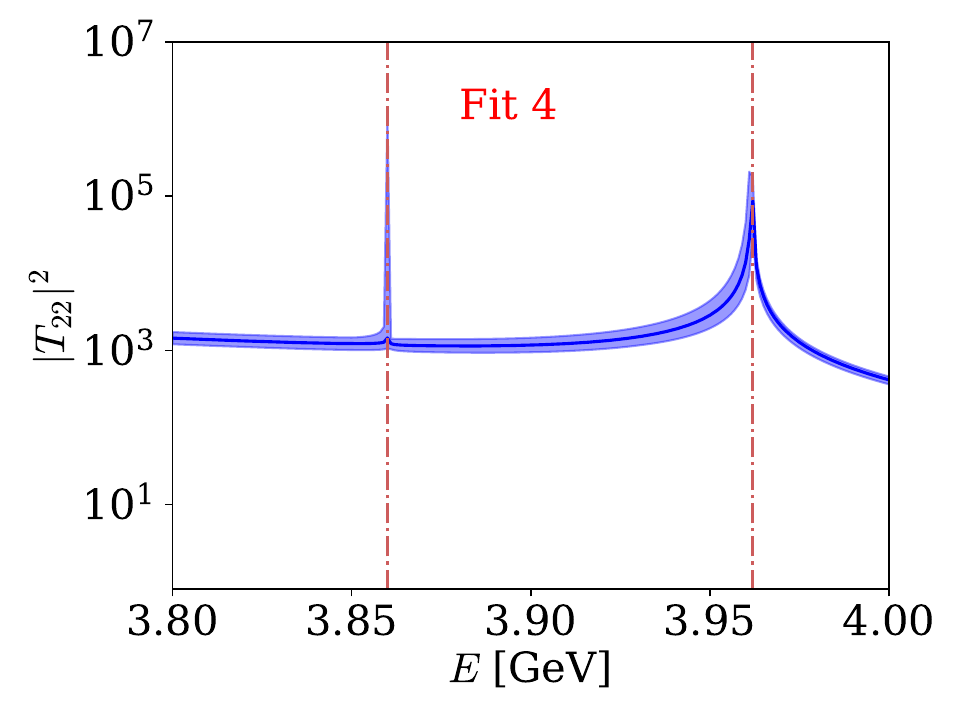}\\
    \caption{Line shapes of the $D\olsi{D}$--$D_s\olsi{D}_s$ coupled-channel $T$-matrix elements from Fits 1, 2, 3, and 4. The vertical dash-dotted lines mark the $D\olsi{D}$ and $D_s\olsi{D}_s$ thresholds. }
    \label{fig:Tsq}
\end{figure*}
{It is seen more clearly from the line shapes of $|T_{ij}|^2$, with $T_{ij}$ the $T$-matrix elements of the $D\olsi{D}$--$D_s\olsi{D}_s$ coupled channels, shown in Fig.~\ref{fig:Tsq}.} The dip in the line shape of $|T_{11}|^2$ near the $D_s\olsi{D}_s$ threshold is a result of the strong attraction in the $D_s\olsi{D}_s$ channel. Similarly, the dip in the line shape of $|T_{22}|^2$ near the $D\olsi{D}$ threshold is due to the strong attraction in the $D\olsi{D}$ channel.
This behavior is a universal feature of the two-channel scattering~\cite{Dong:2020hxe}. It occurs in $|T_{ii}|^2$ if the absolute value of the single-channel scattering length (when the channel $i$ is switched off) in channel-$j$ ($j\neq i$) is large.

In addition, we find an additional  pole above the $D_s\olsi{D}_s$ threshold, around 4--4.1 GeV,  in the $(-,-)$ RS (see the last columns of Tables \ref{tab:pole-fit1}--\ref{tab:pole-fit4}) and the blue ellipses in Fig.~\ref{Fig:pole_position}) . This broad $0^{++}$ resonance  is also predicted in Ref.~\cite{Prelovsek:2020eiw}. In our scheme, this pole has a sizable coupling to the $D_s \olsi{D}_s$ channel, larger than that in Ref.~\cite{Prelovsek:2020eiw}. Its origin is rooted in the charmonium $\chi_{c0}(2P)$ state. We have checked that the pole position evolves to the bare $\chi_{c0}(2P)$ mass as we decrease the $d$ parameter to zero. 

In Ref.~\cite{Prelovsek:2020eiw} another pole with quantum numbers $J^{PC}=2^{++}$ is also predicted. However, the energy levels in the $B_1$ representation, which receives the contribution of the $D$-wave, are not analyzed in this work. Thus, our analysis neglects the contribution of the $D$-wave and precludes us from predicting such a state.

\section[Results for the $D\olsi{D}{}^\ast$ channel]{\boldmath Results for the $D\olsi{D}{}^\ast$ channel}\label{Sec:1++}

We also explore the low energy scattering in the isoscalar $1^{++}$ sector, excluding the $D^*\olsi{D}{}^\ast$ channel due to its significant separation from the $D \olsi{D}{}^\ast$ threshold. In the $T$-matrix of Eq. \eqref{Eq:T-matrix-coupled}, the couplings $C_{0X}$ and {$d^{\prime}$} of the potential in Eq.~\eqref{Eq:potential_1++} and $\mcov$, the bare mass of $2P$ charmonium, are free parameters which should be fitted to the energy levels provided in the LQCD calculation of Ref.~\cite{Prelovsek:2013cra}. They were obtained by considering  $c\olsi c$, $D \olsi{D}{}^\ast$ and $J/\psi \omega$ LQCD interpolating operators. The volume of the lattice calculation was $V=16^3\times 32$ with $a=0.1239(13)\,\fm$, and therefore the spatial box size was approximately equivalent to $2\,\fm$. The data reported in Ref.~\cite{Prelovsek:2013cra} include three energy levels collected in the Table~\ref{Tab:levels_1++}. As mentioned in Eq. \eqref{Eq:green-coupled}, we use a Gaussian form factor to regularize the UV divergence with two different values of the cutoff $\Lambda=0.5\, \GeV$ and $1\, \GeV$. Note that, in Table \ref{Tab:levels_1++}, we compiled the momenta corresponding to each of the energy levels obtained in Ref.~\cite{Prelovsek:2013cra}, which turn out to be sizable and even of the order of $\Lambda$ for the $n=2$ and $n=3$ ones. 

The $1^{++}$ single channel potential in Eq.~\eqref{Eq:potential_single}, as it stands, has three unknown parameters, $C_{0X}$, $d'$, and $\mcov$. As assumed in this latter equation, HQSS predicts $d'=d$ at LO, but in the present analysis we will allow for HQSS breaking corrections. Furthermore, we extend the formalism in the spirit of Ref.~\cite{Cincioglu:2016fkm}, to include an energy-dependent term in the potential, so that it becomes:
\begin{equation}\label{Eq:potential_1++_new}
V(1^+ \lvert 2^+) = C_{0X} + b\frac{p^2}{4\mu^2} + \frac{{d'}^2}{E - {\mathring{m}_{\chi_{c1}|\chi_{c2}}}}\,,
\end{equation}
and hence the number of unknown parameters is four. However, the number of energy levels in Table~\ref{Tab:levels_1++} is only three, which complicates the determination of the unknown parameters. To overcome this challenge, we employ the following strategy. We will explore two different frameworks, $b=0$ and $b\neq 0$. We will fit two or three of the parameters ($C_{0X}$, $\mcov$, and eventually $b$) for different values of $d'$. For the latter, it is possible to estimate its value from quark models. From the $^3P_0$ model of Ref.~\cite{Ortega:2010qq} a value $d'\approx 0.2\,\fm^{1/2}$ was obtained. On the other hand, according to the values of $d$ in Table \ref{tab:LEC-value}, we expect the coupling constant $|d'|<1\,\fm^{1/2}$. Although we are not imposing HQSS, we can still use this estimate to constrain the value of the coupling $d'$. Therefore, it is reasonable to take the range $ 0 < d' < 1\,\fm^{1/2}$.\footnote{Note that the coupling $d'$ appears in the potential as $d'^2$, so the sign of $d'$ is irrelevant for our purposes.}

The interacting energy levels $E_n(L)$ are computed as the poles of the $\widetilde{T}$-matrix defined in Eq. \eqref{Eq:T-matrix-coupled-FV}, \textit{i.e.} $\widetilde{T}^{-1}(E_n(L),L)=0$. The LQCD calculation of Ref.~\cite{Prelovsek:2013cra} involves unphysical $D$ and $D^\ast$ meson masses, as well as nonrelativistic energy-momentum relations. To make a closer comparison with Ref.~\cite{Prelovsek:2013cra}, in our finite volume calculation we replace the $\omega(\vec q)$ in Eq. \eqref{Eq:two-point-rest} by the modified dispersion relation used in the latter reference,
\begin{align}\label{Eq:omegalat}
\omega^\text{lat}(\vec q)  = \drmd{1} + \drmds{1} + \frac{\drmd{2}+\drmds{2}}{2\drmd{2}\drmds{2}}\vec q\,^2 -\frac{\drmd{4}^3+\drmds{4}^3}{8\drmd{4}^3\drmds{4}^3} \vec q\,^4~,
\end{align}
where the values of the parameters $m_{D^{(\ast)},j}$ ($j=1,2$ and $4$) are taken from Table~VI of Ref.~\cite{Lang:2014yfa}. To summarize, in our finite-volume $\widetilde{T}$-matrix, on the one hand, instead of the $D$ and $D^\ast$ physical masses we use $m_{D,1}$ and $m_{D^\ast,1}$, respectively; on the other hand, the nonrelativistic energy-momentum relation in Eq.~\eqref{Eq:two-point-rest} is replaced by the expression in Eq.~\eqref{Eq:omegalat} to compute the finite-volume loop function in Eq.~\eqref{Eq:green-coupled-FV}.

\begin{figure*}
	\centering
	\includegraphics[scale=0.35]{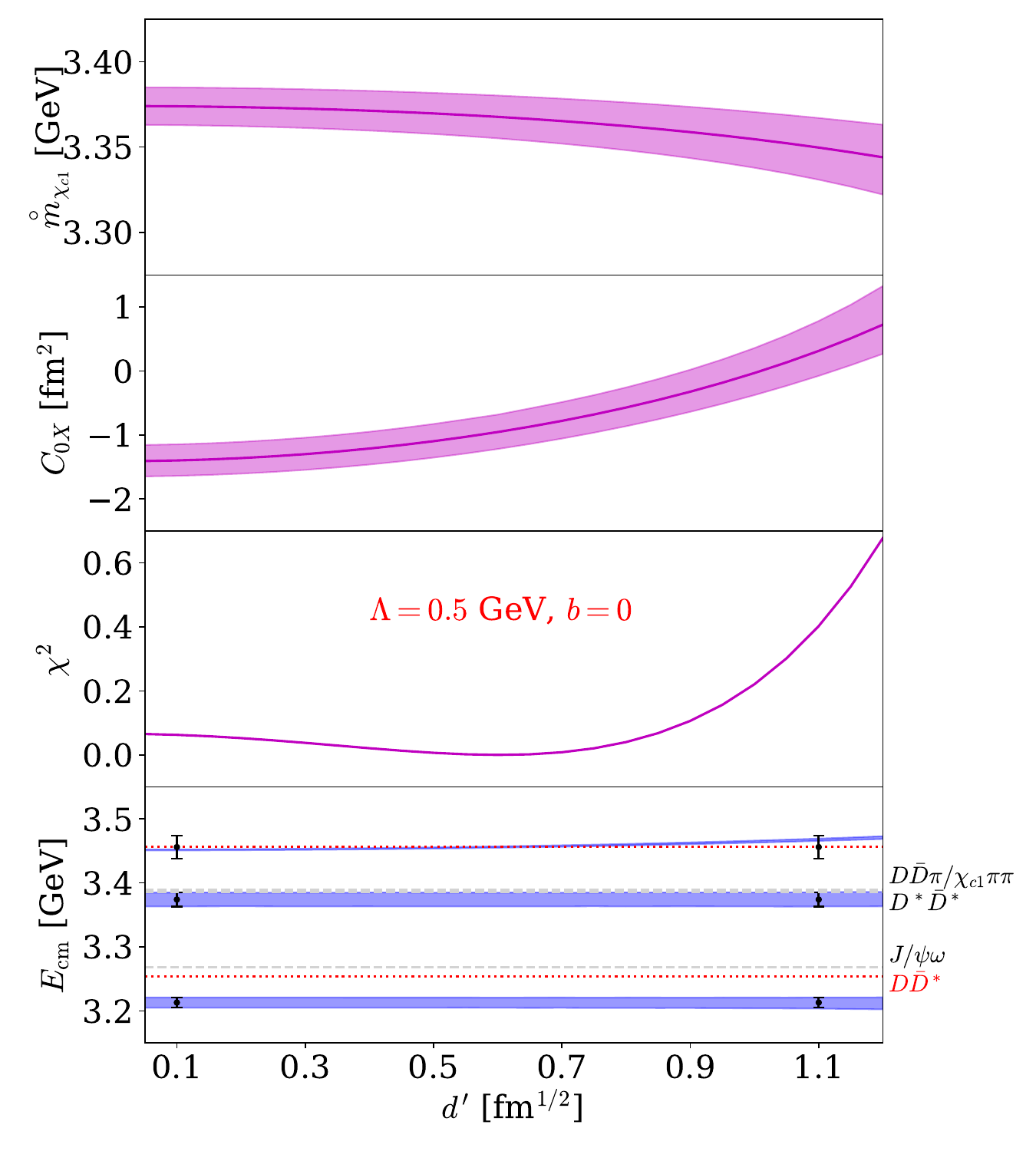}\hspace{0.5cm}
\includegraphics[scale=0.35]{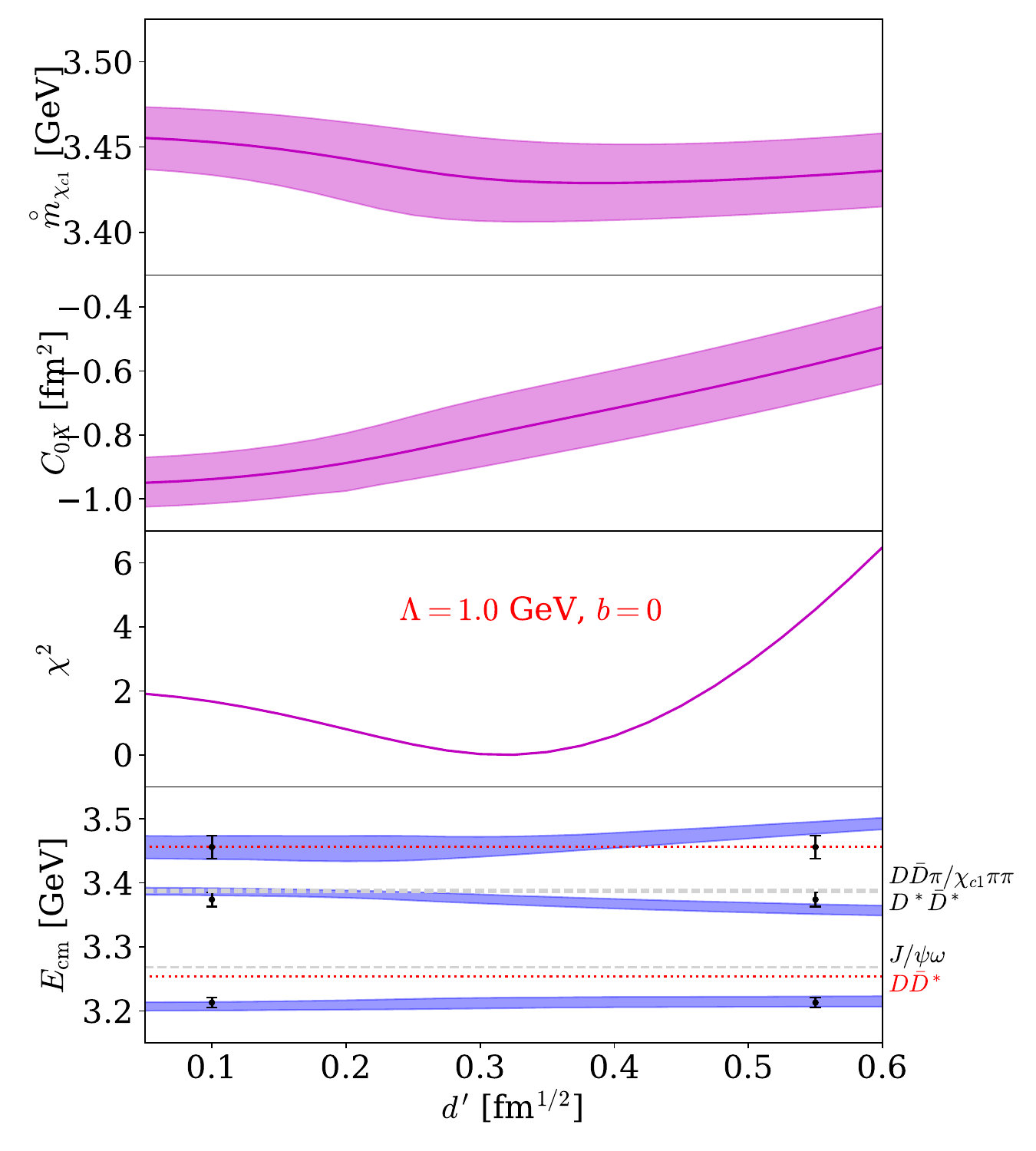}
	\caption{Results obtained from the fit to the LQCD energy levels of Ref.~\cite{Prelovsek:2013cra} in the case $b=0$, as a function of the LEC $d'$. The left (right) panel correspond to $\Lambda =0.5\ \text{GeV}$ ($\Lambda =1.0\ \text{GeV}$). From top to bottom, we show the value of the bare mass $\mcov$, the constant $C_{0X}$, the $\chi^2$ value of the fit, and, in the bottom panel, the comparison between the energy levels computed within our approach and the lattice results. In the bottom plots, the solid coloured bands stand for our results, $E_n$ ($n=1$, $2$ and $3$), while the LQCD levels of Ref.~\cite{Prelovsek:2013cra} ($L=1.98$ fm) are shown as black points. The latter data are displayed twice for convenience of comparison, at the beginning and at the end of the explored $d'$-range. In these two plots, the red dash lines denote the two free energy levels of $D\olsi{D}{}^\ast$ system. The gray lines denote the thresholds of channels other than $D\olsi D{}^*$ (the lattice $\chi_{c1}$ mass is taken from  the lowest energy level in Fig.~1(b) of Ref.~\cite{Prelovsek:2013cra}). Note that the $\Lambda = 0.5\,\GeV$ results (left panel) are shown up to $d' \simeq 1.2\ \text{fm}^{1/2}$, whereas those for $\Lambda = 1.0\,\GeV$ (right panel) are shown only up to $d' \simeq 0.6\,\fm^{1/2}$. In both cases, there is a range of $d'$ in which we obtain a fair description of the LQCD results, but the lattice information is not enough to constrain the free parameters of the present scheme. The bands reflect the $1\sigma$ uncertainties in the fitted LECs and in the predicted energy levels. 
 \label{fig:bEq0}}
\end{figure*}

\subsection{\boldmath Parameters in $1^{++}$ sector}

Unlike the analysis in the $D\olsi{D}$--$D_s \olsi{D}_s$ coupled-channel system, we consider both the potential with and without the $p^2$-dependent term. To perform a comprehensive analysis, we explore four scenarios: the cases with $b=0$ and $b\neq 0$ in the potential, along with cutoff values $\Lambda=0.5$ and $1.0\,\GeV$.

Let us first start with the simplest case, $b=0$. This is the scenario where we have fewer parameters to fit the LQCD energy levels of Ref.~\cite{Prelovsek:2013cra}. In Fig.~\ref{fig:bEq0} we show our results for this case as a function of $d'$, organized in two vertical panels. The left (right) panel corresponds to the case $\Lambda=0.5\,\GeV$ ($1.0\,\GeV$). The bottom plots show the LQCD levels of Ref.~\cite{Prelovsek:2013cra} as black points, and the energy levels obtained with our approach with solid colored bands. Note that here we show the absolute energies $E_n$, not relatives to $M_{\text{av}}^L$ (in~Table~\ref{Tab:levels_1++}). The red dotted lines stand for free energies. For the case $\Lambda=0.5\,\GeV$ we have a good agreement with the LQCD data in the $d'$ range up to $d'\approx 1\,\fm^{1/2}$. For the case $\Lambda=1.0\,\GeV$ this range reaches up to $d'\approx 0.6\,\fm^{1/2}$. The main differences between the $\Lambda=0.5$ and $1.0\,\GeV$ cases are:
\begin{itemize}[leftmargin=0pt]
\setlength{\labelwidth}{0pt}
\setlength{\labelsep}{5pt}
\setlength{\itemsep}{0pt}
\setlength{\itemindent}{10pt}
\item The inclusion of higher momentum interactions when increasing the cutoff from $\Lambda = 0.5\,\GeV$ to $\Lambda = 1.0\,\GeV$ worsens the description of the levels: the $\chi^2$ is generally smaller and the range of $d'$ in which the description of the data is acceptable is larger for the lowest cutoff case. For $\Lambda = 0.5\,\GeV$, as we can see in the left panel of Fig. \ref{fig:bEq0}, the influence of the parameters $\mcov$ and $C_{0X}$ on the $n=3$ level is negligible. For these energies the interaction is suppressed enough by the form factor in~Eq.~\eqref{Eq:form-factor-coupled}, which explains why the $n=3$ energy level is always found close to the free energy. This is not the case in the right panel, where higher momenta interactions are considered due to the larger cutoff $\Lambda = 1.0\,\GeV$. The band associated with the third energy level is much more sensitive to the values of the parameters, although this sensitivity does not translate into a better quantitative agreement with the LQCD energy levels. This can be seen in the $\chi^2$ value for each fit, shown also in Fig.~\ref{fig:bEq0}. Since this case $\Lambda = 1.0\,\GeV$ is accompanied with a poor description of the LQCD data, it could be questioned if the low energy description of $V(E)$ for $b = 0$ in~Eq.~\eqref{Eq:potential_1++_new} is sufficient to describe such high energies. The dependence on the cutoff value also indicates the necessity of including higher order terms in the potential.
This fact also motivates our extension of $V(E)$ to $b \neq 0$ case.
    
\item The value of $C_{0X}$ found in both cases is negative and of the same order, revealing a $D \olsi{D}{}^{\ast}$ attraction near the threshold. But there is a remarkable difference concerning the value of $\mcov$ which turns out to be more than $50\,\MeV$ larger for $\Lambda = 1.0\,\GeV$, as can be seen in Fig.~\ref{fig:bEq0}. Such dependence of $\mcov$ on $\Lambda$ is similar to that observed in Table \ref{tab:LEC-value} for $\mcos$.
\end{itemize}

\begin{figure*}
	\centering
 \includegraphics[width=0.45\textwidth]{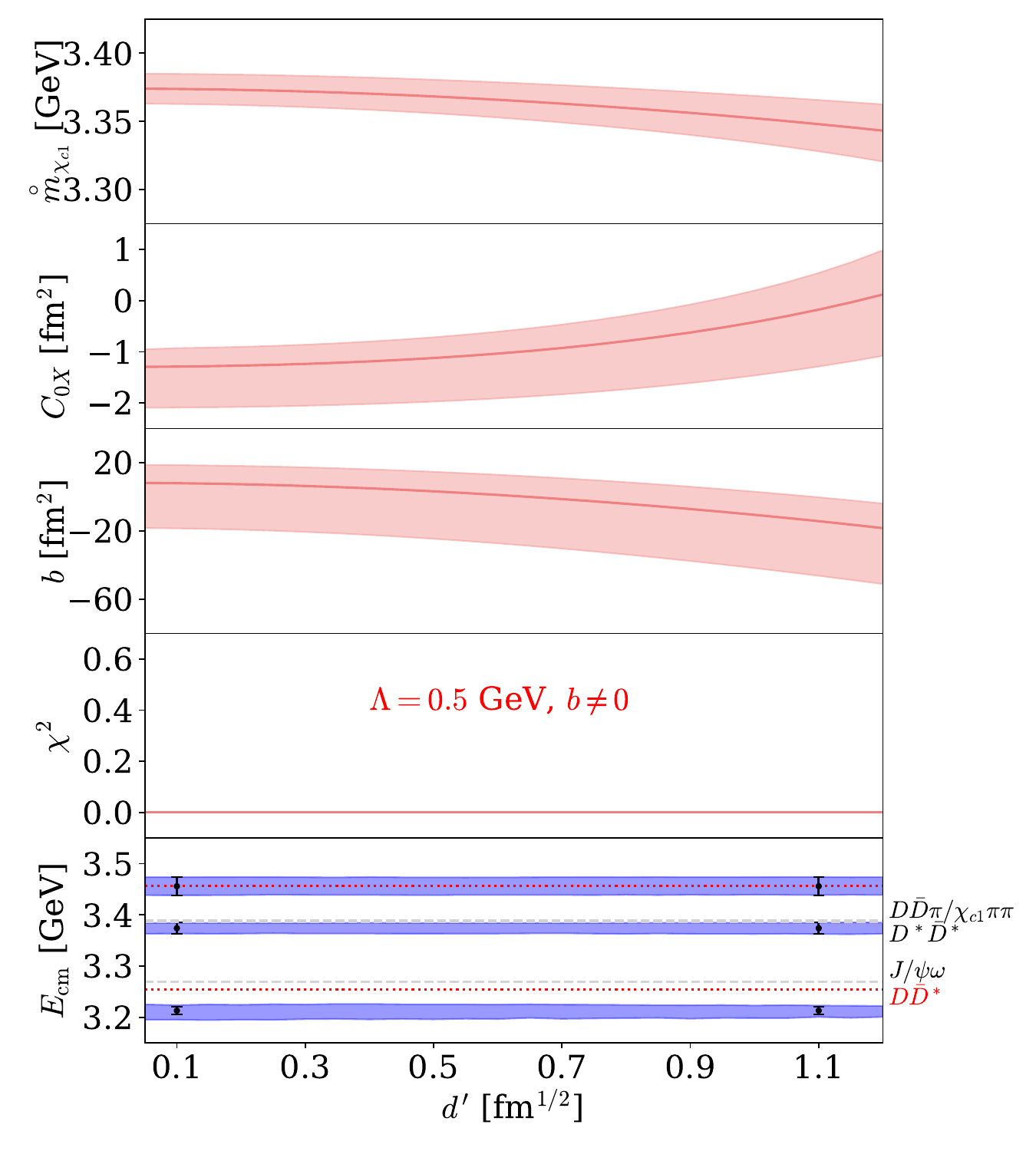}\hspace{0.5cm}
    \includegraphics[width=0.45\textwidth]{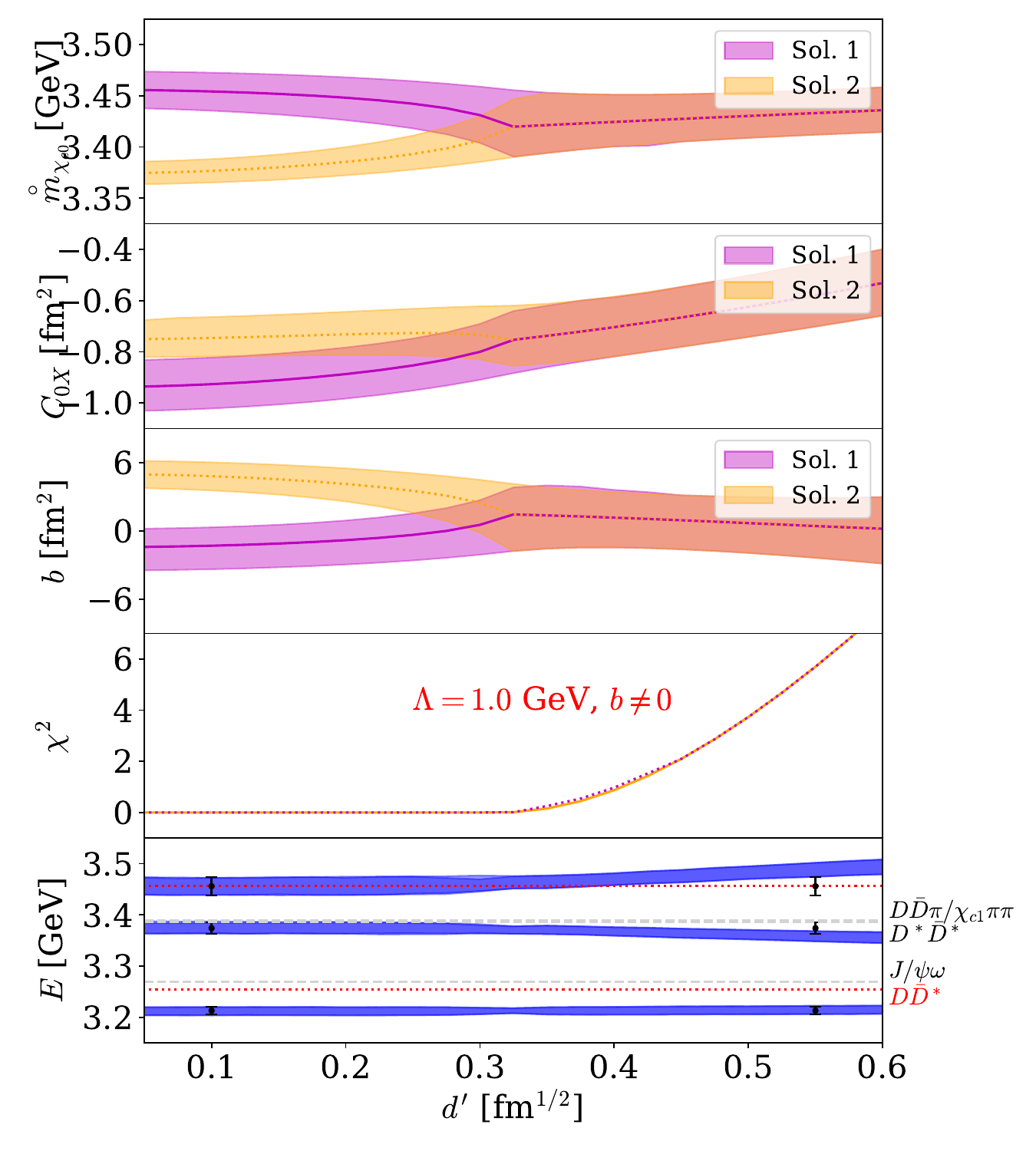}
	\caption{Same as Fig. \ref{fig:bEq0} in the case $b\neq 0$. The values of the LEC $b$ are included in the middle panel. Note that for $\Lambda=1.0\,\GeV$ (right panel), there is a range of values of the $d'$ for which the merit function has two minima, denoted as Sol.~1 and Sol.~2. }
	\label{fig:bneq0}
\end{figure*}

As mentioned above, we have performed the analysis extending the low energy description of the effective potential including a $p^2$-dependent term, see Eq. \eqref{Eq:potential_1++_new}. The results considering an additional parameter $b$ are shown in Fig. \ref{fig:bneq0}. We will discuss the results obtained for $\Lambda = 1.0\,\GeV$ (right panel), which are the ones that change the most. In the bottom plots, we find a better agreement (compared to the case $b=0$) with the LQCD energy levels, which is expected because we have now three parameters to fine-tune. Nevertheless, the description remains valid only for values of $d'$ up to $0.6\,\fm^{1/2}$, as in the case $b=0$. As can be seen in the right panel of Fig.~\ref{fig:bneq0}, the parameter space has two configurations which give the same energy levels, labeled Sol. 1 and Sol. 2. They are different for $d' \lesssim 0.3\,\fm^{1/2}$, and collapse to the same solution above that value. For small $d'$, we find that Sol. 2 gives values of $C_{0X}$ and $\mcov$ similar to those obtained in the case $\Lambda=0.5\,\GeV$. We note here that the value of $b$ is significantly different from zero only for $0 < d' \lesssim 0.25\,\fm^{1/2}$. For $d'$ values above that point, the results are basically those obtained for $b =0$, and the sought improvement of the energy levels is not achieved.

For $\Lambda = 0.5\,\GeV$, compared to the $b=0$ case, the inclusion of the $b$ parameter matches the LQCD uncertainties, allowing larger error bands for the parameters. Indeed, the range of $b$ is almost compatible with $b=0$ in the explored $d'$ range, which is natural, since the $\chi^2$ value was already small in the $(b=0)$ fit. Nonetheless, $b$ varies in a range $-60 < b < 20\,\fm^2$, which has absolute values larger than in the case $\Lambda = 1.0\,\GeV$ (where $b$ is in the range $-6 < b < 6\,\fm^2$), to compensate the stronger form factor suppression. The features of $C_{0X}$ and $\mcov$ remain unchanged but with bigger uncertainties. That means that the results from the $(b=0)$ fit are reasonable.

\subsection{\boldmath Predictions in the infinite volume}
\label{sec:predictionsIV}

\begin{figure*}
	\centering
	\includegraphics[scale=0.5]{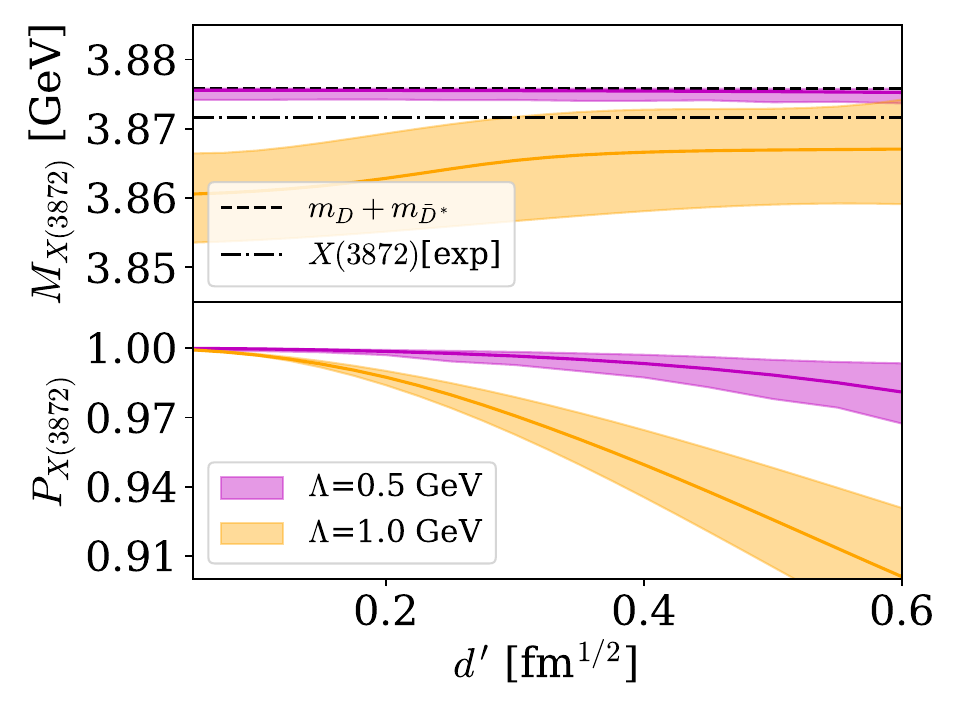}
    \includegraphics[scale=0.5]{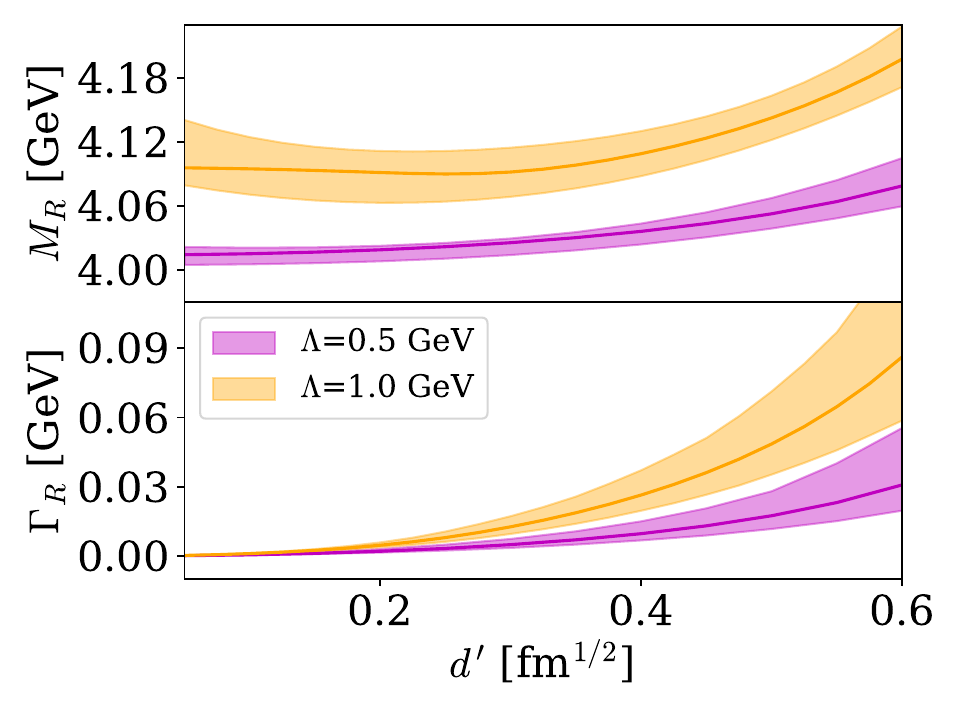}
	\caption{%
Infinite volume predictions for the $X(3872)$ mass ($M_{X(3872)}$) and molecular probability ($P_{X(3872)}$) as a function of the LEC $d'$ for the $(b=0)$ fit (top left and bottom left panels respectively). 
In the right panels, the mass $M_R$ and width $\Gamma_R$ of the dressed charmonium state are also shown.
The bands reflect the uncertainties propagated from the LECs (see Fig.~\ref{fig:bEq0} for details). 
The results correspond to the bound state ($M_{X(3872)}$) and resonance $(M_R,\Gamma_R/2)$ pole positions obtained from the $T$-matrix in Eq.~\eqref{Eq:T-matrix-coupled} with the use of physical meson masses.%
\label{fig:predictions1}}
\end{figure*}
\begin{figure*}
	\centering
	\includegraphics[scale=0.5]{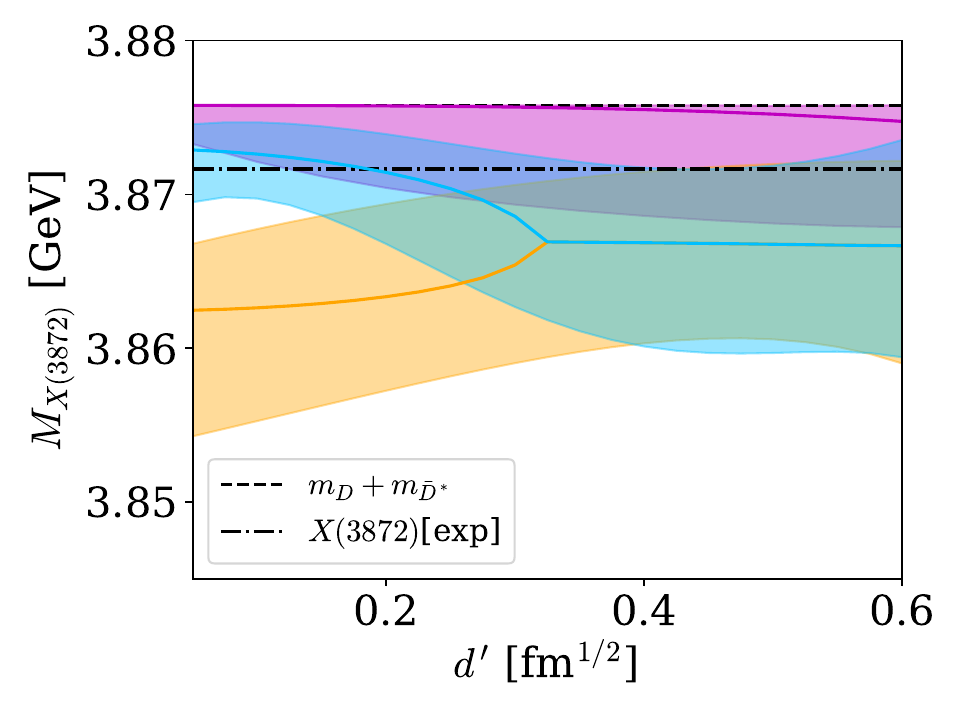}
    \includegraphics[scale=0.5]{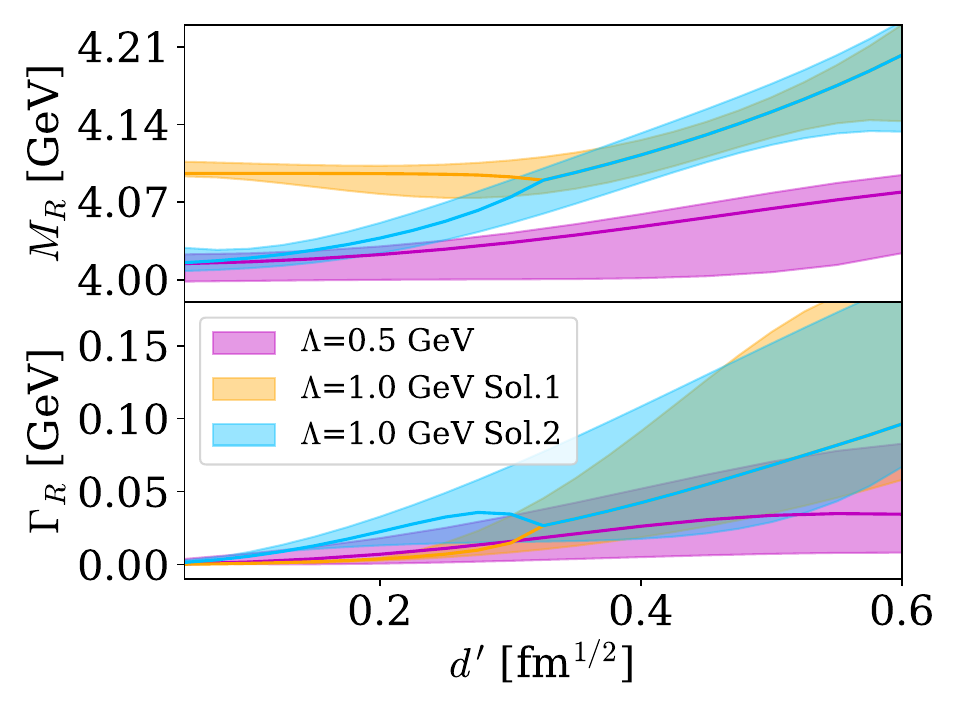}
	\caption{Same as Fig. \ref{fig:predictions1} but in the $b\neq 0$ scenario. In this case, the derived quantities correspond to the LECs shown in Fig.~\ref{fig:bneq0}.%
\label{fig:predictions2}}
\end{figure*}

In the previous subsection we have explored different parameter configurations, summarized in Figs.~\ref{fig:bEq0} and \ref{fig:bneq0}, allowed by the LQCD $1^{++}$ energy levels reported in  Ref.~\cite{Prelovsek:2013cra}. We now use this information to make a spectroscopic analysis of our $T$-matrix, in the infinite volume limit and employing physical isospin-averaged heavy-meson masses.
Since we are dealing with heavy mesons, the dynamics should not be extremely affected by the light quark masses. This is supported by the following reasoning. The lattice energies in Ref.~\cite{Prelovsek:2013cra} are given relative to the spin averaged mass $M_{\text{av}}$, and they are obtained in a calculation with the lattice setup of Ref.~\cite{Mohler:2012na}. The charmonium spectrum obtained in this latter reference is in good agreement with the experimental information, after considering the mass splitting $\Delta M_{\text{av}}=M_{\text{av}}^{\text{exp}}-M_{\text{av}}^{L}= 640(20)\,\MeV$. When applying this shift, the LQCD $D \olsi{D}{}^{\ast}$ threshold, which should be more affected by the unphysical light quark masses, lies only about $20\,\MeV$  above the physical one. Thus, besides changing the masses of the mesons, we apply this shift to the values fitted for the $c\ac$ bare mass $\mcov$, leaving the $C_{0X}$, $d'$, and $b$ LECs unchanged.

The information concerning the poles that we find in our analysis for the $(b = 0)$ fit is collected in Fig.~\ref{fig:predictions1}, where the LEC $d'$ has been varied in the $[0,0.6]\,\fm^{1/2}$ range. The results for the case $b \neq 0$ are shown in Fig.~\ref{fig:predictions2}, and do not require a separate discussion, since they share the same qualitative features that we have discussed for $b = 0$. We have restricted the analysis to the values of $d'$ which provide a good agreement with the LQCD energy levels. In all the scenarios we find a pole below the $D \olsi{D}{}^{\ast}$ threshold energy on the physical Riemann sheet and its mass is shown in the top-left plot of Fig.~\ref{fig:predictions1}. We identify this bound state with the $X(3872)$. In the case $\Lambda = 0.5\,\GeV$ the mass of the bound state is compatible with threshold.\footnote{Recall that we use here isospin averaged masses so that the threshold is located at $m_D + m_{D^\ast} = 3875.81(5)\,\MeV$.} 
We also find another pole, lying on the unphysical Riemann sheet, located at $E=M_R-\textrm{i}\Gamma_R/2$, shown in the top- and bottom-right panels of Fig.~\ref{fig:predictions1}. The mass $M_R$ lies always well above the $D\olsi{D}{}^\ast$ threshold. We shall refer to this pole as a charmonium, since it originates mostly from the $c\olsi{c}$ seed with bare coupling and mass $d'$ and $\mcov$, respectively.\footnote{In Refs.~\cite{Deng:2023mza,Wang:2024ytk}, such a state is also predicted to be the $P$-wave charmonium $\chi_{c1}$.} The $D\olsi{D}{}^\ast$ final state interactions change (renormalize) these parameters to their final values, $M_R - i\Gamma_R/2$ and $g_R$, shown in Figs.~\ref{fig:predictions1}-\ref{fig:predictions2} and Fig.~\ref{fig:effective_coup_1++}, respectively. As $d'\to 0$, we see that the width of this charmonium resonance vanishes, and its mass approaches to the bare value, since its coupling to the $D\olsi{D}{}^\ast$ channel is being turned off. Such a $1^{++}$ hidden-charm state, with a mass in the range $4.0$--$4.1\,\GeV$ and a width of a few tens of MeV, has 
recently been observed by the LHCb Collaboration in ${B}^{+} \rightarrow {D}^{* \pm} {D}^{\mp} {K}^{+}$ decays~\cite{LHCb:2024vfz}. Indeed in that work, a mass of $4012.5_{-3.9-3.7}^{+3.6+4.7}\,\MeV$ and a width of $62.7_{-6.4-6.6}^{+7.0+6.4}\,\MeV$ was reported.

\begin{figure*}
	\centering
	\includegraphics[scale=0.5]{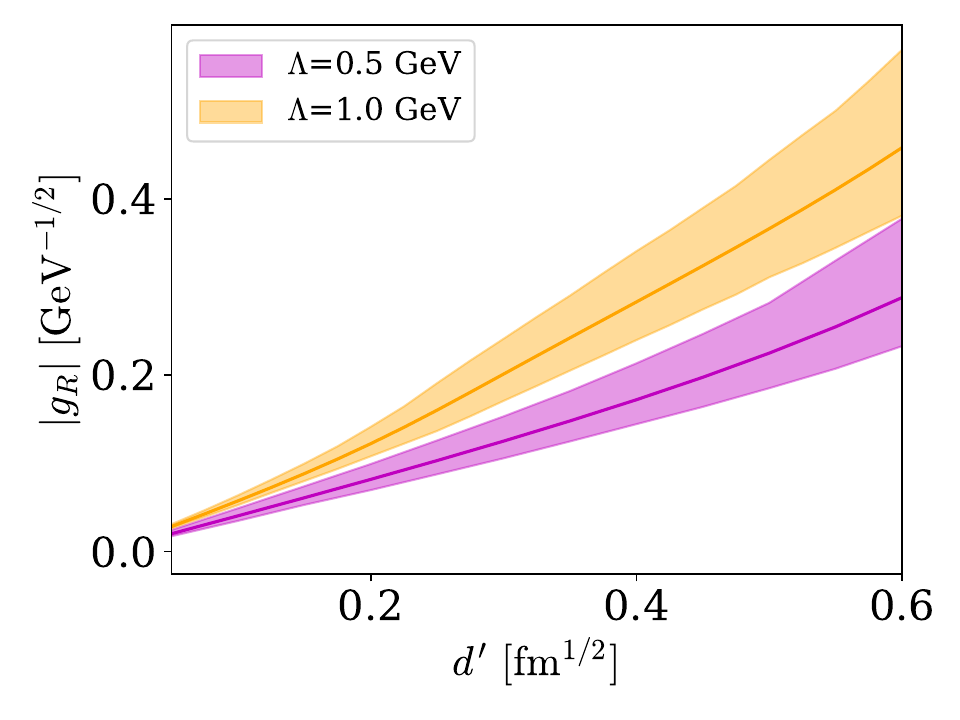}
    \includegraphics[scale=0.5]{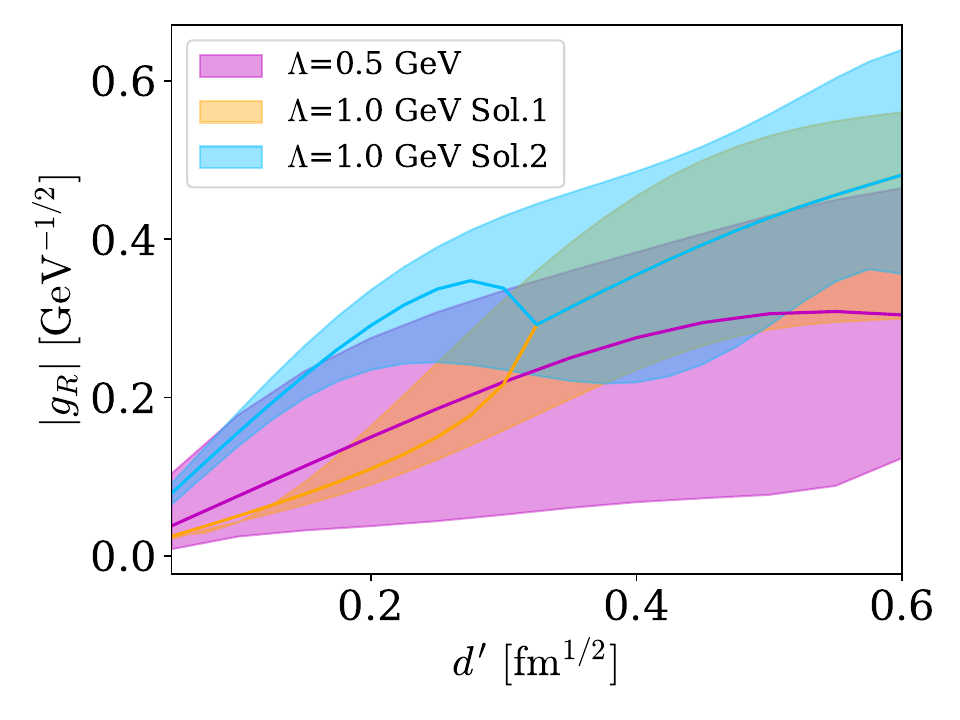}
	\caption{Coupling of the generated charmonium resonance to the $D\olsi{D}{}^*$ pair, determined from residue at the pole, as a function of the the LEC $d'$. The couplings displayed in the left and right panels correspond to the resonances predicted in the right panels of Figs.~\ref{fig:predictions1} and \ref{fig:predictions2}, respectively.
\label{fig:effective_coup_1++}}
\end{figure*}

In the bottom-left panel of Fig.~\ref{fig:predictions1}, we have represented the $X(3872)$ molecular component,\footnote{Note that $P_{X(3872)}$ is not shown in Fig.~\ref{fig:predictions2} because, when $b> 2\mu d^{\prime\, 2}/({E - \mcov})^2$, it becomes greater than one, and hence loses its full probabilistic interpretation. This is the case for Sol. 2 and $d'< 0.3$ fm. For Sol. 1, $P_{X(3872)}$ is always comprised between 0.9-1, as in the $(b=0)$-fit.} $P_{X(3872)}$, computed from the single-channel equivalent of Eq.~\eqref{Eq:posibility-coupled}. It can be seen that the molecular content of the state decreases for increasing values of $d'$. Nevertheless, the values of $P_{X(3872)}$ are never smaller than $0.9$, revealing that the charmonium component in the $X(3872)$ is negligible, according to our analysis. This seems natural given the large energy gap (around $100\,\MeV$) between the two-meson threshold energy and the bare charmonium mass.
It might seem that our findings contradict the claim made in Ref.~\cite{Prelovsek:2013cra}, where the presence of a $c\bar c$ operator was found to be essential to produce the $X(3872)$. However, there is in fact no contradiction. Indeed, the aim of Ref.~\cite{Prelovsek:2013cra} was ``not to choose between interpretations ($\bar{c} c$ state accidentally aligned with $D \olsi{D}{}^\ast$ threshold or $D \olsi{D}{}^\ast$ molecule, etc.), but rather to find a candidate for $X(3872)$ on the lattice and determine its mass.''
In addition, there is no one-to-one correspondence between the overlap of interpolating operators with a physical state and the relative importance of various components in its wave function.
This is, for instance, illustrated in the upper plot of Fig.\,1(d) in Ref.~\cite{Prelovsek:2013cra}, where one can see the $D\olsi{D}{}^\ast$ operator produces the ground state $\chi_{c1}(1P)$ almost at the same energy as obtained by using the $c\bar c$ operators, shown in the upper plot of Fig.\,1(c) in the same reference. Nevertheless, in an unquenched calculation, any operator with the right quantum numbers will produce all corresponding QCD eigenstates for sufficiently large time evolution. In Ref.~\cite{Prelovsek:2013cra}, for the time-length (about $4\,\fm$) used in the lattice calculation, the full spectrum of energy levels are only obtained when both $c\bar c$ and $D\olsi{D}{}^\ast$ operators are used. In addition, the spatial size of the lattice volume $L\simeq2\,\fm$ is relatively small compared with the expected size of the $X(3872)$. Future improved simulations with larger spatial and time volumes can shed light on this issue from the LQCD perspective.

\subsection{Remarks about considering the isospin limit}

\begin{figure}[t]
\centering
	\includegraphics[scale=0.5]{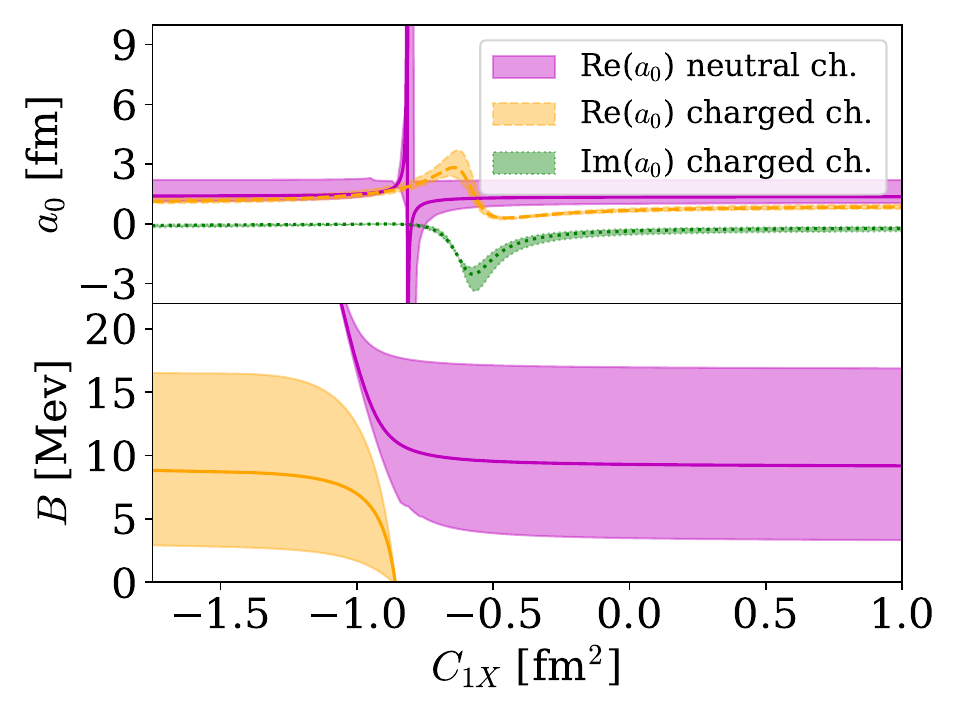}
\caption{(Top panel) Scattering lengths of the  $D\olsi D{}^*$ charged and neutral channels as a function of the LEC $C_{1X}$. In addition, $\Lambda =1.0$ GeV, $b=0$, $d=0.2$ fm$^{1/2}$ and the fitted $C_{0X}$ and $\mcov$, as can be read off in the right panel of Fig. \ref{fig:bEq0}. (Bottom panel) The binding energy of the $X(3872)$, mixed isoscalar and isovector state, relative to the neutral threshold. As can be seen, when attractive interaction for the $I=1$ channel increases, the binding also grows (magenta line) and for $C_{1x}< -0.85$ fm$^2$ generates a new bound state (orange line).%
\label{Fig-isospin-breaking}}
\end{figure}

In this subsection, we discuss the importance of the isospin breaking corrections using physical $D$ and $D^*$ masses. Within a more detailed scheme, the $X(3872)$ turns out not to be an isospin eigenstate~\cite{Gamermann:2009uq,Guo:2014hqa} and it is better described in terms of particle states. This is not only due to its isospin breaking branching ratios and decay properties but also to its small binding energy. When using physical masses it is found that the neutral $D^0\olsi{D}{}^{\ast 0}$ and charged $D^+D^{*-}$ thresholds differ by {$8\,\MeV$}, and the $X(3872)$ has a very small binding energy relative to the neutral threshold, being this binding energy compatible with zero. In this situation, the scattering length $a_0$ is very sensitive to the exact values of the used masses. It is possible to illustrate this statement in our approach. When we write the interaction of Eq. \eqref{Eq:potential_1++_new} in terms of charged $D^+D^{\ast -}+\text{c.c.}$ and neutral $D^0\olsi{D}{}^{\ast 0}+\text{c.c.}$ components, it also appears the isovector LEC $C_{1X}$~\cite{Hidalgo-Duque:2012rqv}. The effective potential is then expressed as
\begin{align}
	V_{ij}(E)=\frac{1}{2}\begin{pmatrix}C_+ & C_- \\ C_- & C_+ \end{pmatrix}+\frac{d^2/2}{E-\mcov}\begin{pmatrix} 1 & 1 \\ 1 & 1 \end{pmatrix},
	\label{Eq:vpot-particle-basis}
\end{align}
where in Eq. \eqref{Eq:vpot-particle-basis} the $i$ and $j$ indexes run over the neutral and charged channels and $C_\pm=C_{0X}\pm C_{1X}$. The $T$-matrix in Eq. \eqref{Eq:T-matrix-coupled}, now in the particle channels space, is related to the $S$-wave scattering matrix~\cite{Albaladejo:2013aka}
\begin{align}
  S_{ij}(E)=\delta_{ij}-i\sqrt{\mu_i k_i}\sqrt{\mu_j k_j}T_{i,j}(E)/\pi.
\label{Eq:s-matrix-particle}
\end{align}
With the identity of Eq. \eqref{Eq:s-matrix-particle} it is possible to define the scattering lengths of the neutral and charged channels making use of the effective range expansion (Eq. \eqref{Eq:ERE}) above each threshold $E^{\rm th}_i$,
\begin{align}
	a_0 = \lim_{E\to E^{\rm th}_i}\frac{\mu_i}{2\pi}T_{i,i}(E).
\label{Eq:scattering-lengths}
\end{align}
For the set of parameters corresponding to $d'=0.2\,\fm^{1/2}$ we have obtained the $C_{1X}$ dependence of the $X(3872)$ pole position and the charged and neutral scattering lengths. The results are shown in Fig.~\ref{Fig-isospin-breaking}. In the upper panel we show the values of the scattering lengths. Note that the charged channel scattering length becomes a complex number. In the lower panel, the binding energy of the $X(3872)$, mixed isoscalar and isovector state, relative to the neutral threshold is shown. As we can see in this latter panel, when attraction in the isovector sector increases the binding energy becomes larger (magenta line). Furthermore, when $C_{1X}$ approaches $-0.85\,\fm^2$, a new bound state is generated (orange line), {which was previously as a virtual state for $C_{1X} \gtrsim -0.85\,\fm^2$.}\footnote{In Ref.~\cite{Zhang:2024fxy}, an isovector virtual state near the  $D\olsi D{}^*$ threshold, called $W_{c1}$, was predicted using chiral effective field theory with the experimentally observed $X(3872)$ properties as inputs, which include the mass and the isospin-breaking ratio of the $X(3872)\to J/\psi\rho^0$ the $X(3872)\to J/\psi\omega$ couplings (see Ref.~\cite{Dias:2024zfh} for an update).
Such a virtual state was supported by the LQCD calculation in Ref.~\cite{Sadl:2024dbd}.
} For these values of $C_{1X}$ presenting a very near-threshold state, the situation would be quite similar to the physical case of the $X(3872)$. It is interesting to note that in this region, when a near-threshold bound state is being generated, the values of the neutral scattering length run from $-\infty$ up to $\infty$ in a way that, for a very shallow bound state the scattering length would be a large positive number. 
Consequently, calculations of the $X(3872)$ in the isospin limit will not provide realistic predictions for this scattering length at the physical mass of the exotic $X(3872)$ state.

\subsection{\boldmath  {Remarks on the dressed} $\chi_{c1}(2P)$ charmonium state}
\label{sec:chi12P}

\begin{figure*}
\centering
    \includegraphics[scale=0.5]{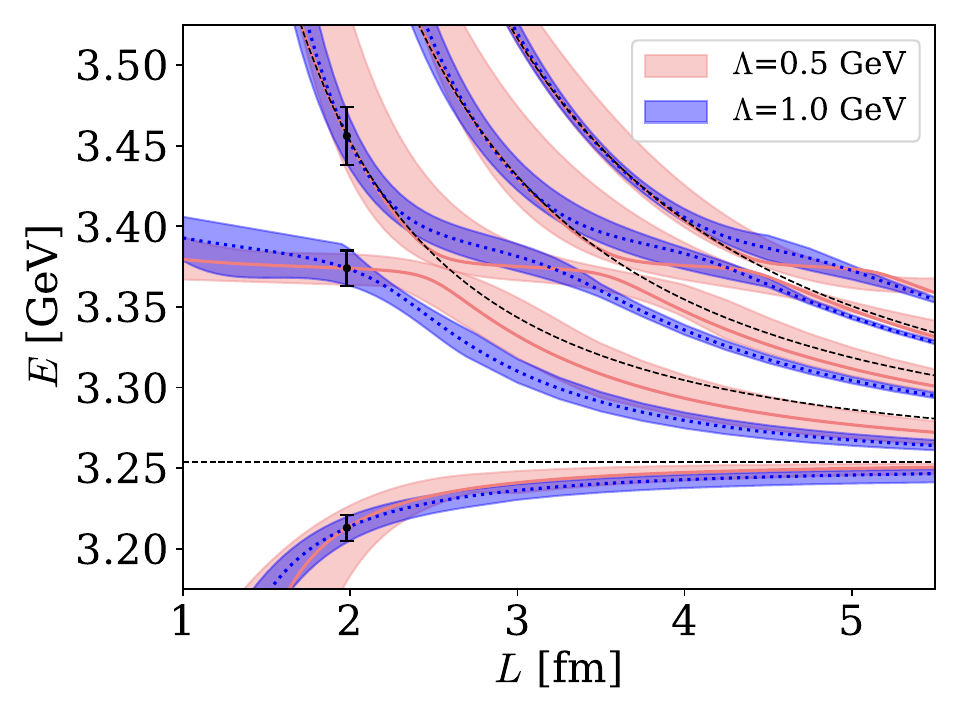}
    \includegraphics[scale=0.5]{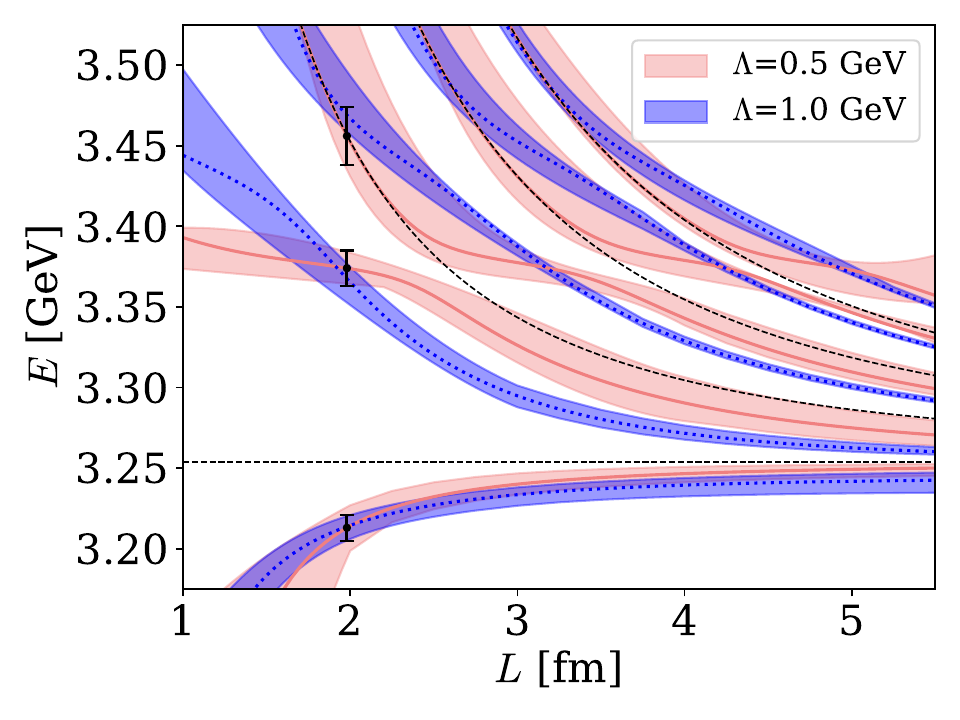}
\caption{Prediction of the volume dependence of the $1^{++}$ energy levels for two different values of $d'$ within $(b \neq 0)$ fitting scenario.
The blue bands are associated to the Sol. 2 LECs (see right panel of Fig. \ref{fig:bneq0}). Two values of $d'=0.2$ and $0.4$ fm$^{1/2}$ are considered in the left and right panels, respectively. Note that LQCD (unphysical) $D^{(\ast)}$ meson masses are used (see the text for further details). The LQCD energy levels correspond to those collected in Table~\ref{Tab:levels_1++}. %
\label{Fig-levels}}
\end{figure*}

As mentioned in Sec.~\ref{sec:predictionsIV}, the predicted $1^{++}$ state with a mass slightly above 4~GeV likely corresponds to the new observed state $\chi_{c1}(4010)$ by LHCb~\cite{LHCb:2024vfz}. In this subsection, we discuss more about the relation of such a state with the radially excited charmonium $\chi_{c1}(2P)$.

The charmonium spectroscopy~\cite{Workman:2022ynf} reveals several states that have successfully been associated to quark model predictions. The lowest lying $\chi_{cJ}(1P)$ with $J=0,1,2$ have been measured and some of them have also been found in LQCD calculations~\cite{Biddick:1977sv,Mohler:2012na, HadronSpectrum:2012gic}. Experimentally, their masses are found to be $3415$, $3511$, and $3556\,\MeV$, respectively, with a splitting around $140\,\MeV$ between the $J=0$ and $J=2$ states. However, the case of the $\chi_{cJ}(2P)$ states is not so clear. In this multiplet only the $J=2$ state has been reported  experimentally~\cite{Uehara:2005qd}, with a mass $m_{\chi_{c2}(3930)}=3922.5(1.0)\,\MeV$ and a width $\Gamma_{\chi_{c2}(3930)}=35.2(2.2)\,\MeV$~\cite{ParticleDataGroup:2024cfk}.\footnote{The state $X(3915)$ has been previously identified with the $\chi_{c0}(2P)$ state~\cite{Liu:2009fe,Duan:2020tsx}, which was questioned in Refs.~\cite{Guo:2012tv, Olsen:2014maa, Zhou:2015uva}. For a recent discussion about the experimental signatures that were assigned to the $X(3915)$ in the RPP~\cite{ParticleDataGroup:2024cfk}, to the $\chi_{c2}(3930)$~\cite{Zhou:2015uva} and a $D_s\olsi D_s$ molecular state, we refer to Ref.~\cite{Ji:2022vdj}. } 
However, the presence of the $D\olsi{D}$, $D\olsi{D}{}^{\ast}+c.c.$ and $D^\ast\olsi{D}{}^{\ast}$ thresholds may induce important mixing between charmonium and meson components. These charmonium states are supposed to decay mainly into $D\olsi{D}{}^{(*)}$ channels. Under these considerations, it is natural to expect that the hypothetical experimental $\chi_{c1}(2P)$ state would have a mass similar to, probably smaller than that of the $\chi_{c2}(2P)$. Concerning the pole above the threshold that we find within our approach, see the right panels in Figs. \ref{fig:predictions1} and \ref{fig:predictions2}, naively it is hard to consider it as a candidate for the $\chi_{c1}(2P)$, since the mass obtained here lies more than $80\ \text{MeV}$ above the $\chi_{c2}(3930)$ ($\chi_{c2}(2P)$ state). The mass of the hidden-charm state reported in our work, $M_R$, is obviously sensitive to the value of the bare charmonium mass $\mcov$ used in the physical-mass limit. Indeed, in such a case, $M_R$ lies always close (within tens of $\text{MeV}$) to the value of the bare charmonium mass $\mcov$. For the latter we use the value obtained in our analysis of the LQCD levels (shown in Figs.~\ref{fig:bEq0} and \ref{fig:bneq0}), shifted to a higher value by $\Delta M_{\text{av}} = 640(20)\,\MeV$, as we have already explained. For example, for the case $\Lambda = 0.5\,\GeV$ and $d'=0.5\,\fm^{1/2}$, $\mcov \simeq 3.37\,\GeV$ when lattice masses are used (see Fig.~\ref{fig:bEq0}), and thus $\mcov$ is shifted to $4.01\,\GeV$ when physical masses are employed.\footnote{Note that the bare charmonium masses obtained here have a cutoff dependence, and the correspondence between them and charmonium masses in quark model predictions is unclear.}

This puzzle can be understood as follows. The bare $\chi_{c1}(2P)$ mass is indeed lower than the $\chi_{c2}(2P)$ mass. However, the mixing between the charmonium and hadronic molecular state has different effects in the $1^{++}$ and $2^{++}$ sectors. 
For a definite discussion, let us take the charmonium masses predicted in Ref.~\cite{Godfrey:1985xj} for the $2P$ states, which are $m_{\chi_{c1}(2P)}=3.95\,\text{GeV}$ and $m_{\chi_{c2}(2P)}=3.98\,\text{GeV}$. Then the $1^{++}$ $D\olsi D{}^*$ and $2^{++}$ $D^*\olsi D{}^*$ molecular states, with masses close to the corresponding thresholds, lie below and above the $\chi_{c1}(2P)$ and $\chi_{c2}(2P)$ masses, respectively.
Since a mixing between two states with different masses will always push the heavier state up and the lighter state down, the $1^{++}$ charmonium state will be pushed up due to its mixing with the $D\olsi D{}^*$ molecule and the $2^{++}$ charmonium will be pushed down due to its mixing with the $D^*\olsi D{}^*$ molecule.
Consequently, it is plausible that the $1^{++}$ state originated from the $\chi_{c1}(2P)$ has a mass larger than the $2^{++}$ state originated from the $\chi_{c2}(2P)$.

Knowing the volume dependence of the lattice energy levels would be very important to shed light on this issue. To illustrate this point we have predicted the volume dependence of our energy levels in different scenarios which lead to different resonance pole positions. They are shown in Fig.~\ref{Fig-levels} where we have chosen the case $b\neq0$ and two different values of $d'$. As can be seen, in the two cases the higher energy levels present different box-size dependencies and that is what makes LQCD calculations at several volumes so interesting and valuable. In particular within our scheme, calculations for different volumes would help to reduce the parameter space and provide a more constrained set of predictions. Due to the lack of more inputs from LQCD and the exploratory nature of the present work, we should not take at face value our analysis of the LQCD simulation carried out in  Ref.~\cite{Prelovsek:2013cra}. However, we would like to recall here that the case $d' \to 0$ gives always a fair description of the energy levels, {\it i.e.}, the $\chi^2$ is small.

\section{\boldmath Predictions in the $2^{++}$ sector}\label{Sec:2++}

\begin{figure*}
\centering
   \includegraphics[scale=0.52]{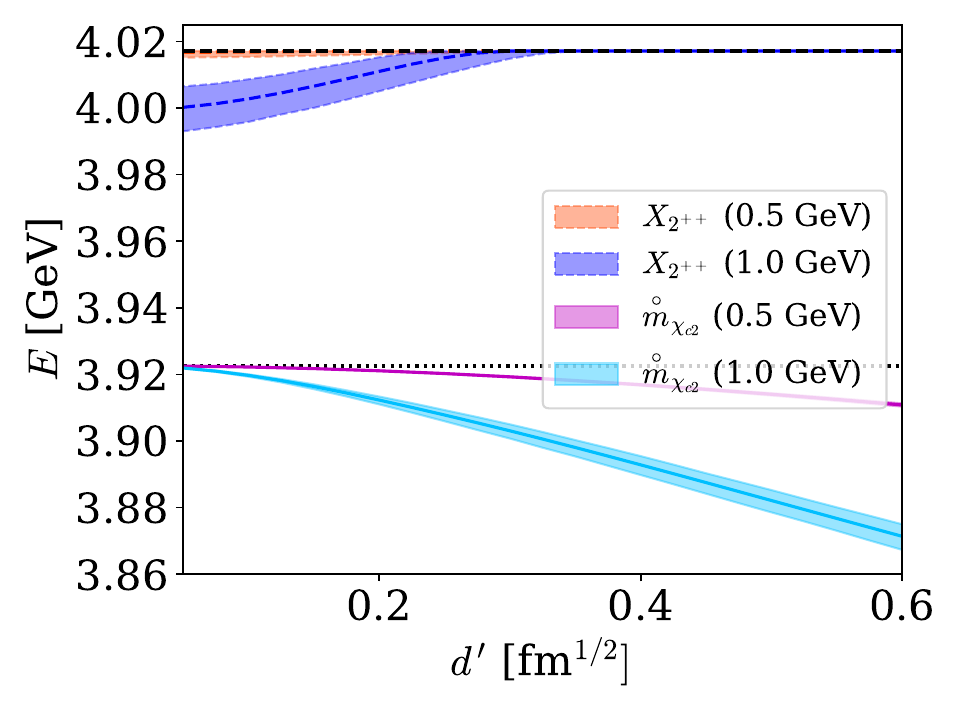}
   \includegraphics[scale=0.52]{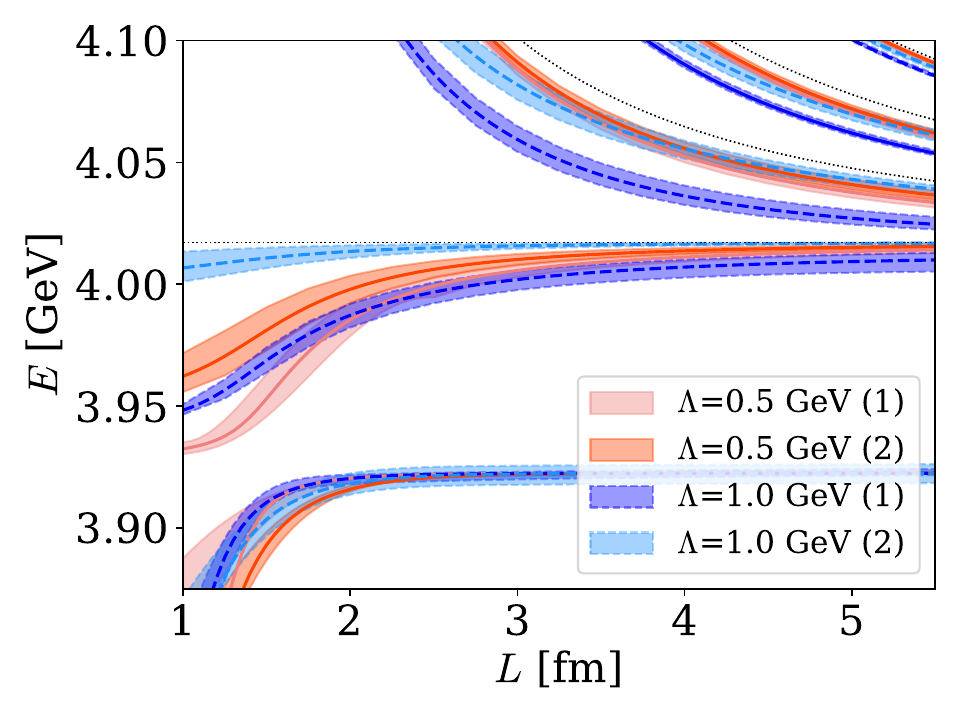}
\caption{Left: Bare charmonium masses ($\mcot$) which generate a bound state at the $\chi_{c2}(2P)$ physical mass as a function of $d'$, keeping fixed $C_{0X}$ to the previously fitted values in the $b=0$ case and for $\Lambda=0.5 $ GeV and 1 GeV (see Fig.~\ref{fig:bEq0}). {For $d'\lesssim 0.3$ fm$^{1/2}$ an additional near-threshold bound state is produced} and its mass is also shown in the plot. The dashed and dotted horizontal black lines correspond to the $D^{\ast}\olsi{D}{}^{\ast}$ threshold and the experimental $\chi_{c2}(2P)$ mass, respectively. Right: Prediction of the $2^{++}$ energy levels as a function of the box-size. The energy-levels are obtained using the corresponding bare masses shown in the left panel for two values of $d'$, $0.2\,\fm^{1/2}$  and $0.4\,\fm^{1/2}$, which are labeled as (1) and (2) in the plot, respectively. Here, the physical charmed meson masses are used.} %
\label{Fig-bare-mass-2++}
\end{figure*}

At LO, HQSS predicts the same contact potentials for the isoscalar scattering of $D^\ast \olsi{D}{}^\ast$ and $D \olsi{D}{}^\ast$ systems with quantum numbers $J^{PC}=2^{++}$ and $1^{++}$, respectively. Hence, we can study the $D^\ast\olsi{D}{}^\ast$ system using the same values for $C_{0X}$, $b$, and $d'$ as for the $1^{++}$ case. The value of the $2^{++}$ charmonium bare mass, denoted as $\mcot$, could in principle have also HQSS symmetry breaking corrections. At first, we employ the previous values of $\mcov$ in order to obtain $\mcot$, shifting this value by the $D^\ast$ and $D$ mass difference,\footnote{We set the $D^\ast$ mass to its physical isospin averaged value.} $\mcot = {} \mcov {} + (m_{D^\ast} - m_D$). When this bare mass is used, no trace of a {dressed $\chi_{c2}$} state, which one would expect to appear as a pole below the $D^\ast \olsi{D}{}^\ast$ threshold, is found. This is not surprising because we expect the magnitude of the HQSS breaking in $c\olsi c$ to be smaller than in the $D-D^*$ system. The next natural step is to adjust the value of $\mcot$ so that a pole is always found at the experimental $\chi_{c2} \left( 3930\right)$ mass. The results thus obtained are shown in the left panel of Fig.~\ref{Fig-bare-mass-2++}, for values of $d'$ in the range $0.05\, {\rm fm}^2 < d' < 0.6\, {\rm fm}^2$. The adjusted bare mass $\mcot$ is shown with solid lines, whereas the mass of an additional bound state that we find, closer to the threshold, is represented with dashed lines. This near-threshold bound state would be the spin-2 HQSS partner of the $X(3872)$. It is more bound for small values of $d'$, and as $d'$ increases it approaches the threshold and it becomes virtual. The location of this state is rather close to the mass of the $X_2$ (around $4014\,\MeV$) reported recently by Belle~\cite{Belle:2021nuv}. 
Notice that the bare mass $\mcot$ lies below the $D^{\ast} \olsi{D}{}^{\ast}$ threshold, and then the interaction between the charmonium and charmed mesons in Eq.\,\eqref{Eq:potential_single} is repulsive, {as discussed in the Sec.~\ref{sec:chi12P}}. This repulsion gets bigger when the coupling $d'$ increases and therefore it is natural that this loosely bound state disappears as $d'$ becomes larger since the interaction close to the threshold is more repulsive (or less attractive) for a given $C_{0X}$. Note that this is the opposite situation to that observed in the $X(3872)$ case already presented, since in the latter, the location of a bare component above the threshold produces an attractive piece in the potential. This different role of the interplay between two-meson and bare charmonium degrees of freedoms in the $1^{++}$ and $2^{++}$ sectors was already pointed out in \cite{Cincioglu:2016fkm}. 

Finally, in the right panel of Fig.\,\ref{Fig-bare-mass-2++} we show the $D^{\ast}\olsi{D}{}^{\ast}$ $2^{++}$ energy levels as a function of the box size $L$, obtained using physical masses and two different values of $d'$ ($0.2$ and $0.4\,\fm^{1/2}$). The needed charmonium bare masses and $C_{0X}$ values are taken from the left panel of the Fig.~\ref{Fig-bare-mass-2++} and Fig.~\ref{fig:bEq0}, respectively. As a novelty, we find an additional energy level at the $\chi_{c2}(2P)$ physical mass. The main difference between the two cases of $d'$ is the presence or absence of the bound state for the spin-2 partner of the $X(3872)$.

\section{Conclusions}\label{Sec:Summary}

Isoscalar hidden-charm two-meson states with positive parity and charge conjugation have the same quantum numbers as $P$-wave charmonia. In the Fock space, the $P$-wave charmonium component should be involved in the physical structure of these charmonium-like states. To consider such contributions, we have constructed $D^{(*)}_{(s)}\olsi D{}^{(*)}_{(s)}$ effective interactions including contact and $\chi_{cJ}(2P)$ exchange potentials. Taking into account finite volume effects, we have extracted the undetermined LECs from a reanalysis of the LQCD energy levels calculated for the $D\olsi{D}$--$D_s\olsi{D}_s$ coupled-channel and $D\olsi{D}{}^\ast$ systems in Refs.~\cite{Prelovsek:2020eiw} and \cite{Prelovsek:2013cra}, respectively. 
Detailed information for these lattice calculations is listed in Table~\ref{tab:lat_detail}.
Using these LECs in the infinite volume, we have searched for the possible charmonium-like states in the $J^{PC}=0^{++}$, $1^{++}$, and $2^{++}$ sectors. 

In the $0^{++}$ sector, we have studied four different fitting scenarios to reanalyze the lattice data. In Fits 1 and 2, the rest-frame energy levels of the $D\olsi{D}$--$D_s\olsi{D}_s$ coupled-channel system are used to extract the LECs for two ultraviolet cutoffs $\Lambda=0.5$ and $1.0\,\GeV$, respectively. For Fit 3 and Fit 4, both rest- and moving-frame energy levels are employed, using again two different cutoffs. Using the fitted  LECs (Table \ref{tab:LEC-value}), we have identified three poles in the $D\olsi{D}$--$D_s\olsi{D}_s$ coupled-channel system, which are shown in Fig.~ \ref{Fig:pole_position}. The pole below the $D \olsi{D}$ threshold is identified as a $D\olsi{D}$ molecule. This pole, dominated by the $D\olsi{D}$ channel, is a bound state for the Fit 1, Fit 2 and Fit 3, while it is a virtual state for Fit 4. The second pole, close to the $D_s \olsi{D}_s$ threshold, is a $D_s \olsi{D}_s$ molecule, which is associated with the new structure $X(3960)$~\cite{LHCb:2022aki}. 
The pole on the third [$(-,-)$] RS is above the $D_s\olsi{D}_s$ threshold, and it originates from the scalar $P$-wave charmonium state. Compared with the original analysis carried out in Ref.~\cite{Prelovsek:2020eiw}, the number of poles remains the same though the third pole becomes much heavier than that reported in Ref.~\cite{Prelovsek:2020eiw}.

Among the 4 different fitting scenarios, 
Fits 3 and 4 used more data, and the parameters should be better constrained. Thus, the corresponding results can be regarded as our main ones. The difference between the results from Fits 3 and 4 should account for theoretical uncertainties, as it stems from the remaining scale dependence after having absorbed the leading UV divergence by the contact terms. 
However, we also note the $\chi^2/{\rm d.o.f.}$ values for Fits 3 and 4 are sizeably larger than 1. Whether this can be improved needs to be further investigated.

In the $1^{++}$ sector,  we have adjusted the LECs to reproduce the isoscalar $D\olsi{D}{}^\ast$ energy levels obtained in the LQCD simulation carried out in Ref.~\cite{Prelovsek:2013cra}. With the LECs thus fixed, we have searched for poles in the infinite volume, revealing two poles in the complex-energy plane. The first pole is below the $D\olsi{D}{}^\ast$ threshold and its compositeness exceeds $90\%$, indicating a negligible charmonium contribution. This pole corresponds to the $X(3872)$ which is identified as a shallow bound state. The second pole, located above the $D \olsi{D}{}^\ast$ threshold, originates from the $\chi_{c1}(2P)$ charmonium state and is likely associated with the $\chi_{c1}(4010)$ recently reported by LHCb~\cite{LHCb:2024vfz}. 

With the use of the LECs determined in the $1^{++}$ sector, we have tuned the (unknown) $\chi_{c2}(2P)$ charmonium bare mass and found two poles in the $2^{++}$ sector. One pole close to $D^*\olsi{D}{}^*$ threshold would be a $D^* \olsi{D}{}^*$ shallow bound state, the HQSS partner of the $X(3872)$, while the second pole, originating from the $\chi_{c2}(2P)$ charmonium, is located below the first one and corresponds to the $\chi_{c2}(3930)$. 

In the $1^{++}$ and $2^{++}$ sectors, our investigations suggest the intertwined presence of both hadronic bound states and charmonium resonances in these three sectors.
This is compatible with View III in Ref.~\cite{Hanhart:2019isz} that amounts to the coexistence of hadronic molecular and quark model states. Our exploratory study of these sectors offers valuable insights into their dynamics, but given that the fits that we carry out are underconstrained, more lattice data are required to draw robust conclusions.

\vspace{-10pt}

\begin{acknowledgments}
We acknowledge the work of Pedro Fern\'andez-Soler in early stages of this work. We are grateful to Sasa Prelovsek for sharing the lattice data. This work is supported by the Spanish Ministerio de Ciencia e Innovaci\'on (MICINN) under contracts PID2020-112777GB-I00, PID2023-147458NB-C21 and CEX2023-001292-S; by Generalitat Valenciana under contracts PROMETEO/2020/023 and  CIPROM/2023/59. This work is also supported in part by the Chinese Academy of Sciences (CAS) under Grant No. XDB34030000 and No. YSBR-101; by the National Natural Science Foundation of China (NSFC) under Grants No. 12125507, No. 12361141819, and No. 12047503; and by the National Key R\&D Program of China under Grant No. 2023YFA1606703. M.\,A. acknowledges financial support through GenT program by Generalitat Valencia (GVA) Grant No.\,CIDEGENT/2020/002, Ramón y Cajal program by MICINN Grant No.\,RYC2022-038524-I, and Atracción de Talento program by CSIC Grant No.\, PIE 20245AT019.
\end{acknowledgments}

\appendix

\section{\boldmath Correlation matrix for the $D\olsi{D}$--$D_s\olsi{D}_s$ coupled channels}\label{sec:appendix}

In this appendix, we list the correlation matrix for the $D\olsi{D} - D_s \olsi{D}_s$ coupled channel system for the four fitting scenarios explored in this work. The LECs are listed in Table \ref{tab:LEC-value}. For Fit 1 through Fit 4, the symmetric correlation matrices are, in order: 
\begin{equation}
 \begin{matrix}
 & \mcos &  d  &  C_{0a}  &  C_{1a}  \\
\mcos \ldelim[{4}{0.1cm} & 1.00  & 0.29 & -0.17 & -0.12 & \rdelim]{4}{0.1cm}\\
d & 0.29   & 1.00  & 0.79  & -0.00 \\
C_{0a} & -0.17   & 0.79     & 1.00   & 0.12 \\
C_{1a} & -0.12   & -0.00   & 0.12      & 1.00 \\
\end{matrix}\,,
\end{equation}
\begin{equation}
\begin{matrix}
 & \mcos &  d  &  C_{0a}  &  C_{1a}  \\
\mcos \ldelim[{4}{0.1cm} & 1.00  & 0.44 & -0.00 & -0.33 & \rdelim]{4}{0.1cm}\\
d  & 0.44   & 1.00  & 0.84  & -0.18\\
C_{0a} & -0.00   &  0.84    & 1.00   & 0.08\\
C_{1a} & -0.33   & -0.18     & 0.08      & 1.00 \\
\end{matrix} ~,
\end{equation}
\begin{equation}
\begin{matrix}
& \mcos &  d  &  C_{0a}  &  C_{1a}  \\
\mcos \ldelim[{4}{0.1cm} & 1.00  & 0.41 & 0.25 & -0.01 & \rdelim]{4}{0.1cm}\\
d  & 0.41    & 1.00  & 0.89  & 0.21\\
C_{0a}  & 0.25    & 0.89     & 1.00   & 0.24\\
C_{1a}  & -0.01    & 0.21    & 0.24      & 1.00 \\
\end{matrix}\,,
\end{equation}
\begin{equation}
\begin{matrix}
& \mcos &  d  &  C_{0a}  &  C_{1a}  \\
\mcos \ldelim[{4}{0.1cm}
 & 1.00  & 0.57 & 0.13 & -0.18 & \rdelim]{4}{0.1cm}\\
d  & 0.57    & 1.00  & 0.81  & -0.10\\
C_{0a}  & 0.13    & 0.81     & 1.00   & 0.02\\
C_{1a}  & -0.18    & -0.10     & 0.02      & 1.00 \\
\end{matrix}~.
\end{equation}

\xpatchcmd\bibsection{19}{9}{}{}
\xpatchcmd\bibsection{\begingroup}{\vskip30pt\begingroup}{}{}

\bibliography{refs}

\end{document}